\documentclass[preprint,aoas]{imsart}
\usepackage{amsthm,amsmath,amssymb,bm}
\usepackage{natbib}
\RequirePackage[colorlinks,citecolor=blue,urlcolor=blue]{hyperref}
\usepackage{graphicx,psfrag,epsf}
\usepackage{enumerate}
\usepackage{longtable,lscape}
\usepackage{soul,xcolor}
\usepackage{subfigure}
\usepackage{float}
\usepackage{url} 
\usepackage{rotating,titlesec}
\usepackage{algorithm}
\usepackage[noend]{algpseudocode}
\newcounter{algsubstate}
\renewcommand{\thealgsubstate}{\alph{algsubstate}}
\newenvironment{algsubstates}
{\setcounter{algsubstate}{0}%
	\renewcommand{\State}{%
		\stepcounter{algsubstate}%
		\Statex {\footnotesize\thealgsubstate:}\space}}
{}

\startlocaldefs
\def\0{\mbox{\boldmath{$\mathbf{0}$}}}
\def\1{\mbox{\boldmath{$\mathbf{1}$}}}

\def\bbeta{\mbox{\boldmath$\beta$}}

\def\bepsilon{\mbox{\boldmath$\epsilon$}}

\def\bGamma{\mbox{\boldmath$\Gamma$}}

\def\bSigma{\mbox{\boldmath$\Sigma$}}

\def\calF{\mbox{$\mathcal{F}$}}
\def\calH{\mbox{$\mathcal{H}$}}

\def\I{\mbox{\boldmath{$\mathbf{I}$}}}

\def\X{\mbox{\boldmath{$\mathbf{X}$}}}

\def\v{\mbox{\boldmath{$\mathbf{v}$}}}

\def\y{\mbox{\boldmath{$\mathbf{y}$}}}

\endlocaldefs

\begin{document}
	
	\begin{frontmatter}
		
		\title{RADIOHEAD: Radiogenomic Analysis Incorporating Tumor Heterogeneity in Imaging Through Densities}
		\runtitle{RADIOHEAD: Radiogenomic Analysis Through Densities}
		
		
		\begin{aug}
			\author{\fnms{Shariq} \snm{Mohammed}\thanksref{t1,m1}\ead[label=e1]{shariqm@umich.edu}},
			\author{\fnms{Karthik} \snm{Bharath}\thanksref{m2}\ead[label=e2]{karthik.bharath@nottingham.ac.uk}},
			\author{\fnms{Sebastian} \snm{Kurtek}\thanksref{m3}\ead[label=e3]{kurtek.1@stat.osu.edu}},
			\author{\fnms{Arvind} \snm{Rao}\thanksref{t1,m1}\ead[label=e4]{ukarvind@umich.edu}},
			\and
			\author{\fnms{Veerabhadran} \snm{Baladandayuthapani}\thanksref{t1,m1}
				\ead[label=e5]{veerab@umich.edu}}
			
			\thankstext{t1}{Corresponding authors \url{{shariqm, ukarvind, veerab}@umich.edu}}
			\runauthor{S. Mohammed et al.}
			
			\affiliation{University of Michigan\thanksmark{m1}, University of Nottingham\thanksmark{m2} and Ohio State University\thanksmark{m3}}
			
			\address{Address of Shariq Mohammed\\
				Department of Biostatistics\\
				Department of Computational Medicine \& Bioinformatics\\
				100 Washtenaw Ave\\
				Palmer Commons Building, Room 2035\\
				Ann Arbor, MI 48103, USA.\\
				\printead{e1}}
			
			\address{Address of Karthik Bharath\\
				School of Mathematical Sciences\\
				Room C13\\
				University Park\\
				Nottingham,	NG7 2RD, UK.\\
				\printead{e2}}
			
			\address{Address of Sebastian Kurtek\\
				Department of Statistics\\
				404 Cockins Hall\\
				CH 440B\\
				Columbus, OH 43210, USA.\\
				\printead{e3}}
			
			\address{Address of Arvind Rao\\
				Department of Biostatistics\\
				Department of Computational Medicine \& Bioinformatics\\
				100 Washtenaw Ave\\
				Palmer Commons Building, Room 2035\\
				Ann Arbor, MI 48103, USA.\\
				\printead{e4}}
			
			\address{Address of Veerabhadran Baladandayuthapani\\
				Department of Biostatistics\\
				Department of Computational Medicine \& Bioinformatics\\
				1415 Washington Heights\\
				School of Public Health I, Room 4622\\
				Ann Arbor, MI 48103, USA.\\
				\printead{e5}}
		\end{aug}
		
		\begin{abstract}
			Recent technological advancements have enabled detailed investigation of associations between the molecular architecture and tumor heterogeneity, through multi-source integration of radiological imaging and genomic (radiogenomic) data. In this paper, we integrate and harness radiogenomic data in patients with lower grade gliomas (LGG), a type of brain cancer, in order to develop a regression framework called RADIOHEAD (RADIOgenomic analysis incorporating tumor HEterogeneity in imAging through Densities) to identify radiogenomic associations. Imaging data is represented through voxel intensity probability density functions of tumor sub-regions obtained from multimodal magnetic resonance imaging, and genomic data through molecular signatures in the form of pathway enrichment scores corresponding to their gene expression profiles. Employing a Riemannian-geometric framework for principal component analysis on the set of probability densities functions, we map each probability density to a vector of principal component scores, which are then included as predictors in a Bayesian regression model with the pathway enrichment scores as the response. Variable selection compatible with the grouping structure amongst the predictors induced through the tumor sub-regions is carried out under a group spike-and-slab prior. A Bayesian false discovery rate mechanism is then used to infer significant associations based on the posterior distribution of the regression coefficients. Our analyses reveal several pathways relevant to LGG etiology (such as synaptic transmission, nerve impulse and neurotransmitter pathways), to have significant associations with the corresponding imaging-based predictors.
		\end{abstract}
		
		
		
	\end{frontmatter}
	
	\section{Introduction}\label{sec: intro}
	Gliomas are a group of tumors occurring in the brain and spinal cord, further categorized into sub-groups. Lower grade gliomas (LGG) are characterized as World Health Organization grade II and III tumors, and they come from two different types of brain cells known as astrocytes and oligodendrocytes. The causes of these types of tumors are not well understood, and recent studies have examined their molecular characterization from datasets generated by The Cancer Genome Atlas (TCGA), and have associated disease prognosis with their underlying molecular architecture \citep{verhaak2010integrated}. In the context of gliomas, there has been growing interest in exploring the underlying comprehensive molecular characterization \citep{noushmehr2010identification,verhaak2014comprehensive,venneti2015evolving,fishbein2017comprehensive}. For example, \cite{ceccarelli2016molecular} studied the complete set of genes associated with diffuse grade II-III-IV gliomas from TCGA, to identify molecular correlations by comprehensively analyzing the sequencing and array-based molecular profiling data, and to improve disease classification and provide insights into the progression of the tumor from low- to high-grade.
	
	Gliomas usually contain various heterogeneous sub-regions: edema, non-enhancing and enhancing core, which reflect differences in tumor biology, have variable histologic and genomic phenotypes, and exhibit highly variable clinical prognosis \citep{bakas2017advancing}. This intrinsic heterogeneity in tumor biology is also reflected in their radiographic phenotypes through different intensity profiles of the sub-regions in imaging. Such phenotypes can be obtained from images based on computed tomography (CT), positron emission tomography (PET) and magnetic resonance imaging (MRI), each of which allows integration with other data sources (e.g., genomics). Moreover, imaging and genomic data provide complementary information in terms of tumor heterogeneity and molecular characterization, respectively. Molecular classification of LGGs can be facilitated, and sometimes even validated, through radiogenomic analyses based on non-invasive medical image-derived features. Imaging features have been known to capture physiological and morphological heterogeneity of tumors as they progress from a single cell \citep{marusyk2012intra}. Such studies have an important bearing on the design of personalized therapeutic strategies in cancer, and potentially guide monitoring of disease development or progression for early stage cancers. Thus, examination of inter- and intra-tumor heterogeneity through imaging features, and their potential association with genomic markers, can lead to a better understanding of molecular signatures of LGGs.
	
	In this work, we focus on the MRI modality as it furnishes a wide range of image contrasts at a high resolution, which can be used to exhibit and evaluate the location, growth and progression of tumors. Moreover, improved resolution of MRIs has facilitated the understanding of different aspects of tumor characteristics \citep{just2014improving}. The apparent utility of MRI in studying heterogeneity of sub-regions of gliomas can be seen in Figures \ref{fig: mri} and \ref{fig: densities_gray}, where different intensity profiles disseminated across the multimodal MRI scans appear to exhibit complementary information. Studying heterogeneity in the sub-regions is now feasible due to the availability of their gold standard labeling \citep{bakas2015glistrboost}, which facilitates further radiomic and radiogenomic analyses.
	
	\subsection{Voxel Intensity Densities as an Imaging Feature}
	Using the raw MRI scans as predictors in the modelling is a challenge as we do not have an underlying atlas structure to compare between subjects that is commonly available for other imaging modalities such as neuro-imaging studies \citep{ombao2016handbook}. Diagnostic image-based features using voxel-level data have been utilized for modelling purposes (e.g. to visualize the progression/regression of tumors). However, one of the main drawbacks of existing studies is that only a few chosen summary statistics/metrics represent entire regions of interest. Some of these summary statistics include percentiles, extreme percentiles (e.g. 5th and 95th), quartiles, skewness, kurtosis, histographic pattern, range and mode of MRI-based voxel intensity histograms \citep{baek2012percent,just2011histogram,song2013true}. Although such metrics have clear utility in the assessment of tumor heterogeneity, they generally do not provide a comprehensive representation due to (a) the subjectivity in the choice of the number and location of summary features, and (b) the limitation of these features in terms of capturing the entire information in a voxel intensity distribution. As a result, any statistical analysis based on such an approach is unable to detect potential small-scale and sensitive changes in the tumor due to treatment effects \citep{just2014improving}.
	
	As an alternative to summary statistics, associations between genomic variables and tumor heterogeneity can be examined on different scales of the voxel intensity probability density function (PDF): while significant genomic variation might manifest as markedly distinct aspects of a PDF (e.g., number of modes, large changes in location of mean/mode), genomic variation (relative to the measurement scale) might show up in subtle, small-scale changes in overall shape of the PDF (e.g. slopes between modes), and sometimes in the tails. Indications of such a behavior were evident in an unsupervised clustering setting in earlier work that considered entire voxel intensity PDFs as data objects \citep{saha2016demarcate}. Including such small-scale changes without summarizing the entire PDF through coarse summary statistics could result in better correlative and predictive power of models associating genomic variables to radiographic phenotypes \citep{yang2020quantile}.
	
	In this article, we propose to examine variations in the genomic signature of a tumor through changes, both large and subtle, in overall shape\footnote{`shape' is used in a non-technical sense} of the PDF of voxel intensities, using a Riemannian-geometric framework on the space of PDFs. This space is a nonlinear, infinite-dimensional manifold, and the lack of a global linear structure brings about non-trivial challenges in their analyses. Here, we develop a regression framework called RADIOHEAD (RADIOgenomic analysis incorporating tumor HEterogeneity in imAging through Densities) to model associations between genomic variables characterizing the molecular signature of tumors and voxel intensity PDFs from multimodal MRI scans. In what follows, we use PDFs and densities interchangeably.
	
	\subsection{RADIOHEAD Modelling Outline}
\begin{figure}[!t]
		\centering
		\resizebox{\textwidth}{!}{
			\includegraphics[trim=1.5cm 1.9cm 1.5cm 2cm, clip, scale=0.5]{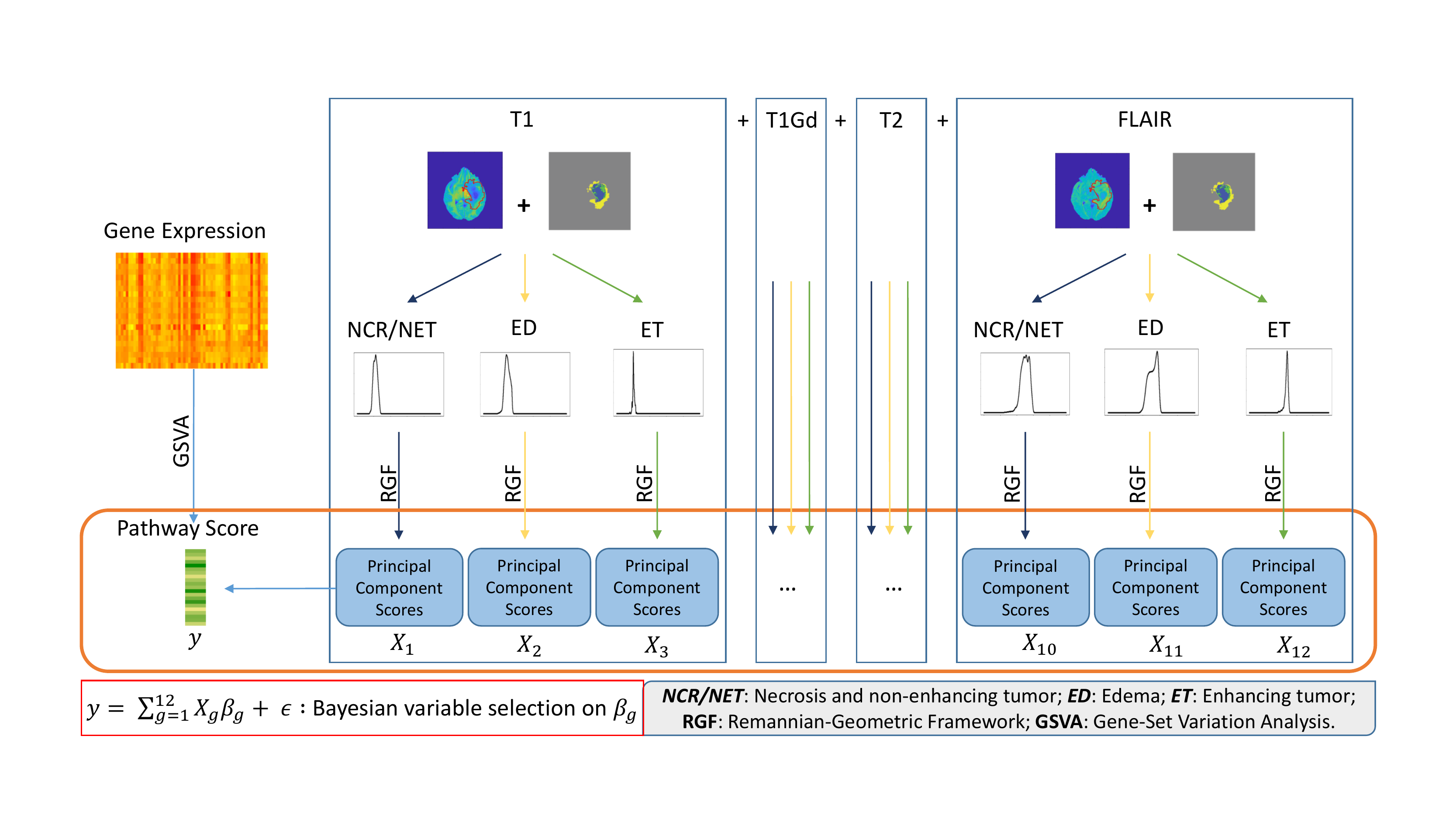}
		}
		\caption{Schematic representation of the RADIOHEAD modelling approach. Pathway scores are constructed from gene expression using gene-set variation analysis (GSVA). From each MRI sequence, we construct densities for each of the three tumor sub-regions and use them to construct principal component scores under a Reimannian-geometric framework. Pathway scores are used as a response and the principal component scores as predictors in the downstream analysis.}
		\label{fig: workflow}
	\end{figure}

	We propose an integrated end-to-end method: from MR images to evaluation of voxel-level density-based radiomic features, gene expression to associated pathway-level enrichment, and subsequent statistical modelling framework. Figure \ref{fig: workflow} shows the schematic workflow diagram for our method. For each patient, we generate PDFs corresponding to three heterogeneous tumor sub-regions: (i) necrosis and non-enhancing, (ii) edema, and (iii) enhancing core.  The expression/activation of the pathways is evaluated by computing pathway enrichment scores through gene-set variation analysis (GSVA); these scores are subsequently used as a univariate response variable. We apply the proposed RADIOHEAD approach to the TCGA dataset of LGGs. 
	
	Fitting a model by regressing enrichment scores against multiple PDFs (one from each combination of tumor sub-region and MRI sequence) poses two main challenges:
	\begin{enumerate}
		\item Each PDF is a non-negative function which integrates to one, and hence cannot be treated as a standard functional predictor;
		\item The grouping structure between tumor sub-regions needs to be incorporated while examining the functional relationship between a pathway score and its corresponding PDFs.
	\end{enumerate}
	The first challenge is addressed by mapping each PDF to a finite-dimensional vector of principal component (PC) scores by carrying out Riemannian principal component analysis (PCA) on the sample of PDFs corresponding to each tumor sub-region. These PC scores corresponding to the multiple PDFs act as imaging meta-features and are incorporated as individual predictors, which leads to a $p\gg n$ situation wherein the number of radiomic meta-features ($p$) is higher than the number of subjects ($n$). In the presence of uncertainty in the actual effects of small changes in the PC scores on the enrichment scores, it is natural to employ a Bayesian model for variable selection. To this end, we address the second challenge by using a group-structured continuous spike-and-slab prior \citep{ishwaran2005spike,andersen2014bayesian} on the total set of PC scores, in an effort to capture information on the biological structure in the data, and to provide analyses that are more amenable to interpretation. The prior formulation also simplifies the computation by allowing for simple (conditional posteriors from standard distributions) and fast MCMC sampling (via Gibbs sampling). Other existing prior formulations incorporating group structure \citep{zhang2014bayesian,xu2015bayesian,yang2018consistent} could also be used. Furthermore, to address the issue of multiple comparisons, a Bayesian false discovery rate-based approach is used to build inference based on error rates.
	
	Section \ref{sec: data} describes the data along with the acquisition process and pre-processing steps. We describe the algorithm to compute the density-based PC scores in Section \ref{subsec: density}; the computation of GSVA-based enrichment scores is outlined in Section S2 of the supplementary material. Section \ref{subsec: regression} describes the regression setup with densities as covariates. In Section \ref{subsec: regression_pc}, we describe the regression in terms of PC scores and the modelling approach based on Bayesian variable selection using the group spike-and-slab prior. The estimation and inference strategies follow in Sections \ref{subsec: estimation} and \ref{subsec: fdr}. In Section \ref{sec: results}, we present our results and describe the identified radiogenomic associations in LGG. We close with a brief discussion and some directions for future work in Section \ref{sec: discussion}.
	
	\section{Dataset Description}\label{sec: data}
	We describe the data acquisition and pre-processing steps involved for the imaging and genomic data separately.
	
	\subsection{Imaging Data}\label{subsec: imaging}
	To conduct our analyses, we use MRI scans that include reliable tumor segmentations along with identified tumor sub-regions. We consider pre-operative multi-institutional scans in the TCGA LGG collection, publicly available in The Cancer Imaging Archive (TCIA - \cite{clark2013cancer}). We obtain segmentation labels for these MRI scans using an automated method called GLISTRboost \citep{bakas2015glistrboost,bakas2017advancing}. Segmentation labels generate a mask for each subject's MRI scan, which distinguishes between necrotic and non-enhancing tumor  (NCR/NET or NC), peritumoral edema (ED) and enhancing tumor (ET).
	
	\begin{figure}[!t]
		\centering
		\includegraphics{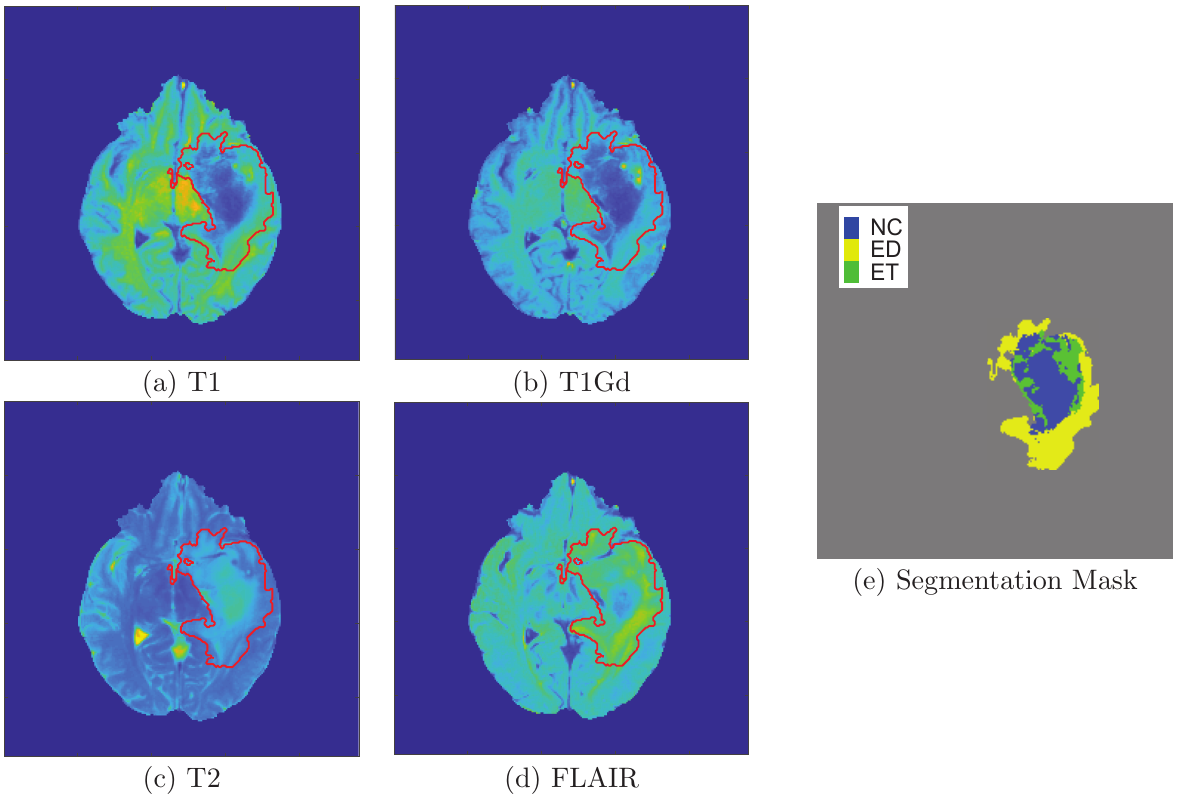}
		\caption{Figures (a)-(d): Axial slice of a skull-stripped brain MRI for a subject with LGG, shown for the four sequences T1, T1Gd, T2 and FLAIR, respectively. The segmented tumor region is displayed using a red boundary overlaid on the images. Figure (e): The corresponding sub-region segmentation mask with the NC, ED and ET regions marked in different colors.}
		\label{fig: mri}
	\end{figure}
	
	MRI provides a wide range of imaging contrasts through multimodal images. The primary MRI sequences include (a) native (T1), (b) post-contrast T1-weighted (T1Gd), (c) T2-weighted (T2), and (d) T2 fluid attenuated inversion recovery (FLAIR). Each of these sequences identifies different types of tissue and displays them using varying contrasts based on the tissue characteristics. We use LGG data for 65 subjects, obtained from \cite{bakas2017lggsegmentation}, which contain (a) MRI scans based on all four sequences (T1, T1Gd, T2 and FLAIR), and (b) corresponding segmentation masks generated by GLISTRboost.
	
	The structure of the data under study is as follows: each MRI scan is a three-dimensional array with the third axis representing different axial slices. For each subject, we have four sequences, as described above, corresponding to four different 3D arrays accompanied by a unique segmentation mask that has a one-to-one correspondence with the voxels in the MRI scans. That is, there is a voxel-to-voxel correspondence across all four MRI sequences and the segmentation mask. An example of a single axial slice from a brain MRI for a subject with LGG, for the four aforementioned sequences, is shown in the left panel in Figure \ref{fig: mri}. The segmented tumor region is indicated by a red boundary overlaid on the images, and is further classified into the tumor sub-regions NC, ED and ET, as shown in the right panel in Figure \ref{fig: mri}. The voxel intensity values of MRI scans are difficult to interpret and compare as they are sensitive to the configuration of the MRI scanner. These values are not comparable either between study visits within a single subject or across different subjects, which necessitates pre-processing of the images in terms of intensity value normalization. We address this issue through a biologically motivated normalization technique using the \texttt{R} package \emph{WhiteStripe} \citep{shinohara2014statistical}.
	
	\subsection{Genomic Data}\label{subsec: genomic}
	The genomic data was obtained from LinkedOmics\footnote{\url{www.linkedomics.org}} \citep{vasaikar2017linkedomics}, which is a publicly available portal that includes multi-omics data for LGG among many other cancer types. We consider the normalized gene-level RNA sequencing data from the primary solid tumor tissue using the Illumina HiSeq system (high-throughput sequencing) with expression values in $\log_2$ scale. The entire dataset contains gene expression data for 516 samples and 20086 genes; we consider a subset of 65 matched samples corresponding to the imaging data described in Section \ref{subsec: imaging}. We consider the enriched pathways in LGG as identified by \cite{ceccarelli2016molecular}, hereafter referred to as \emph{C-Pathways}.
	
	We obtain the mapping from genes to pathways and use them along with the gene expression data to obtain \emph{pathway scores}. These scores are numerical estimates of the relative enrichment of a pathway of interest across a sample population using a non-parametric, unsupervised method called GSVA. It estimates a value per sample and pathway for the variation in the activity of a pathway within an entire gene expression set. In other words, it assesses the relative variability of gene expression in the pathway as compared to expression of genes not in the pathway. The computation details of the pathway scores can be found in Section S2 of the supplementary material. For the C-Pathways (such as ion transport and synaptic transmission) considered in this paper, the genes to pathway mappings are obtained from the molecular signature database \citep{liberzon2011molecular}. For each gene-set within the collection, we construct the pathway score using gene-set variation analysis \citep{hanzelmann2013gsva}. Of the 22 C-Pathways, we only include 21 of them as the gene membership for one of the pathways was not available. The pathway scores are computed using the \emph{GSVA} package in \texttt{R} obtained from Bioconductor \citep{gentleman2004bioconductor}. Summary statistics for the pathway scores are shown in Table S1 of the supplementary material.
	
	\section{Statistical Framework}\label{sec: statistical}
	
	Our main goal is to identify associations between imaging meta-features and gene expression-based pathway scores. In this section, we first describe the Riemannian-geometric approach to construct the voxel PDF-based PC scores for each subject corresponding to a certain tumor sub-region. We also define a formal regression model based on the group spike-and-slab prior as well as associated estimation and variable selection procedures.
	
	\subsection{Density-based Principal Component Scores}\label{subsec: density}
\begin{figure}[!t]
		\centering
		\resizebox{\textwidth}{0.814\textheight}{
			\includegraphics{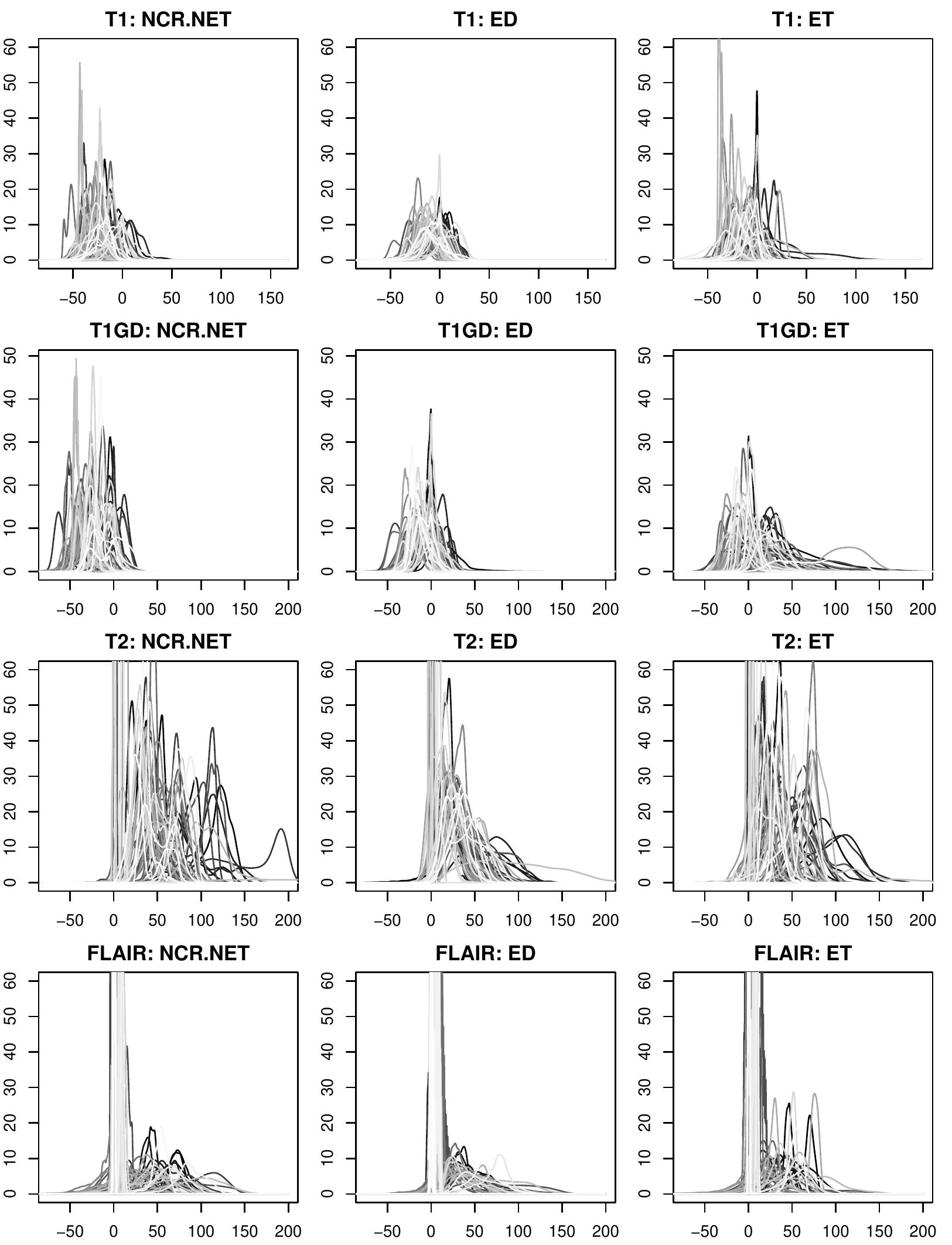}
		}
		\caption{Kernel densities $f_i^M(R)$ for all subjects across all four MRI sequences and three tumor sub-regions. For visual convenience, the $y$-axes are truncated for each of the subplots. The x-axis shows the voxel-intensity values; however, we transform them to $[0,1]$ for each imaging sequence to compute the KDEs. Supplementary Figure S1 shows similar plots in color and supplementary Figure S2 shows similar plots without truncation of the $y$-axis.}
		\label{fig: densities_gray}
	\end{figure}

	We use $R$ to index tumor sub-regions and $M$ for the different MRI sequences. Consider MRI scans for $n$ subjects from four sequences with the tumor masks containing the segmented tumor region and indicating the sub-regions. For a given sequence $M$, we construct the kernel density estimate $f^M_{i}(R),\ i=1,\ldots,n$ for the tumor sub-region $R$ in subject $i$ based on the voxel intensity values in the MRI scan at the array locations of region $R$ obtained from the segmentation. Hence, for each subject $i$ and each sequence $M$, we have PDF estimates denoted by $f^M_{i}(\text{NC}),\ f^M_{i}(\text{ET})$ and $f^M_{i}(\text{ED})$ corresponding to the necrotic and non-enhancing tumor core (NC), the peritumoral edema (ED) and the enhancing tumor (ET) sub-regions, respectively. Thus, we consider univariate kernel-density estimates for all tumor sub-regions and all subjects across the four imaging sequences. The density plots are displayed in Figure \ref{fig: densities_gray}, where each row corresponds to a specific imaging sequence while each column corresponds to a tumor sub-region. We compute the PC scores for each sequence $M$ separately. For brevity, we shall drop the sequence indicator $M$ from the densities and use $f_{iR}$ instead of $f^M_{i}(R)$ for the remainder of this section.
	
	The kernel density estimates ($f_{iR}$ for all $i=1,\ldots,n$ and $R \in \mathcal{T} = \{NC, ET, \allowbreak ED\}$) are proper PDFs and belong to the Banach manifold of all PDFs. The following description focuses on PDFs with domain $[0,1]$; however, the methods apply to more general domains with small adjustments. PDFs are elements of the space $\calF = \Big\{f:[0,1] \rightarrow \mathbb{R}_{>0} \Big| \int_0^1 f(x) dx = 1 \Big\}$. To make $\calF$ a Riemannian manifold, and to facilitate computation on this space, we endow it with the Fisher-Rao (F-R) Riemannian metric \citep{rao1992information,kass2011geometrical,srivastava2007riemannian}. For brevity, we omit the specific formula for this metric and simply mention that it is closely related to the Fisher information matrix and has useful statistical properties, e.g., invariance to bijective and smooth transformations of the PDF domain \citep{cencov1982statistical}. Unfortunately, the F-R metric is difficult to use in practice as the computation of geodesic paths and distances between PDFs is cumbersome and requires numerical methods for approximation. Thus, for simplification, we further transform the kernel density estimates using a square-root transformation \citep{bhattacharyya1943measure,kurtek2015bayesian}. As a result, the space of PDFs becomes the positive orthant of the unit sphere in $\mathbb{L}^2:=\mathbb{L}^2([0,1])$, the geometry of which is well-known, and the F-R metric flattens to the standard $\mathbb{L}^2$ metric enabling the computation of geodesic paths and distances in analytical form \citep{kurtek2015bayesian}. Briefly, this result provides simple tools for the statistical tasks of interest including (a) definition of a distance between two densities, (b) computation of a Karcher mean of a sample of densities, and (c) PCA of a sample of densities. We elaborate on these procedures next.
	
	\noindent\paragraph{Distances between PDFs and their Karcher Mean.} Let $h_{iR}=+\sqrt{f_{iR}}$ denote the (positive) square-root densities (SRDs) corresponding to the kernel density estimates $f_{iR}$ for all $i=1,\ldots,n$ and $R \in \mathcal{T}$. Each $h_{iR}$ is an element of $\calH = \Big\{h:[0,1] \rightarrow \mathbb{R}_{> 0} \Big| \int_0^1 h^2(x) dx = 1 \Big\}$, the positive orthant of a unit sphere in $\mathbb{L}^2$, i.e., $\calH$ is the collection of SRDs corresponding to all PDFs in $\calF$. Equipped with the standard $\mathbb{L}^2$ metric, $\calH$ becomes a Riemannian manifold (recall that the $\mathbb{L}^2$ metric on $\calH$ corresponds to the F-R metric on $\calF$). Under this setup, the geodesic distance between two densities $f_1,f_2 \in \calF$, represented by their SRDs $h_1,h_2 \in \calH$, is defined as the shortest great circle arc connecting them on $\calH$: $d(f_1,f_2)=d(h_1,h_2)_{L^2}:=\cos^{-1}(\langle h_1,h_2\rangle) = \cos^{-1}(\int_0^1 h_1(x)h_2(x)dx):= \theta$. We can now compute the mean of a sample of SRDs using a generalized version of a mean on a metric space, called the Karcher mean \citep{karcher1977riemannian,dryden1998statistical}. The sample Karcher mean $\bar{h}$ on $\calH$ is defined as the minimizer of the variance functional $\calH \ni h \mapsto \sum_{i=1}^n d(h,h_i)^2_{L^2}$. An algorithm for computing the Karcher mean is given in Section S3 of the supplementary material; Figure S3 shows the Karcher mean of the densities across all of the subjects for all tumor sub-regions and imaging sequences, overlaid within each subplot. The computations require two tools from differential geometry called the exponential and inverse-exponential maps. Let $T_h(\calH)=\Big\{\delta h|\langle\delta h, h\rangle=0\Big\}$ denote the tangent space at $h$. For $h \in \calH$ and $\delta h \in T_h(\calH)$, the exponential map at $h$, $\exp: T_h(\calH) \rightarrow \calH$ is defined as $\exp_h(\delta h) = \cos(\|\delta h\|)h + \sin(\|\delta h\|)\delta h/\|\delta h\|$, where $\|\delta h\|=\sqrt{\int_0^1 \delta h^2(x) dx}$ is the $\mathbb{L}^2$ norm \citep{billioti2017riemannian}. The inverse-exponential map is denoted by $\exp^{-1}_h: \calH \rightarrow T_h(\calH)$ and for any $h_1, h_2 \in \calH$, it is defined as, $\exp^{-1}_{h_1} (h_2) = \theta[h_2 - \cos(\theta)h_1]/\sin(\theta)$, where $\theta = d(h_1,h_2)_{L^2}$ as before\footnote{For the unit sphere in $\mathbb{L}^2$, strictly speaking, although the exponential map is well-defined on the entire tangent space \citep{billioti2017riemannian}, the inverse-exponential map may not be. We eschew handling of this technical detail since this is not an issue when computing using the map in practice.}.
	
	\noindent\paragraph{Principal Component Analyses on a Sample of PDFs.} To perform PCA of a sample of SRDs (equivalently PDFs), we utilize the linear tangent space at the sample Karcher mean SRD. That is, we first project all SRDs onto this tangent space using the inverse-exponential map. The sample covariance matrix is then computed in the tangent space at the mean SRD, and PCA is applied through singular value decomposition (SVD) of this matrix. In practice, the densities and their corresponding SRDs are approximated using $m$-dimensional vectors, which specify the functional values at a set of $m$ discrete points on the domain $[0,1]$ resulting in $m\times m$-dimensional covariance matrices, where $m\gg n$. We describe the above step-by-step process in Algorithm \ref{algo: pca}.
	\begin{algorithm}[h]
		\caption{PCA on $T_{\bar{h}}(\calH)$}\label{algo: pca}
		\begin{algorithmic}[1]
			\State Compute $h_{iR}$ from $f_{iR}$ (at $m$ discrete points).
			\State Compute the Karcher mean of $h_{iR}$ for each tumor sub-region $R \in \mathcal{T}$ as $\bar{h}_R$ (see Section S3 in the supplementary material).
			\State Use the inverse-exponential map to compute $\v_{iR} = \exp^{-1}_{\bar{h}_R} (h_{iR})\ \in T_{\bar{h}_R}(\calH)$.
			\State Evaluate the sample covariance matrix $K_R = \frac{1}{n-1} \sum_{i=1}^n \v_{iR}\v_{iR}^\top \in \mathbb{R}^{m \times m}$ for each $R \in \mathcal{T}$.
			\State Compute the SVD of $K_R = U_R\Sigma_R U_R^\top$.
		\end{algorithmic}
	\end{algorithm}
	
	The first $L$ columns of $U_R$, denoted as $\tilde{U}_R \in \mathbb{R}^{m \times L}$, span the $L$-dimensional principal subspace of the given sample of densities. We can compute the principal coefficients as $X_R = V_R\tilde{U}_R $, where $V_R^\top = [\v_{1R}~ \v_{2R}~ \ldots~ \v_{nR}] \in \mathbb{R}^{m \times n}$ for each $R \in \mathcal{T}$. These principal coefficients $X_R^M$, referred to as PC scores, act as Euclidean coordinates corresponding to the kernel density estimates $f_i^M(R)$ generated from each MRI sequence $M$, and will be used as predictors in our model. This procedure accomplishes two major goals: (1) it estimates orthogonal directions of variability in a sample of PDFs along with the amount of variability explained by each direction via the covariance decomposition, and (2) it performs dimension reduction by effectively exploring variability in the sample of PDFs through the primary modes of variation in the data.
	 
		\subsection{Regression with Densities}\label{subsec: regression}
		The PDFs $f^M_i(R)$ are representations of the heterogeneity in the tumor voxels from the imaging sequence $M$ and the tumore sub-region $R$ for subject $i$. To identify radiogenomic associations, we build regression models with PDFs $f^M_i(R) ~ \forall M,R$ as covariates. For ease of exposition, we drop the indices $M$ and $R$, and explicate the model for one density $f_i(t)$ for $t \in [0,1]$ for subject $i$ as the covariate. Let $h_i(t)$ denote the corresponding SRD. If $y_i$ corresponds to the pathway score for subject $i$, $h_i$ can be related to $y_i$ using the data-driven model
		\begin{equation}
		y_i = \beta_0 + \int_0^1\exp_{\bar{h}}^{-1}\big(h_i(t)\big)\beta(t)dt+\epsilon_i, \quad i=1,\ldots,n,
		\label{eq: density_model}
		\end{equation}
		where $\bar{h}\in \calH$ is the Karcher mean of SRDs $h_1,\ldots,h_n \in \calH$. Here $t \mapsto \beta(t)$ is the real-valued coefficient function, $\beta_0$ is a real-valued intercept, and $\epsilon_i$ are i.i.d. $N(0,\sigma^2)$. We specify the model on the tangent space at the data-dependent Karcher mean $\bar h$. That is, $\bar h$ is the reference SRD for the inverse-exponential map. While this choice influences the model specification (as $\bar h$ changes with changing sample composition), it removes the arbitrariness associated with choosing the reference SRD. Effectively, $\exp_{\bar{h}}^{-1}\big(h_i(t)\big)$ is the Riemannian-geometric equivalent of `centering' the functional covariate $h_i$. The amount of dependence of the model on $\bar h$ is directly dependent on the variability of the sample $h_1,\ldots,h_n$, which can be quantified using the geodesic distances between $h_i$ and $\bar{h}$.
		
		When it exists, the range of $\calH \ni h \mapsto \exp_{\bar{h}}^{-1}(h)$ is contained within a linear subspace of $\mathbb{L}^2$, and we can thus express $\exp_{\bar{h}}^{-1}\big(h_i\big)=\sum_k \alpha_{ik} \phi_k$ for some sequence $(\alpha_{ik},\ k \geq 1)$ with $\sum_k|\alpha_{ik}|^2 < \infty$, where $\{\phi_k,\  k=1,2,\dots\}$ is an orthonormal set of basis functions for $\mathbb{L}^2$. Similarly, we can write $\beta=\sum_k \beta_k \phi_k$ for some sequence $(\beta_k,\ k \geq 1)$ with $\sum_k|\beta_k|^2 < \infty$. Hence, the model in Equation (\ref{eq: density_model}) reduces to
		\begin{equation}
		y_i = \beta_0 + \sum_{k=1}^\infty \alpha_{ik} \beta_k + \epsilon_i,
		\label{eq: simplified_model}
		\end{equation}
		since $\langle \phi_i,\phi_j \rangle = 1$ if $i=j$, and 0 otherwise. For a given gene-set, we denote the pathway scores as $\y = (y_1,\ldots,y_n)^\top$, where $y_i$ corresponds to the score for subject $i$. Having chosen $\bar h$, we truncate the number of basis functions at some positive integer $r_n< \infty$. The model in Equation (\ref{eq: simplified_model}) is then further simplified as
		\begin{equation}
		\y_{n \times 1} = \beta_0 \1_n+ A \bm \beta +\bm \epsilon,
		\label{eq: simplified_model_2}
		\end{equation} 
		where $\1_n \in \mathbb{R}^n$ is the vector with all entries as 1, row $i$ of $A \in \mathbb{R}^{n \times r_n}$ is given as $(\alpha_{i1},\ldots,\alpha_{ir_n})^\top \in \mathbb{R}^{r_n}$ and $\bbeta = (\beta_1,\ldots,\beta_{r_n})^\top\in \mathbb{R}^{r_n}$. Let $A^\top A=PDP^\top$, where $P\in \mathbb{R}^{r_n \times r_n}$ is an orthogonal matrix of eigenvectors of $A^\top A$ and $D$ is diagonal with $\lambda_1 \geq \lambda_2 \geq \cdots\lambda_{s_n}>0=\lambda_{s_n+1}= \cdots \lambda_{r_n}$. If every $\lambda_k>0\ \forall\  k$, then $\bm y$ is regressed on the principal components of $A$, which is $AP$.
		
		The model in Equation (\ref{eq: simplified_model_2}) depends on the choice of the orthonormal basis $\{\phi_k\}$ of $\mathbb{L}^2$, or in other words the matrix $A$ and its eigenvectors in $P$. We use a PC basis for two reasons:
		(i) it is the optimal empirical orthogonal basis (see e.g., Chapter 6 of \cite{ramsay2005functional}) for data on $\mathbb{L}^2$, of which $\mathcal{T}_{\bar h}(\calH)$ is a linear subspace, and (ii) the map $\exp_{\bar{h}}^{-1}\big(h_i\big)\mapsto  (\alpha_{i1},\alpha_{i2},\dots)$ is an isometry, and for a fixed positive integer $r_n$, the corresponding full isometry group is $O(r_n)$ (the set of square orthogonal matrices in dimension $r_n$). From the perspective of (ii), choosing another orthornormal basis and truncating at $r_n$ amounts to an orthogonal transform of the corresponding coefficients. Thus, we are effectively regressing the score $y_i$ of the $i$th subject on the `optimal' $r_n$-dimensional linear representation of the SRD $h_i$ in the tangent space $\mathcal{T}_{\bar h}(\calH$) of the sample Karcher mean $\bar h$.
		
		The model in Equation (\ref{eq: simplified_model_2}) corresponds to one PDF as a covariate for each subject $i$. However, from our imaging data, we have twelve PDFs, from four imaging sequences and three tumor sub-regions, as covariates for each subject. Hence, the model in Equation (\ref{eq: density_model}) can be extended as 
		\begin{equation}
		y_i = \beta_0 + \sum\limits_M \sum\limits_R \int_0^1\exp_{\bar{h}_R^M}^{-1}\big(h_{iR}^M(t)\big)\beta_M^R(t)dt+\epsilon_i,
		\label{eq: density_model_2}
		\end{equation}
		where $h_{iR}^M(t)$ is the SRD for the PDF $f_{iR}^M(t)$, and $\beta_M^R(t)$ is the coefficient function corresponding to the tumor sub-region $R$ in imaging sequence $M$. Here $\bar{h}^M_R$ is the sample Karcher mean of $h_{1R}^M(t), \ldots, h_{nR}^M(t)$. Each of the integrals in Equation (\ref{eq: density_model_2}) can be reduced to the PC regression form in Equation (\ref{eq: simplified_model_2}). In Section \ref{subsec: regression_pc}, we directly work with the PC regression form with the twelve groups of PCs as covariates.
		
		\subsection{Regression with PC Scores}\label{subsec: regression_pc}
		The PDFs belong to a function space and they carry rich information of the voxel density of different tumor sub-regions at different scales. As a consequence, they also result in a large number (greater than the number of subjects) of principal components across sequences and tumor sub-regions. This $p\gg n$ situation necessitates the use of variable selection approaches that can induce sparsity as well as regularization. As the PC scores are surrogates for the entire density, it is natural to model the aspects of the density not captured through the scores (such as information on the tumor sub-regions) using a Bayesian approach by appropriately placing a prior on the high-dimensional feature space. Consequently, this allows us to construct and assess posterior distributions of coefficients for inference.
	
	Our goal is to identify the density-based principal components across tumor sub-regions that are significantly associated with the expression levels in the gene-set considered. We address this problem from a Bayesian perspective and use the continuous spike-and-slab prior \citep{george1997approaches, ishwaran2005spike}, which has inherent variable selection properties. We model the pathway scores $\y$ using principal component scores obtained from all of the tumor sub-regions and MRI sequences as the predictors. In other words, we assume
	\begin{equation}
	\y = \X\bbeta + \bepsilon \text{ with } \bepsilon \sim N(\0,\sigma^2\I),
	\label{eq: model_y}
	\end{equation}
	where 
	\begin{eqnarray*}
		\X & = & \Big[X^{T1}_{NC} ~X^{T1}_{ED} ~X^{T1}_{ET} ~X^{T1Gd}_{NC} ~X^{T1Gd}_{ED} ~X^{T1Gd}_{ET} \\
		& & \qquad \qquad \qquad \qquad ~X^{T2}_{NC} ~X^{T2}_{ED} ~X^{T2}_{ET} ~X^{FLAIR}_{NC} ~X^{FLAIR}_{ED} ~X^{FLAIR}_{ET}\Big]
	\end{eqnarray*} 
	corresponds to the $n \times L$ matrix of predictors containing the principal component scores. The normality assumption is reasonable here since the pathway scores are unimodal and approximately normal by construction \citep{hanzelmann2013gsva}. The model can also be adapted to categorical or survival response types by incorporating latent variable approaches. Here, $L$ is defined as the total number of principal components considered across all sequences and tumor sub-regions: $L = \sum_M\sum_R L^M_R$, where $L^M_R$ corresponds to the number of columns in $X^M_R$ for $R \in \mathcal{T}$ and $M$ belongs to the four different sequences. We choose $L^M_R$ based on a threshold for the total variation explained by the chosen number of principal components. In the coefficient vector $\bbeta \in \mathbb{R}^L$, each component is the coefficient corresponding to the principal component from each tumor sub-region $R$ and each MRI sequence $M$; $\sigma^2$ is the variance parameter.
	
	\paragraph{Group Spike-and-Slab Prior:} Our aim is to identify the tumor sub-regions in a specific sequence (through the principal components) influencing the pathway scores. This translates to identifying the nonzero coefficients of the model in Equation (\ref{eq: model_y}). However, the PC scores within each $X_R^M$ contain rich information about the small-scale variability in the densities for region $R$ in sequence type $M$. The number of principal components to include is dictated by the cumulative amount of variability explained by them. As these densities belong to a function space, capturing variability requires including a large number of PC scores. Moreover, each of these principal components captures different aspects of the variability for the same group, i.e., $(M,R)$ pair, and hence they will need to be evaluated as a group. Incorporating this grouping structure into the modelling framework, we rewrite the model in Equation (\ref{eq: model_y}) as
	\begin{equation}
	\y \sim N\Bigg(\sum\limits_{g = 1}^G X_g\bbeta_g, \sigma^2\I_n\Bigg),
	\label{eq: model_group}
	\end{equation}
	where $G = 4\times 3$, as we have 12 groups arising from four MRI sequences and three tumor sub-regions. Here, $\bbeta_g = (\beta_{g1},\ldots,\beta_{gL_g})^\top$, where $L_g$ is the number of principal components included for the $g$-th group of covariates $X_g$. Note that our covariates have a clear grouping structure, where each group corresponds to the principal components of a tumor sub-region within an imaging sequence. We now introduce a group spike-and-slab prior onto the coefficients $\bbeta_g$ to identify the groups $X_g$ influencing the pathway scores. Consider the following prior structure
	\allowdisplaybreaks{\begin{eqnarray}
		\beta_{gk} & \stackrel{ind}{\sim} & N(0,\sigma^2\zeta_g\nu^2_{gk}), \nonumber\\
		\zeta_g & \stackrel{iid}{\sim} & (1-w)\delta_{v_0}(\zeta_g) + w \delta_1(\zeta_g), \nonumber \\
		w & \sim & U(0,1), \label{eq: group_ss} \\
		\nu^{-2}_{gk} & \stackrel{iid}{\sim} & Gamma(a_1,a_2), \nonumber\\
		\sigma^{-2} & \sim & Gamma(b_1,b_2), \nonumber
		\end{eqnarray}}
	where $\zeta_g\nu^2_{gk}$ is the hypervariance of $\beta_{gk}$ with $\zeta_g$ acting as the group-level indicator variable taking values $1$ or $v_0$ (a small number $> 0$) with probability $w$ or $1-w$, respectively. If $\zeta_g = 1$, the hypervariance is dictated by the Inverse-Gamma prior on $\nu^2_{gk}$; if $\zeta_g = v_0$, the prior on $\beta_{gk}$ is concentrated at $0$ allowing for shrinkage of the coefficient parameter $\beta_{gk}$. The choice of hyperparameters $a_1$ and $a_2$ should be such that we have a continuous bimodal prior on $\beta_{gk}$. Further, $w$ acts as the complexity parameter, indicating the proportion of groups with nonzero coefficients, and has a continuous uniform prior on $(0,1)$. We consider an Inverse-Gamma prior on the variance parameter $\sigma^2$. Note that the group structure is incorporated into the variable selection through the indicator $\zeta_g$, which impacts the variance of the parameter $\beta_{gk}$. That is, if a specific group is not selected, the hypervariance for the coefficients corresponding to all columns in $X_g$ is small, leading to the prior on $\beta_{gk}$ being concentrated at zero, and vice-versa.
	
	\subsection{Estimation}\label{subsec: estimation}
	For the model in Equation (\ref{eq: model_group}), and the group spike-and-slab prior structure in Equation (\ref{eq: group_ss}), the full posterior distribution is provided in Section S4 of the supplementary material. Let us define $\bGamma_g = \text{diag}(\gamma_{g1},\ldots,\gamma_{gL_g})$ and $\bGamma = \text{block-diag}(\bGamma_1,\ldots,\bGamma_G)$, where $\gamma_{gk} = \zeta_g\nu^2_{gk}$. The conditional posteriors for all of the parameters arise from standard distributions, and hence, we can use Markov Chain Monte Carlo (MCMC) sampling procedures such as Gibbs sampling. Details of the Gibbs sampling approach along with the conditional posteriors for the parameters $\bbeta_g, \zeta_g, \nu^{-2}_{gk}, w$ and $\sigma^{-2}$ are given in Algorithm \ref{algo: gibbs}. Since we are modelling data from each gene-set separately, the estimation can be run in parallel across all pathways making the analysis computationally feasible.
	\begin{algorithm}[!h]
		\caption{Gibbs Sampling for Estimation}\label{algo: gibbs}
		\begin{algorithmic}[1]
			\For {$T$ iterations}
			\State Sample $\bbeta_g$ from $\bbeta_g|\zeta_g, \nu^{-2}_{gk}, \sigma^{-2} \sim N(\bSigma \X^\top\y,\sigma^2\bSigma)$, where $\bSigma = ( \X^\top \X + \bGamma^{-1})^{-1}$.
			\State Sample $\zeta_g$ from $\zeta_g|\beta_{gk}, \nu^2_{gk}, w, \sigma^{-2} \sim \frac{w_{1g}}{w_{1g}+w_{2g}}\delta_{v_0}(.) + \frac{w_{2g}}{w_{1g}+w_{2g}}\delta_1(.)$, where
			\begin{eqnarray*}
				w_{1g} = (1-w)v_0^{-\frac{L_g}{2}}\exp\Big(-\sum\limits_{k=1}^{L_g} \frac{\beta_{gk}^2}{2\sigma^2v_0\nu^2_{gk}} \Big) & \text{and} & w_{2g} = w \exp\Big( -\sum\limits_{k=1}^{L_g}\frac{\beta_{gk}^2}{2\sigma^2\nu^2_{gk}} \Big).
			\end{eqnarray*}
			\State Sample $\nu^{-2}_{gk}$ from $\nu^{-2}_{gk}|\beta_{gk}, \zeta_{g}, \sigma^{-2} \stackrel{ind}{\sim} \text{Gamma}\Big(a_1+\frac{1}{2}, a_2+\frac{\beta^2_{gk}}{2\sigma^2\zeta_{g}}\Big)$.
			\State Sample $w$ from $w|\zeta_g \stackrel{ind}{\sim} Beta(1+\#\{\zeta_g = 1\}, 1+\#\{\zeta_g = v_0\})$.
			\State Sample $\sigma^{-2}$ from
			\begin{equation*}
			\sigma^{-2}|\bbeta_g, \zeta_g, \nu^{-2}_{gk} \stackrel{ind}{\sim} \text{Gamma}\Big(b_1+\frac{n+\sum_{g = 1}^G L_g}{2}, b_2+\frac{1}{2} \big[(\y-\X\bbeta)^\top (\y-\X\bbeta)+\bbeta^\top\bGamma^{-1}\bbeta\big]\Big).
			\end{equation*}
			\EndFor
		\end{algorithmic}
	\end{algorithm}
	
	\subsection{False Discovery Rate-based Variable Selection}\label{subsec: fdr}
	The MCMC samples explore the distribution of the coefficients corresponding to the principal components of each of the subgroups as guided by the data. There are different ways of summarizing the information from these MCMC samples. We could use the posterior mode (maximum a-posteriori or MAP estimate) of the coefficients $\beta_{gk}$ and conduct conditional inference based on these point estimates. While this approach provides interpretable point estimates, it does not yield exact zero values as estimates for the coefficients corresponding to principal components not associated with the response; it also does not make use of the complete posterior samples. We use Bayesian model averaging \citep{hoeting1999bayesian}, which builds inference based on various configurations visited by the MCMC sampler. This approach adequately accounts for the uncertainty in the data, and allows for variable selection through downstream inference based on error rates. In this paper, we use a multiplicity-adjusted inference for regression on each pathway separately, since in each of these regressions we are trying to infer from the estimates of $\beta_{gk}$ if they are zero or not. The variable selection also contributes to the multiplicity correction by inducing sparsity. In our discussion, we present results of the false discovery rate (FDR)-based variable selection approach using Bayesian model averaging combined with the MAP estimates.
	
	From the model in Equation (\ref{eq: model_group}), for each $\beta_{gk}$, we obtain $S$ samples $\beta_{gk}^{(1)},\ldots,\beta_{gk}^{(S)}$ from the posterior distribution. For any given threshold $c>0$, we can empirically compute $p_{gk} = \frac{1}{S} \sum_{s=1}^S I(|\beta_{gk}^{(s)}| \leq c)$, which can be interpreted as the local FDR \citep{morris2008bayesian}; then, $(1-p_{gk})$ is the probability that the principal component $k$ from group $g$ significantly impacts the pathway score. Owing to the inherent variable selection property of the group spike-and-slab prior in Equation (\ref{eq: group_ss}), for some $g$ and $k$ it is almost certain that the corresponding $\beta_{gk}$ is close to zero. The value of $p_{gk}$ for such a $\beta_{gk}$ is large, and it is almost certain that including such a nonzero coefficient is an inferential error. We also expect some of the coefficients to have moderate values for $p_{gk}$. Furthermore, we expect to have some coefficients $\beta_{gk}$ such that the corresponding $p_{gk}$ are close to zero and they almost certainly influence the pathway score.
	
	Based on this discussion, we assume that the principal component $k$ from group $g$ will be included in the estimation as a significant coefficient if $p_{gk} < \phi$. Note that $p_{gk}$ is a Bayesian $q$-value or an estimate of the local FDR \citep{storey2003positive}. This threshold $\phi$ can be determined based on different criteria such as Bayesian utility considerations \citep{muller2004optimal}, or by controlling false-positive/false-negative errors, or the average Bayesian FDR. We determine a threshold $\phi_\alpha$, which controls the overall average FDR at some level $\alpha$, so that we expect only $100\alpha\%$ of the elements of the set $\{(g,k)|p_{gk} < \phi_\alpha\}$ to actually be false-positive inclusions in terms of associations with the pathway scores. To compute the threshold $\phi_\alpha$, we sort the posterior inclusion probabilities $p_{gk}$ across all principal components $k = 1,\ldots,L_g$ and groups $g = 1,\ldots,G$, and denote the sorted probabilities as $p_{(l)}$ for $l=1,\ldots,L = \sum_{g = 1}^G L_g$. We then compute $\phi_\alpha = p_{(u)}$, where $u = \max\{l^*|\frac{1}{l^*}\sum_{l=1}^{l^*} p_{(l)} \leq \alpha \}$. The set of principal components $k$ from group $g$ with $p_{gk} < \phi_\alpha$, that is, $\{(g,k)|p_{gk} < \phi_\alpha\}$, can then be claimed to be significantly associated with the pathway score based on an average Bayesian FDR of $\alpha$.
	
	In summary, we start with the MRI scans for each patient and identify the three tumor sub-regions. Based on these sub-regions, we construct imaging-based meta-features through PCA on the space of voxel intensity PDFs using a Riemannian-geometric framework. The resulting PC scores are used as predictors in a regression model with the pathway score as a response. The pathway scores act as genomic markers capturing the enrichment activity in a gene-set. We then use a group structured spike-and-slab prior, which captures the natural grouping of the principal components arising from various tumor sub-regions to identify radiogenomic associations. We use Gibbs sampling for estimation and an FDR-based criterion for variable selection. The complete approach is outlined in Algorithm S1 in Section S1 of the supplementary material.
	
	\section{Radiogenomic Analyses of Lower Grade Gliomas}\label{sec: results}
	
	
	We consider the imaging and matched genomic data described in Sections \ref{subsec: imaging} and \ref{subsec: genomic}, respectively, which comprises 65 samples. However, four of the 65 samples do not posses segmentation labels for all three tumor sub-regions and hence are dropped from the analysis, resulting in a final sample size of 61. For each patient, we have $G = 12$ groups arising from four MRI sequences (T1, T1Gd, T2 and FLAIR) and three tumor sub-regions (NC, ED and ET). First, the density estimates are obtained using the \textit{ksdensity} function in \texttt{MATLAB} software, which uses an optimal value for estimating normal densities using Silverman's rule as the default bandwidth \citep{silverman1986density}. We present a sensitivity analysis to assess the differences in the density estimates based on the choice of bandwidth in Section S11 of the supplementary material. The results indicate reasonable consistency in the computed density estimates. We then compute the PC scores for these $61$ subjects for each of the $12$ groups. The number of principal components included within each group is decided such that the included principal components cumulatively explain $99.99\%$ of the total variance. For each of the four imaging sequences, we display the cumulative percentage of variance explained by the principal components in supplementary Figure S4. Note that this cut-off of $99.99\%$ results in choosing a different number of principal components across each group $g$. Although the choice of this cut-off could include a large number of PCs, any overfitting concerns are addressed by regularization via the spike-and-slab prior that incorporates explicit shrinkage on the regression coefficients \citep{morris2006wavelet,scheipl2012spike}. We include a total of 143 covariates across all of the 12 groups for the LGG data. We discuss results of a sensitivity analysis to assess the effect of sample composition on the computation of PC bases in Section S10 of the supplementary material. We consider only the C-Pathways \citep{ceccarelli2016molecular}, and the corresponding pathway scores are computed for the 61 subjects as described in Section S2 of the supplementary material. We provide an \texttt{R} package, \emph{RADIOHEAD}\footnote{\url{www.github.com/bayesrx/RADIOHEAD}}, which includes all relevant code, including the data under consideration, i.e., the pathway scores corresponding to C-Pathways and PC scores along with their grouping labels for the 61 LGG subjects. 
	
	\noindent\paragraph{Prior Elicitation and MCMC Settings:} In our model, we have shape ($a_1, b_1$) and rate ($a_2, b_2$) hyper-parameters corresponding to $\sigma^2$ and $\nu^2_{gk}$ in Equation (\ref{eq: group_ss}). We choose these hyper-parameters so as to have non-informative/vague priors with $a_1=a_2=0.001$ and $b_1=b_2=0.001$: the mean is $1$ with a large variance. The other hyper-parameter is $v_0$, one of the two possible values of the indicator $\zeta_g$. We choose $v_0 = 0.005$ to be close to zero, which generates continuous bimodal priors for $\beta_{gk}$. We perform a sensitivity analysis based on different values for $v_0$. These results are included in Section S9 of the supplementary material. We run the MCMC chain for $10^5$ iterations and discard the first $20,000$ samples as burn-in. The final estimates are based on MCMC samples with a thinning of $125$ iterations to reduce auto-correlation. In supplementary Figures S9 and S10, we show the posterior densities and trace plots corresponding to randomly chosen $\beta_{gk}$s for the transmission of the nerve impulse pathway showing good convergence of the parameters. In supplementary Figure S11, we present boxplots for the potential scale reduction factors \citep{gelman1992inference} computed based on the MCMC samples of $\beta_{gk}$ from seven different chains. This plot indicates convergence of the MCMC samples across multiple chains.
	
	\begin{figure}[!t]
		\centering
		\resizebox{\textwidth}{!}{
			\includegraphics[trim = 0cm 0cm 0cm 0.65cm, clip, page=1]{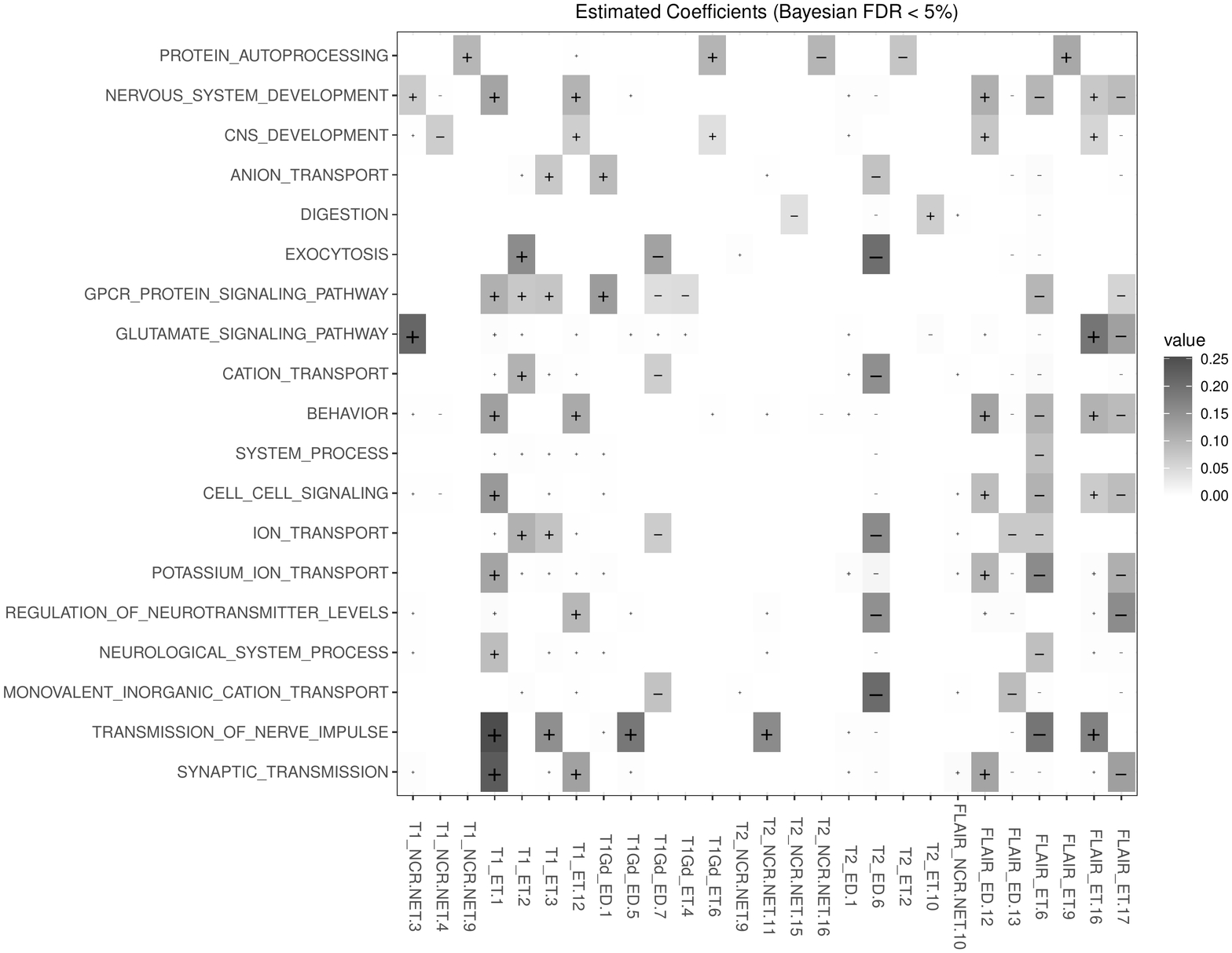}
		}
		\caption{Posterior estimates of $\beta_{gk}$, after FDR-based variable selection, corresponding to different PC scores across MRI sequences and tumor sub-regions. Each row corresponds to a pathway from the C-Pathways. The average Bayesian FDR is controlled at the level $\alpha=0.05$. Values on the gray-scale indicate the magnitude of $\hat{\beta}_{gk}$ and the overlaid symbol (+/-) indicates its sign. The size of the symbol is proportional to the magnitude of $\hat{\beta}_{gk}$. Lack of a symbol denotes a zero estimate indicating no significant association.}
		\label{fig: lgg}
	\end{figure}
	
	The results from the regression of these pathway scores on the imaging predictors through the corresponding PC scores are shown in Figure \ref{fig: lgg}. We display only the gene-sets that have at least one significantly associated covariate among all of the gene-sets in the C-Pathways. Hence, any pathway not shown indicates no significant association between that pathway and the imaging predictors. Similarly, any principal components for any of the 12 groups not listed in this figure are not significantly associated with any of the C-Pathways. Each cell in Figure \ref{fig: lgg} represents the magnitude of the estimated (MAP) coefficients $\hat{\beta}_{gk}$ and the overlaid symbol denotes its sign; the significantly associated PCs are determined using FDR-based variable selection on the MCMC samples as described in Section \ref{subsec: fdr}. For example, we see that the scores of the first principal component of enhancing tumor (\texttt{T1\_ET.1}) sub-region have a significant association with the transmission of nerve impulse gene-set. The average Bayesian FDR is controlled at the level $\alpha=0.05$; we use a threshold $c=0.001$ to compute the values for $p_{gk}$ across all of the pathway score regressions. The value of $c$ is chosen such that it is comparable to the bandwidth used to compute the kernel density estimates, which in turn is essential in computing the MAP estimate from the MCMC chain. Diagnostics for the linear model in Equation \ref{eq: model_group} reveal no obvious violations of modelling assumptions (Figures S5-S8 in the supplementary material). 
	
\begin{figure}[!t]
		\centering
		\resizebox{0.85\textwidth}{!}{%
			\includegraphics{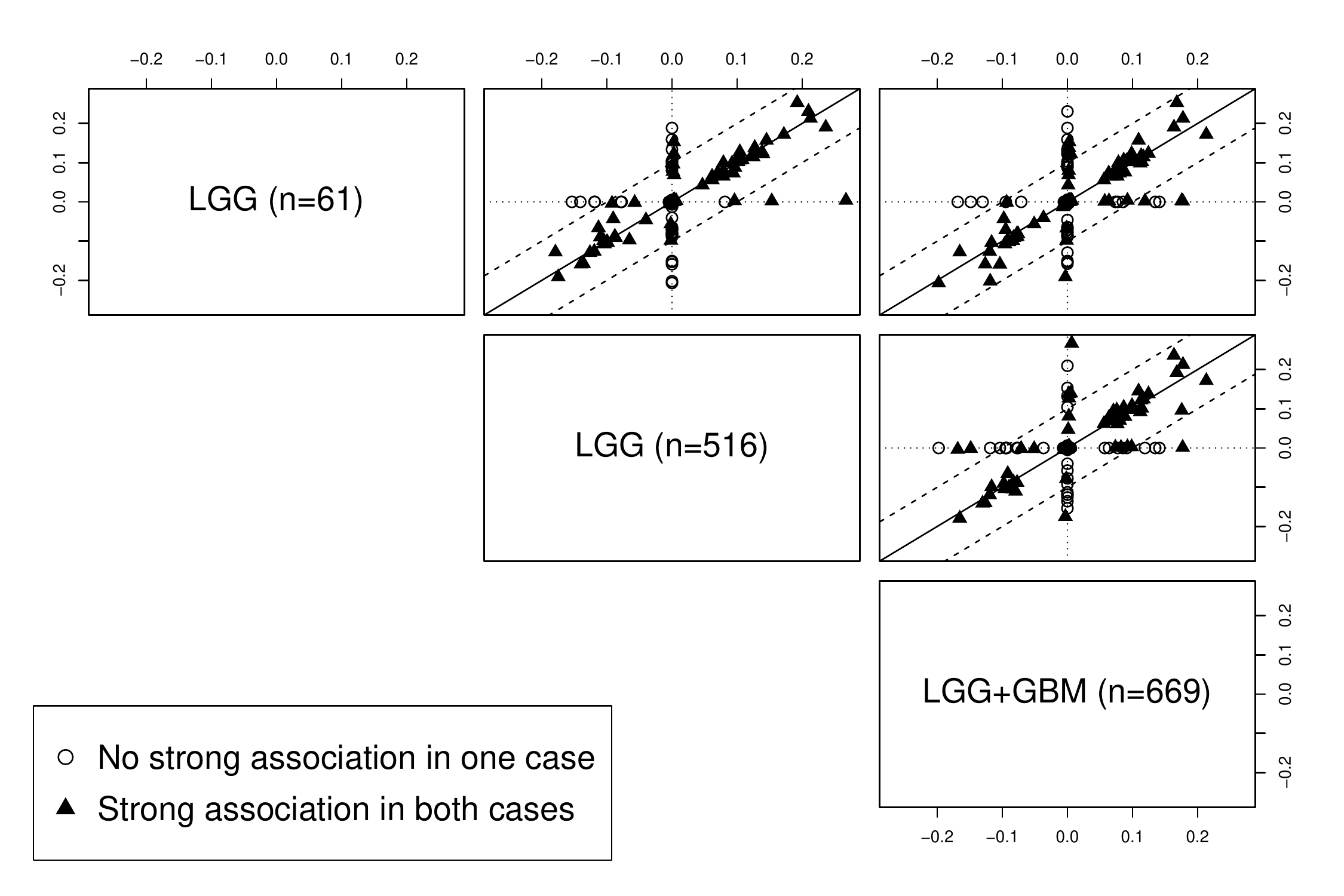}
		}
		\caption{Scatterplots of the estimated coefficients when the pathway scores are constructed using (a) 61 LGG subjects for which imaging data was available, (b) 516 LGG subjects from TCGA, and (c) 516 LGG and 153 GBM subjects from TCGA.}
		\label{fig: scatter}
	\end{figure}

	\noindent\paragraph{Effect of Sample Composition:} As the computation of the pathway scores can be sensitive to the samples in the patient cohort, the associations identified by our model are dependent on the sample composition. A visual illustration of the distribution of the pathway scores (using violin plots) is provided in the supplementary Figures S12-S18. To address this issue, we calibrate the results from our model by computing the pathway scores corresponding to the $61$ subjects in three different scenarios. For the calibration, we include genomic data from TCGA for additional glioma patients (including glioblastoma multiforme (GBM)). The three scenarios include computing the pathway scores with (a) the 61 LGG subjects, (b) 516 LGG subjects, and (c) 516 LGG and 153 GBM subjects. We build the model in Equation (\ref{eq: model_group}) for all three cases, and carry out the estimation and inference as described in Section \ref{sec: statistical}. The results presented earlier in Figure \ref{fig: lgg} correspond to the first case, where the pathway scores were computed with the $n=61$ LGG subjects only. However, in supplementary Figures S19-S21, we present plots for the estimated coefficients (rows and columns are matched in these plots) when the pathway scores are computed as described in cases (a)-(c), respectively. These plots are summarized in Figure \ref{fig: scatter}, with pairwise scatterplots of the estimated coefficients from the three different cases; e.g., the top-right plot in Figure \ref{fig: scatter} corresponds to the scatterplot of estimated coefficients when the pathway scores were computed with 61 LGG subjects versus all 669 glioma subjects (LGG+GBM). The triangles indicate coefficients that are selected as significantly associated in both cases, whereas the circles indicate coefficients that were not selected in one of the two cases. From Figure \ref{fig: scatter}, we see that across all three pairwise comparisons, we estimate many coefficients to be similar (as indicated by the solid line $y=x$).
	
	\begin{figure}[!t]
		\centering
		\resizebox{\textwidth}{6.025in}{
			\includegraphics{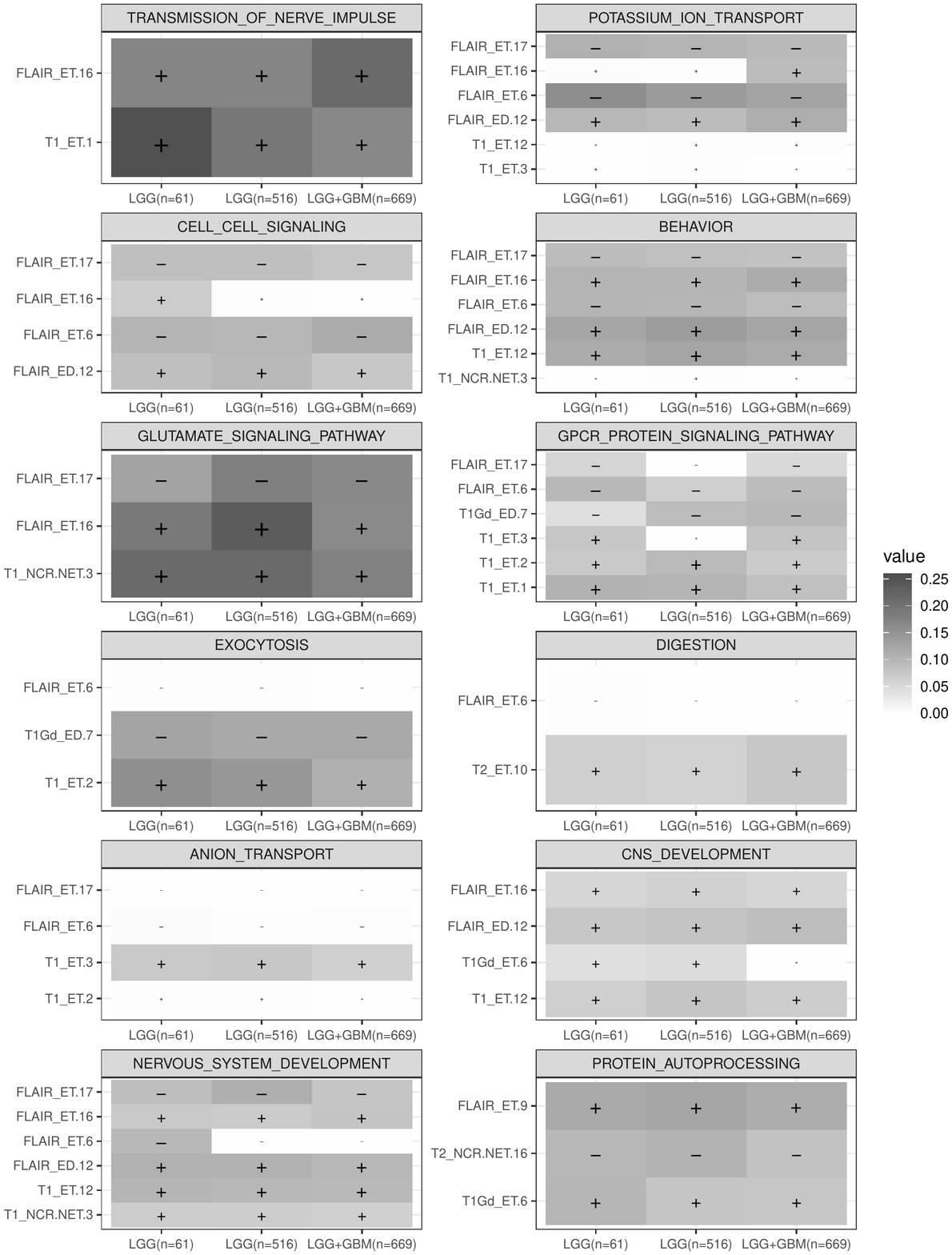}
		}
		\caption{Posterior estimates of $\beta_{gk}$, after FDR-based variable selection, corresponding to different PC scores across MRI sequences and tumor sub-regions. Each panel corresponds to a pathway from the C-Pathways. Each column within the panel corresponds to the sample composition used for calibration to compute the pathway scores. The color on the gray-scale indicates the magnitude of $\hat{\beta}_{gk}$ and the overlaid symbol (+/-) indicates its sign. The size of the symbol is proportional to the magnitude of $\hat{\beta}_{gk}$.}
		\label{fig: lgg_comp}
	\end{figure}
	
	\noindent\paragraph{Biological Associations:} We now focus on those pathways and coefficients whose estimates are consistent across all three cases (within a deviation of $\pm 0.1$); these coefficients are the triangles lying within the dotted lines parallel to $y=x$ in Figure \ref{fig: scatter}. We plot these estimated coefficients across the three cases in Figure \ref{fig: lgg_comp}. These plots include pathways related to synaptic transmission, ion transport, glutamate signaling, G protein receptor signaling, exocytosis, nervous system development and protein autoprocessing. Here, we focus on two major findings in terms of the magnitudes of the different associations:
	\begin{enumerate}
		\item The transmission of nerve impulse pathway is associated with the enhancing tumor region from the T1 and FLAIR imaging sequences. The region enhanced in both of the sequences could potentially indicate demyelination due to glioma invasion, which could in turn lead to disruption in transmission of nerve impulses. This association between the metabolic activity and the infiltrating tumor region is identified by our model. It is also known that neuronal activity promotes glioma growth \citep{venkatesh2015neuronal}, which is supported by the associations of the transmission of nerve impulse pathway with these imaging predictors.
		\item The association of glutamate signaling pathway with the enhancing tumor region from the FLAIR sequence and the necrotic and non-enhancing region from the T1 sequence highlights metabolic activity related to the infiltration of the tumor. In the mammalian central nervous system (CNS), glutamate is a major excitatory neurotransmitter, and experimental evidence suggests that glutamate receptor antagonists may limit tumor growth \citep{brocke2010glutamate}.
	\end{enumerate}
	The aforementioned associations indicate that a deeper validation of these phenotypes is essential to better understand tumor etiology, which may illuminate more specific nuances. Accordingly, we list some of our other findings:
	\begin{enumerate}
		\item Ion channels are important regulators in cell proliferation, migration and apoptosis, and play an important role in the pathology of glioma. Biological processes can be disrupted, or cancer progression can be influenced, by malfunction and/or aberrant expression of ion channels \citep{wang2015ion}. Our model identifies these connections via associations of the imaging predictors with ion transport pathways such as potassium ion transport, cell signaling, behavior and anion transport.
		\item G protein-coupled receptor (GPCR) signaling affects tumor growth, metastasis and angiogenesis \citep{cherry2014g}. Our model identifies this association with the pathway score for GPCR protein signaling.
		\item The inhibition of lysosome exocytosis from glioma cells is known to play an important modulatory role in their migration and invasion \citep{liu2012inhibition}. Such influences are identified through the radiogenomic association with the exocytosis pathway.
	\end{enumerate}
	
	\section{Discussion}\label{sec: discussion}
	In this paper, we propose a statistical framework for integrating multimodal data from both radiological images and genomic profiles. This model aims to identify underlying radiogenomic associations, that is, associations between the radiological characteristics extracted from MRI images and molecular underpinnings encoded in gene expression data. Toward this end, from the transcriptomic profiling data, we have constructed pathway scores corresponding to those pathways that are known to have influence specifically in LGGs; from the radiological imaging data, we have constructed meta-features based on voxel intensities of tumor sub-regions through PDF-based approaches, which effectively capture tumor heterogeneity. These meta-features, constructed from multiple MR sequences, are then used as covariates in a model with pathway scores as responses. We use a Bayesian variable selection strategy by employing a continuous spike-and-slab prior with a grouping structure, which accounts for the inherent grouping in the imaging meta-features. This approach identifies many underlying associations between gene pathway activations and image-based tumor characteristics.
	
	We note that, although we incorporate the grouping structure in the RADIOHEAD framework, we are not (explicitly) interested in the  associations with the entire PDF. That is, our inference is not based only on groups where $\hat{\beta}_{gk} \neq 0$ for all $k$ in a given group $g$. Instead, our focus is to identify associations with \textit{any} aspect of the PDFs. The imaging meta-features (PC scores) facilitate evaluation of any underlying associations of the genomic markers with various aspects of the PDFs. Furthermore, inference on the group-level indicator is not feasible in our model setup as $\zeta_g$ is not identifiable. Such an inference is inferior in performance under cases with high within-group sparsity, even under a model which has  identifiability of the group-level indicator \citep{yang2018consistent}. We demonstrate this using a simulation study described in Section S8 of the supplementary material.
	
\begin{figure}[!h]
		\centering
		\resizebox{\textwidth}{!}{
			\includegraphics[trim = 0cm 0cm 0cm 1cm, clip, page=2]{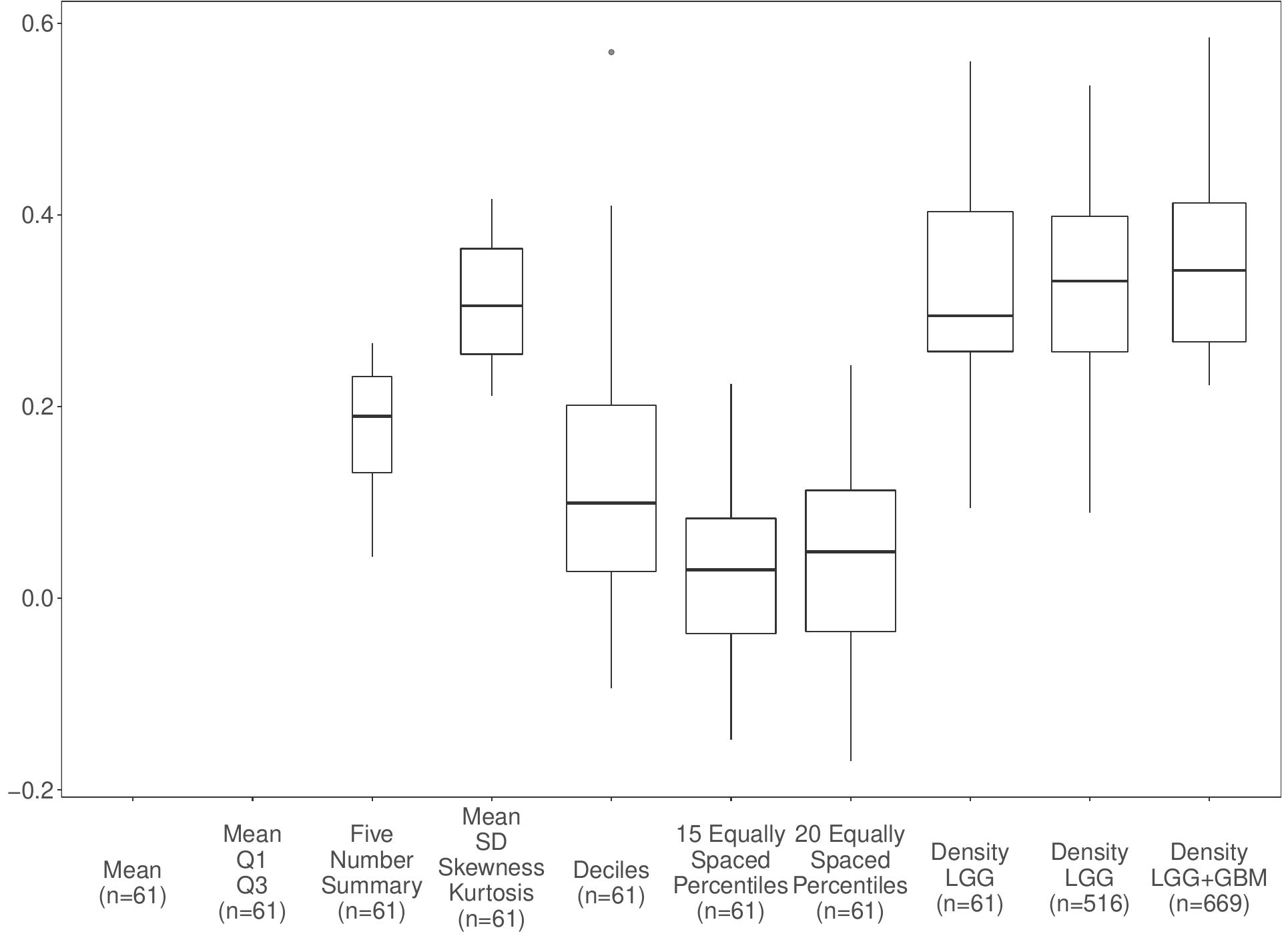}
		}
		\caption{Boxplot of Spearman correlations between computed and fitted pathway scores ($X\hat{\bbeta}$) using RADIOHEAD, while different sets of covariates are considered. The width of the boxplots is proportional to the number of pathways exhibiting significant associations with at least one of the imaging meta-features}
		\label{fig: spearman_boxplot}
	\end{figure}

	\noindent\paragraph{Utility in using densities:} Data integration from multiple modalities comes with computational and modelling challenges. For the imaging data, MRIs facilitate the characterization of tumor sub-regions and are obtained from four different sequences. The tumor sub-regions are represented as voxel intensity values, and standard analyses utilize summaries from the histograms of these voxel intensity values. As an improved alternative, we have used the complete information from the voxel intensities though smoothed histograms (kernel density estimates). Next, we show the benefits of this more comprehensive representation by considering seven different cases as potential predictors: (a) mean, (b) mean, first and third quartiles ($Q_1$ and $Q_3$), (c) five-number summary, (d) mean, standard deviation, skewness and kurtosis, (e) deciles, (f) 15 equally spaced percentiles, and (g) 20 equally spaced percentiles.  The summary statistics are computed across all of the $12$ groups separately. In each of these seven cases, we employ the RADIOHEAD pipeline which uses the group spike-and-slab prior and FDR-based variable selection to identify associations. The issue of multicollinearity within the predictors is handled by the shrinkage properties of the spike-and-slab prior. These seven cases include scenarios where the number of predictors is higher/lower compared to the 143 predictors across groups from the PC scores. The results based on these seven cases are presented in supplementary Figures S22-S25. We see that having just the mean or just the mean, $Q_1$ and $Q_3$, does not identify any associations with the pathway scores. However, adding more summary statistics describing the histogram aids in identifying associations. But, as we will see next, the PC scores offer more relevant information about the densities rather than including a larger number of summary statistics as covariates (cases (f) and (g)). Hence, using the PDF-derived PC scores has a higher utility in terms of understanding the pathway scores. In Figure \ref{fig: spearman_boxplot}, we show the boxplots of the Spearman correlations between computed (observed) and fitted (using estimated coefficients of density-based meta-features/summary statistics from RADIOHEAD) pathway scores, that is, Spearman correlation between $y$ and $X\hat{\bbeta}$, respectively. These correlations are computed separately by considering cases (a)-(g) and density-based PC scores as predictors.
	
	Additionally, since the computation of the pathway score was dependent on the sample composition, for the case with density-basd PC scores as predictors we also include boxplots of Spearman correlations for three different computations of pathway scores as described in Section \ref{sec: results}. The width of these boxplots is proportional to the number of pathways exhibiting significant associations with at least one of the imaging meta-features. In supplementary Figure S25, we also show the Spearman correlations between the computed pathway scores and the fitted pathway scores. This figure demonstrates that we are able to better understand the underlying radiogenomic associations through our modelling approach when the density-based meta-features are considered as covariates. Furthermore, our model can be used in other applications (including other cancers and disease systems) involving imaging and genomic data, as the methodology is readily generalizable to different application domains.
	
	\noindent\paragraph{Future Work:} Although we see promise in the proposed modelling framework to identify radiogenomic associations in LGG, there are certain directions which can be further explored. While using density-based features extracted from multimodal MRI scans does facilitate modelling and provide improved performance, these densities do not explicitly utilize potentially important spatial information in their construction. Incorporating voxel-based spatial information in addition to intensity values is non-trivial and will be explored in our future studies. The current model explores linear relationships between the PC scores and pathway scores only, which could be further extended to investigate non-linear associations as well. Such analyses will better inform the understanding of the inter- and intra- tumor heterogeneity in LGG. Other directions could be to (a) extend the framework to incorporate dependencies between pathways (data-derived or based on canonical topology), or (b) use gene-level data instead of pathways while incorporating cross-correlations between the genes. Our framework could also be explored further with other forms of pan-omic data, such as epigenomic and proteomic data. Furthermore, our findings could be used to build predictive models for clinical phenotypes (such as survival or progression) that include biologically relevant information based on radiogenomic associations. This provides a statistically-informed strategy to incorporate relevant information for the prediction of clinical phenotypes from complex data.
	
	
	\section*{Acknowledgments}
	All of the authors acknowledge support by the NCI grant R37-CA214955. SM was partially supported by Precision Health at The University of Michigan (U-M). SM and AR were partially supported by U-M institutional research funds. SK and KB were partially supported by the NSF grants DMS 1613054 and DMS 2015374. SK was also partially supported by the NSF grant CCF 1740761. VB was supported by NIH grants R01-CA160736, R21-CA220299, P30 CA 46592 and NSF grant 1463233 and start-up funds from the U-M Rogel Cancer Center and School of Public Health. We would like to extend our gratitude to Kirsten Herold from the Writing Lab at the U-M School of Public Health. We acknowledge the efforts of the anonymous Associate Editor and referees, whose comments have strengthened this paper.
%
	\bibliographystyle{imsart-nameyear} 
	\bibliography{bibliography}  

	\newpage
	\setcounter{figure}{0}
	\renewcommand{\thefigure}{S\arabic{figure}}
	\setcounter{section}{0}
	\renewcommand{\thesection}{S\arabic{section}}
	\setcounter{algorithm}{0}
	\renewcommand{\thealgorithm}{S\arabic{algorithm}}
	\setcounter{table}{0}
	\renewcommand{\thetable}{S\arabic{table}}
	\setcounter{equation}{0}
	\renewcommand{\theequation}{S\arabic{equation}}

	\title{Supplement to ``RADIOHEAD: Radiogenomic Analysis Incorporating Tumor Heterogeneity in Imaging Through Densities."}

	\section{Overall Outline of RADIOHEAD}\label{app: overall}
	Here, we describe the algorithm with an outline of the overall approach of this paper to identify the radiogenomic associations by modelling the genomic-based pathway scores using the radiomic-based PC scores.
	\begin{algorithm}[h]
		\caption{Outline of RADIOHEAD}\label{algo: outline}
		\begin{algorithmic}[1]
			\For {each MRI sequence $M=$ T1, T1Gd, T2, FLAIR}
			\For {each tumor sub-region $R=$ NC, ET, ED}
			\For {each subject $i=1,\ldots,n$} Compute the kernel densities $f_i^M (R)$.
			\EndFor
			\State Compute the principal component scores $X_R^M$ using PCA in Algorithm \ref{algo: pca}.
			\EndFor
			\EndFor
			\State Consider a pathway of interest and compute pathway scores $\mathbf{y} = (y_1,\ldots,y_n)^\top$ (as described in Section S2) with the sample $i=1,\ldots,n$ in the cohort.
			\State Bayesian Modelling
			\begin{algsubstates}
				\State Model:
				\begin{eqnarray*}
					\y \sim N\bigg(\sum_{g = 1}^{(4\times 3)} X_g\bbeta_g, \sigma^2\I_n\bigg); & \beta_{gk} \stackrel{ind}{\sim} N(0,\sigma^2\zeta_g\nu^2_{gk}); \nonumber\\
					\zeta_g \stackrel{iid}{\sim} (1-w)\delta_{v_0}(\zeta_g) + w \delta_1(\zeta_g); & w \sim U(0,1); \nonumber \\
					\nu^{-2}_{gk} \stackrel{iid}{\sim} Gamma(a_1,a_2); & \sigma^{-2} \sim Gamma(b_1,b_2). \nonumber
				\end{eqnarray*}
				\State Gibbs sampling for the parameters $\bbeta_g, \zeta_g, \nu^{-2}_{gk}, w, \sigma^{-2}$ as described in Algorithm \ref{algo: gibbs}.
				\State FDR-based variable selection as described in Section \ref{subsec: fdr} to identify non-zero $\beta_{gk}$.
			\end{algsubstates}
		\end{algorithmic}
	\end{algorithm}

	\section{Computation of Pathway Scores}\label{app: pathway}
	Instead of directly including the gene expression profiles in the model, we use the corresponding pathway scores. Pathway-based methods offer a significant benefit in terms of interpretability as gene function is exerted collectively and may vary based on several factors, such as disease state, genetic modification or environmental stimuli. As mentioned in \cite{hanzelmann2013gsva}, using gene-sets obtained by organizing genes provides an intuitive and stable context for assessing biological activity. We compute these gene-set scores using gene-set variation analysis (GSVA) \citep{hanzelmann2013gsva}, which is a gene-set enrichment method that estimates variation of pathway activity over a sample population in an unsupervised manner. We provide a brief overview of the GSVA procedure next.
	
	Let $Z$ denote the $p \times n$ matrix of normalized gene expression values of $p$ genes for $n$ samples ($p \gg n$) and a collection of gene-sets $G = \{g_1,\ldots,g_m\}$. The expression profile for gene $i$ is defined as $z_i = (z_{i1},\ldots,z_{in})$ and each gene-set is a subset of genes with its cardinality being denoted by $|g_k|$. First, GSVA evaluates whether a gene $i$ is highly or lowly expressed in sample $j$ in the context of the sample population distribution. An expression-level statistic is computed so that distinct expression profiles can be compared on the same scale. For each $z_i$, a non-parametric kernel estimation of its cumulative density function is performed using a Gaussian kernel to compute $\hat{F}_{s_i}(z_{ij}) = \frac{1}{n} \sum_{r=1}^n \Phi(\frac{z_{ij}-z_{ir}}{s_i})$, where $s_i$ is the gene-specific bandwidth parameter controlling the resolution of the kernel estimation. These statistics $\hat{F}_{s_i}(z_{ij})$ are converted to ranks $r_{(i)j}$ for each sample $j$ and further normalized using $t_{ij} = |\frac{p}{2}-r_{(i)j}|$. We use these $t_{ij}$ to compute a Kolmogorov-Smirnov (KS)-type random walk statistic for $l = 1,\ldots,p$ as
	
	\begin{equation*}
		\eta_{jk}(l) = \frac{\sum_{i=1}^l |t_{ij}|^\tau I(u_{(i)} \in g_k)}{\sum_{i=1}^p |t_{ij}|^\tau I(u_{(i)} \in g_k)} - \frac{\sum_{i=1}^l I(u_{(i)} \in g_k)}{p - |g_k|},
	\end{equation*}
	\noindent where $\tau$ is a parameter describing the weight of the tail and $I(u_{(i)} \in g_k)$ is an indicator taking the value $1$ if the gene corresponding to the rank $i$ expression-level statistic belongs to the gene-set $g_k$. The statistic $\eta_{jk}(l)$ produces a distribution over the genes by identifying whether the genes in a gene-set are more likely to belong to either tail of the rank distribution. This KS-like statistic is now converted into an enrichment score of the pathway using $S_{jk} = \max_l (0, \eta_{jk}(l)) - \min_l (0, \eta_{jk}(l)) $. \cite{hanzelmann2013gsva} note that $S_{jk}$ has a clear biological interpretation as it emphasizes genes in pathways that are concordantly activated in one direction only, i.e., ones that are either over-expressed or under-expressed relative to the overall population. Low enrichment is shown for pathways containing genes strongly acting in both directions.
	
	\section{Computation of the Karcher Mean}\label{sec: karcher}
	In this section, we provide a gradient-based algorithm to compute the Karcher mean on $\calH$ \citep{dryden1998statistical}. The algorithm can be initialized using one of the densities in the sample or an extrinsic average.
	
	\begin{algorithm}[h]
		\caption{Sample Karcher mean of densities}\label{algo: karcher}
		\begin{algorithmic}[1]
			\State $\bar{h}_0$ (initial estimate for the Karcher mean) $\gets$ any one of the densities in the sample OR the extrinsic average. Set $j \gets 0$ and $\epsilon_1, \epsilon_2 > 0$ be small.
			\State For $i=1,\ldots,n$ compute $u_i = \exp^{-1}_{\bar{h}_j}(h_i)$.
			\State Compute the average direction in the tangent space $\bar{u} = \frac{1}{n} \sum_{i=1}^n u_i$.
			\If {$||\bar{u}||_{L^2} < \epsilon_1$}
			\State \Return $\bar{h}_j$ as the Karcher mean.
			\Else {}
			\State$\bar{h}_{j+1} = \exp_{\bar{h}_j}(\epsilon_2 \bar{u})$.
			\State Set $j \gets j+1$.
			\State Return to step $2$.
			\EndIf
		\end{algorithmic}
	\end{algorithm}
	
	\section{Expression for Joint Posterior Distribution}\label{sec: joint}
	Here, we provide the joint posterior distribution for the proposed Bayesian model. The model is given in Equation (3) in the main manuscript, along with the group spike-and-slab prior structure in Equation (4). The full posterior distribution is given by:
	\allowdisplaybreaks{
		\begin{eqnarray*}
			\pi(\beta_{gk}, \zeta_g, \nu^{-2}_{gk}, w, \sigma^{-2} | \y, \X) & \propto & (\sigma^{-2})^{\frac{n}{2}} \exp\Big( -\frac{1}{2\sigma^2} (\y-\X\bbeta)^\top (\y-\X\bbeta) \Big) \\
			& & \times \prod\limits_{g =1}^G \prod\limits_{k=1}^{L_g} (\sigma^2\zeta_g\nu^2_{gk})^{-1/2} \exp\Big( -\frac{\beta_{gk}^2}{2\sigma^2\zeta_g\nu^2_{gk}} \Big) \\
			& & \times \prod\limits_{g=1}^G \Big[(1-w)\delta_{v_0}(\zeta_g) + w\delta_1(\zeta_g) \Big] \\
			& & \times 1 \times \prod\limits_{g =1}^G \prod\limits_{k=1}^{L_g} (\nu^{-2}_{gk})^{a_1 - 1}\exp( -a_2\nu^{-2}_{gk} ) \\
			& & \times (\sigma^{-2})^{b_1 - 1}\exp( -b_2\sigma^{-2} ).
		\end{eqnarray*}
	}

	\section{Supplementary Figures and Tables}\label{sec: supp_figures}
	
\begin{figure}[H]
		\centering
		\resizebox{\textwidth}{!}{
			\includegraphics{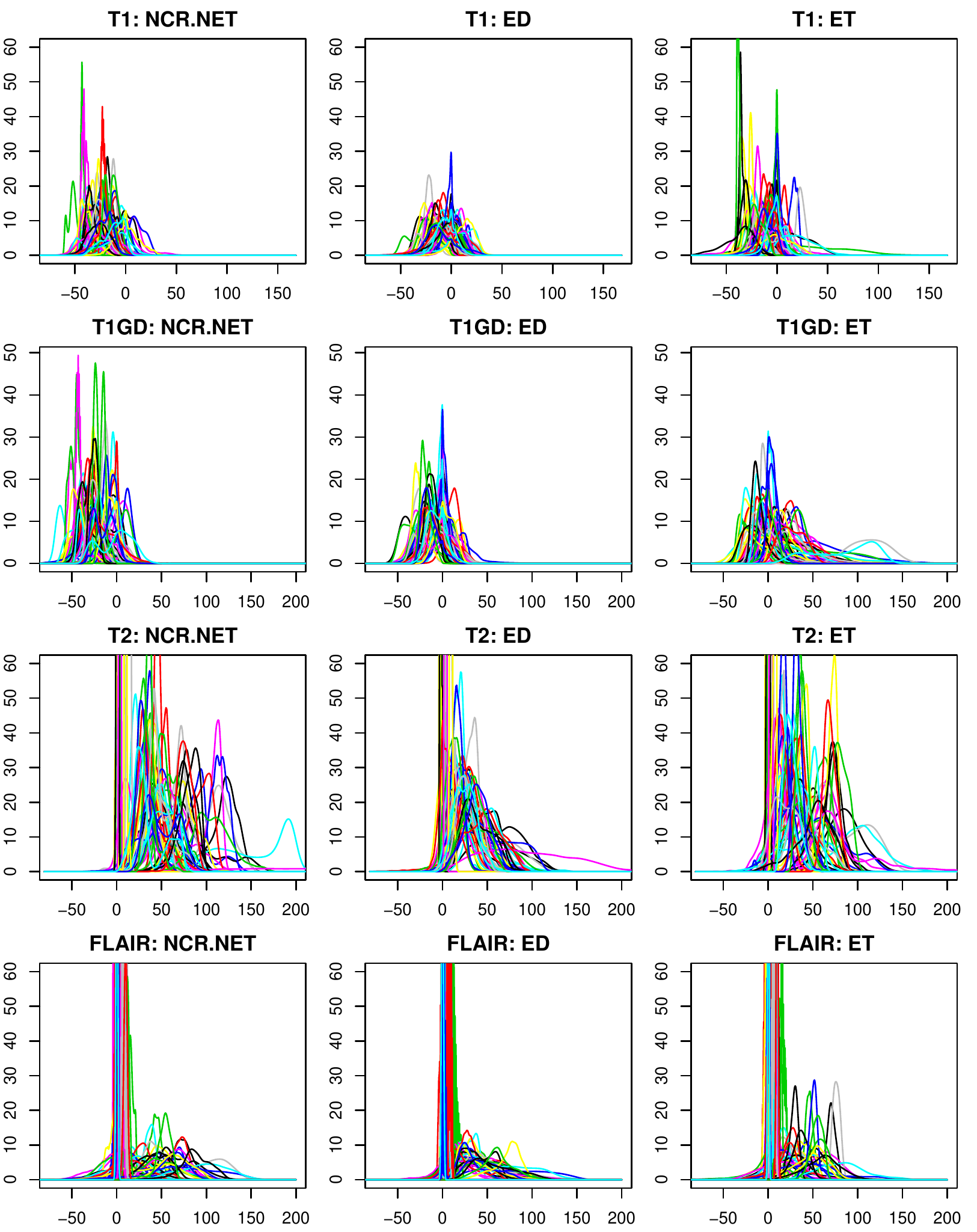}
		}
		\caption{Kernel densities $f_i^M(R)$ for all subjects across all four MRI sequences and three tumor sub-regions. For visual convenience, the $y$-axes are truncated for each of the subplots. Similar plots without truncation of the $y$-axis are shown in Figure S2. The x-axis shows the voxel-intensity values; however, we transform them to $[0,1]$ for each imaging sequence to compute the density estimates.}
		\label{fig: densities_trunc}
	\end{figure}

\begin{figure}[H]
		\centering
		\resizebox{\textwidth}{!}{
			\includegraphics[scale=0.8]{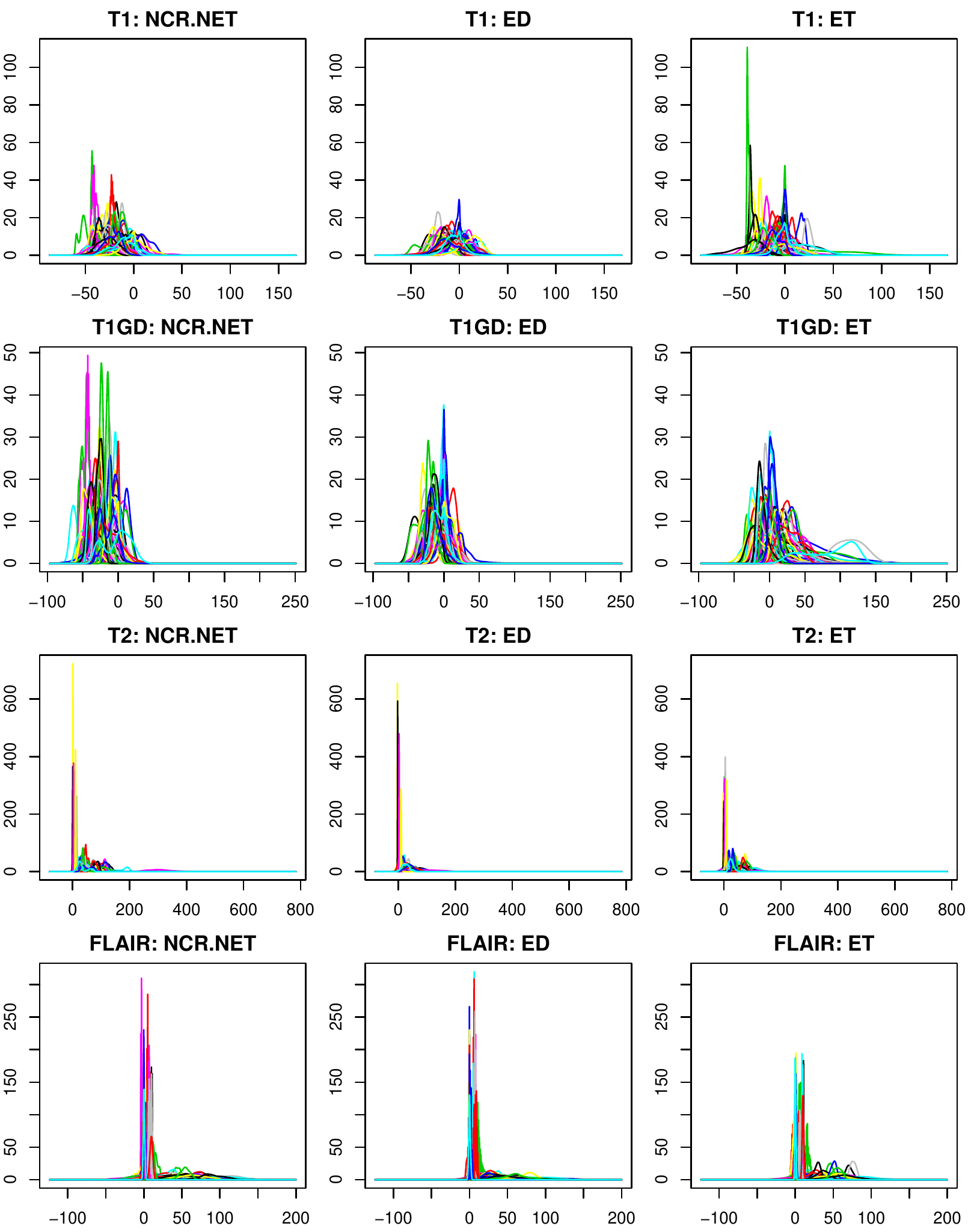}
		}
		\caption{Kernel density estimates $f_{iR}$ for all subjects across all four MRI sequences and all three tumor sub-regions. Each row corresponds to a specific imaging sequence and each column corresponds to a tumor sub-region. The x-axis shows the voxel-intensity values; however, we transform them to $[0,1]$ for each imaging sequence to compute the density estimates.}
		\label{fig: densities}
	\end{figure}

\begin{figure}[H]
		\centering
		\resizebox{\textwidth}{!}{
			\includegraphics{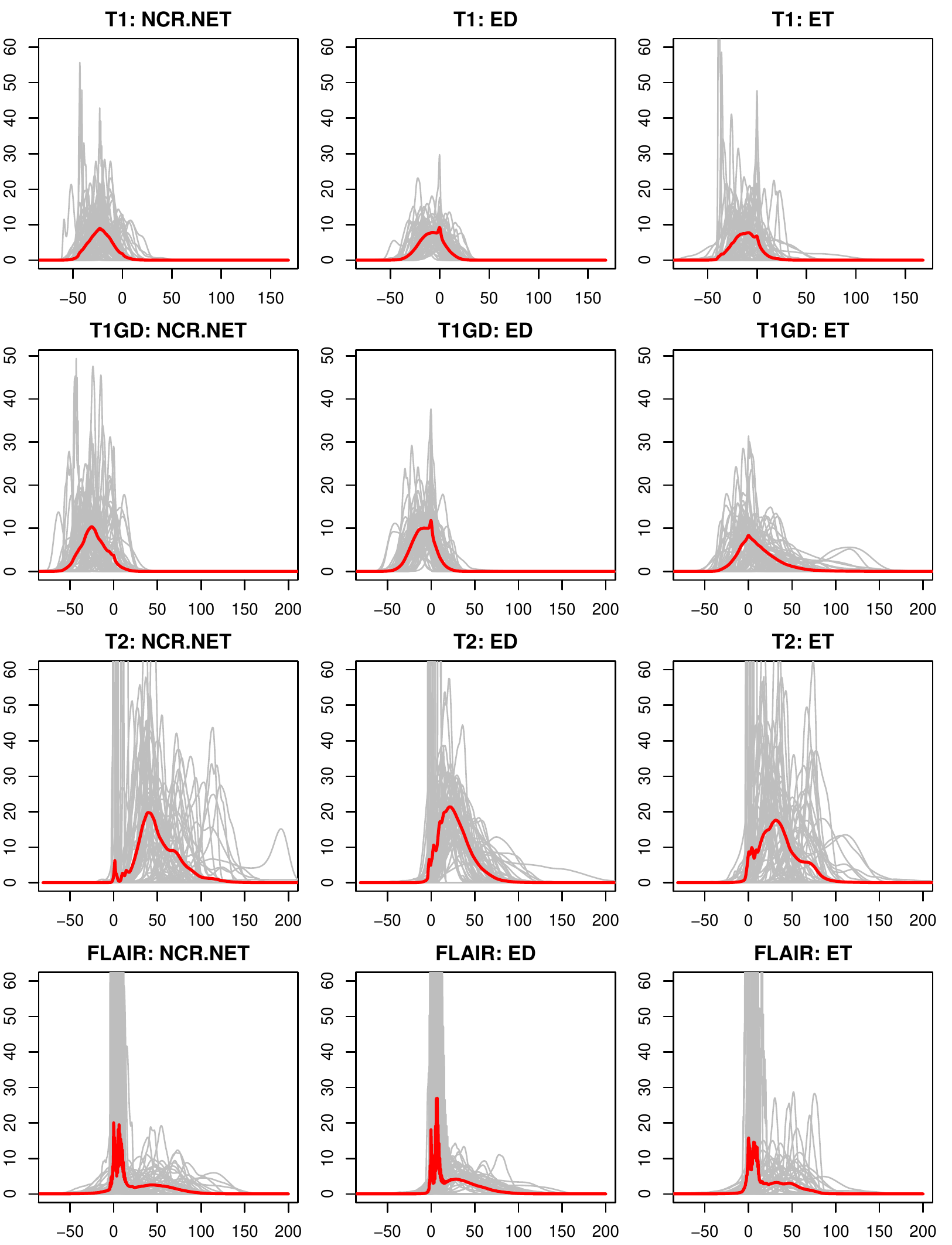}
		}
		\caption{Kernel densities $f_i^M(R)$ for all subjects (grey) across all four MRI sequences and three tumor sub-regions. The solid red curve corresponds to the sample Karcher mean density. The $y$-axes are truncated for each subplot similarly to Figure 4 in the main manuscript. The x-axis shows the voxel-intensity values; however, we transform them to $[0,1]$ for each imaging sequence to compute the density estimates.}
		\label{fig: densities_mean}
	\end{figure}

\begin{figure}[H]
		\centering
		\resizebox{\textwidth}{!}{%
			\begin{tabular}{|c|c|}
				\hline
				\subfigure[T1]{\label{fig: t1_pc}\includegraphics[trim=0cm 0cm 0cm 1.5cm, clip, scale=0.5,page=1]{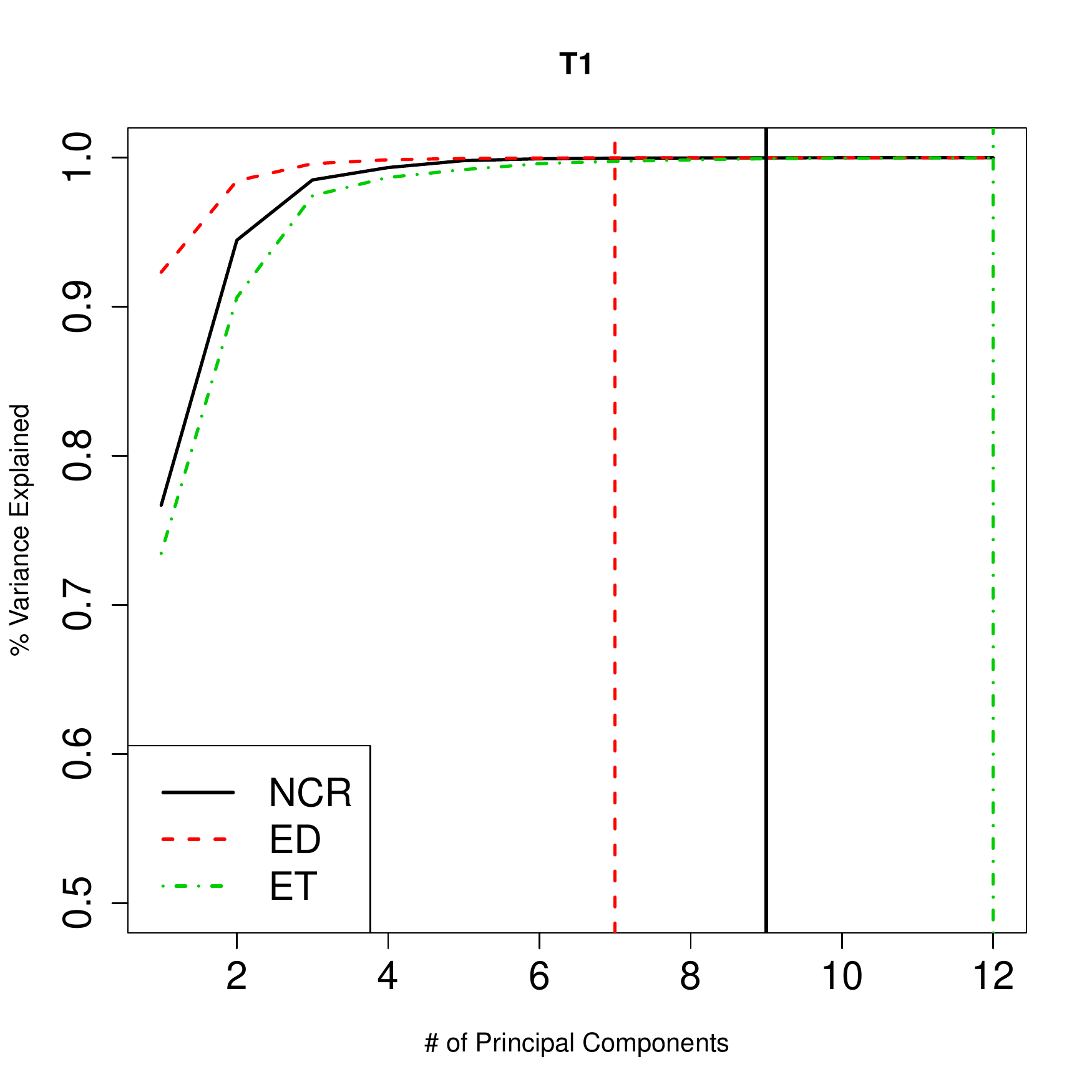}} & \subfigure[T1Gd]{\label{fig: t1Gd_pc}\includegraphics[trim=0cm 0cm 0cm 1.5cm, clip, scale=0.5,page=2]{plots/pc_variance.pdf}} \\ \hline
				\subfigure[T2]{\label{fig: t2_pc}\includegraphics[trim=0cm 0cm 0cm 1.5cm, clip, scale=0.5,page=3]{plots/pc_variance.pdf}} & \subfigure[FLAIR]{\label{fig: flair_pc}\includegraphics[trim=0cm 0cm 0cm 1.5cm, clip, scale=0.5,page=4]{plots/pc_variance.pdf}} \\
				\hline
			\end{tabular}
		}
		\caption{Percentage of overall variance explained by the principal components constructed within each imaging sequence across all three tumor sub-regions. Vertical lines correspond to cutoffs for the number of principal components used in the model as predictors for each sub-region.}
	\end{figure}

\begin{figure}[H]
		\centering
		\resizebox{\textwidth}{!}{
			\includegraphics[page=1]{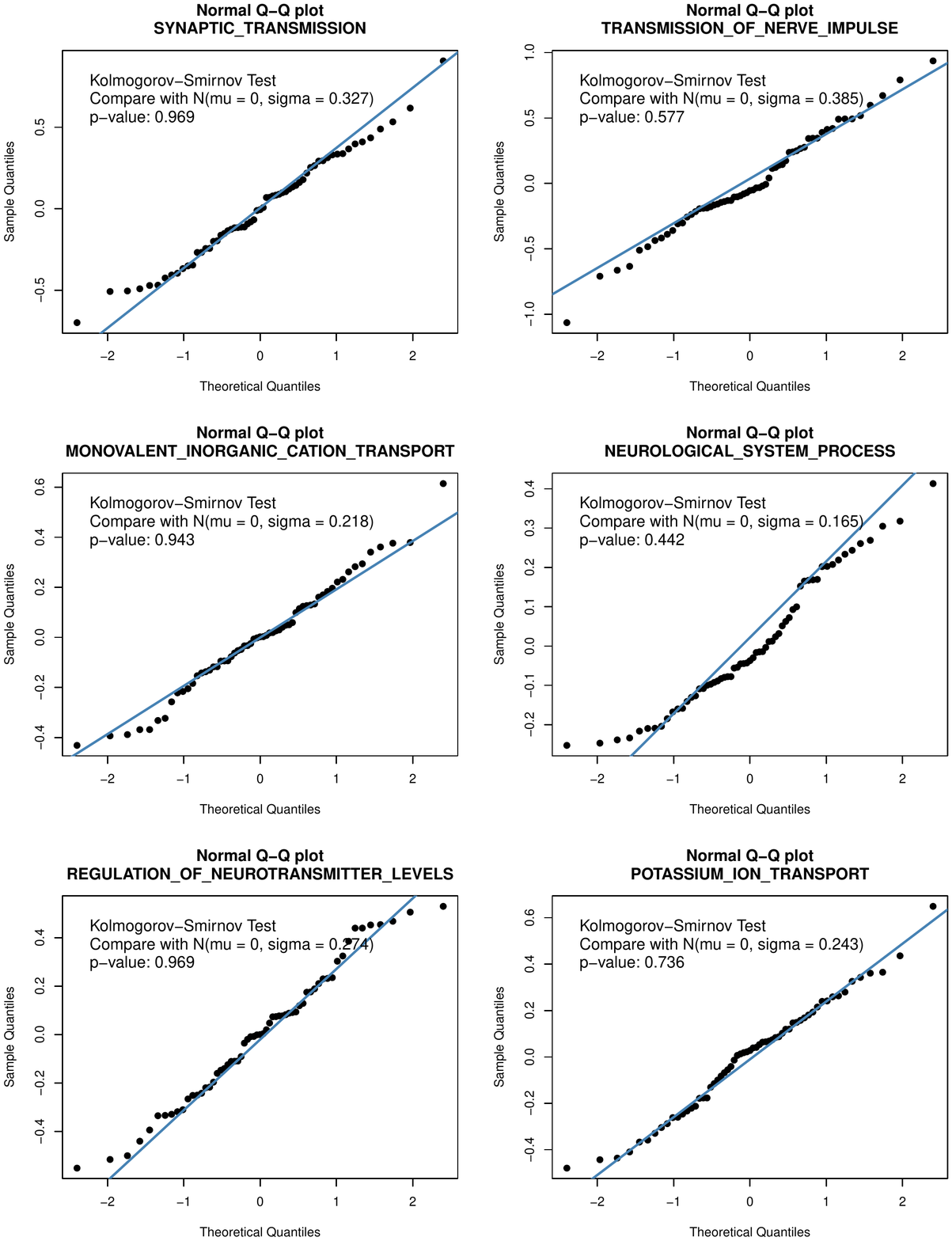}
		}
		\caption{Normal Q-Q plots for the error terms, $\y - X\hat{\bbeta}$, from the RADIOHEAD framework for different C-Pathways.}
		\label{fig: qq1}
	\end{figure}

\begin{figure}[H]
		\centering
		\resizebox{\textwidth}{!}{
			\includegraphics[page=2]{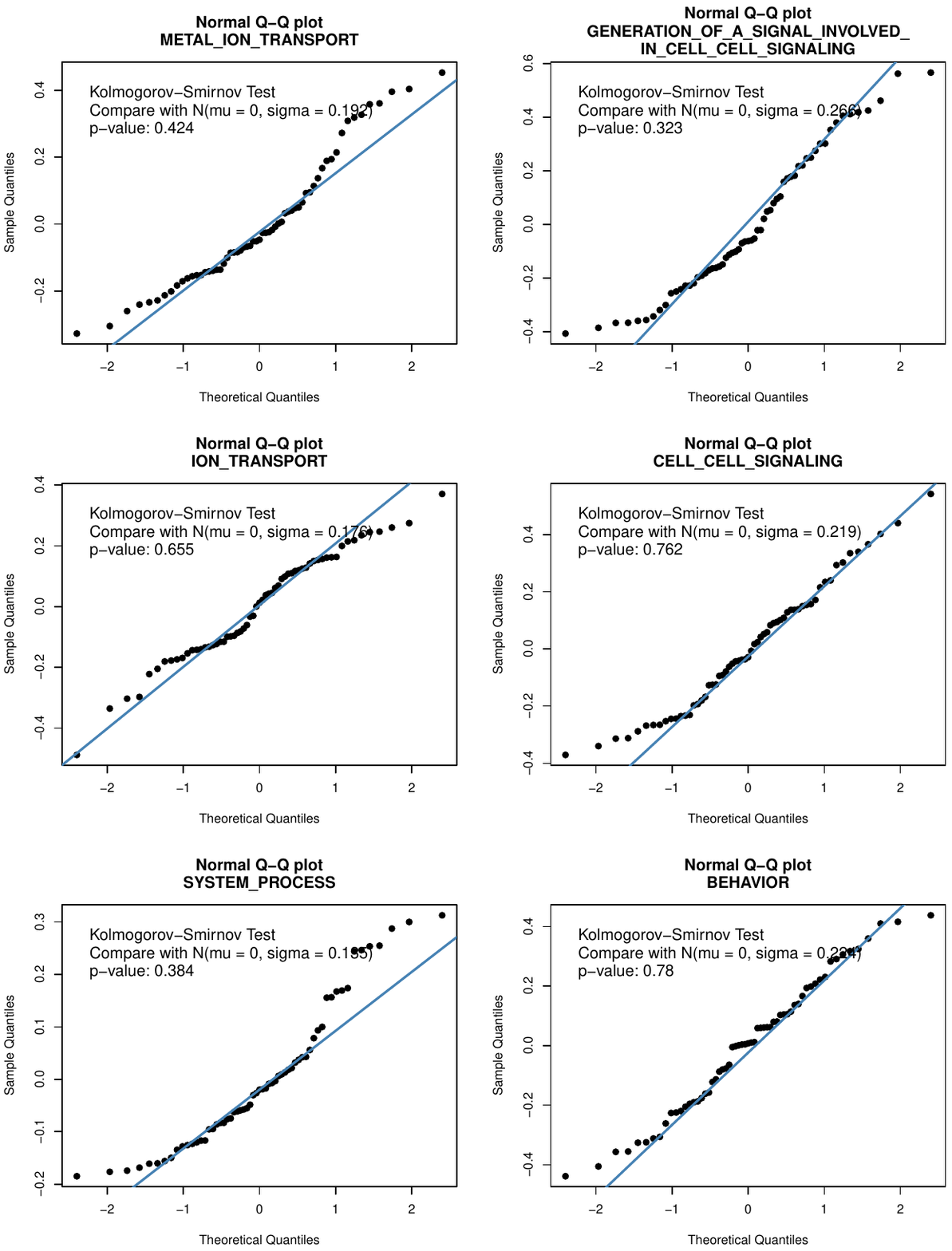}
		}
		\caption{Normal Q-Q plots for the error terms, $\y - X\hat{\bbeta}$, from the RADIOHEAD framework for different C-Pathways.}
		\label{fig: qq2}
	\end{figure}

\begin{figure}[H]
		\centering
		\resizebox{\textwidth}{!}{
			\includegraphics[page=3]{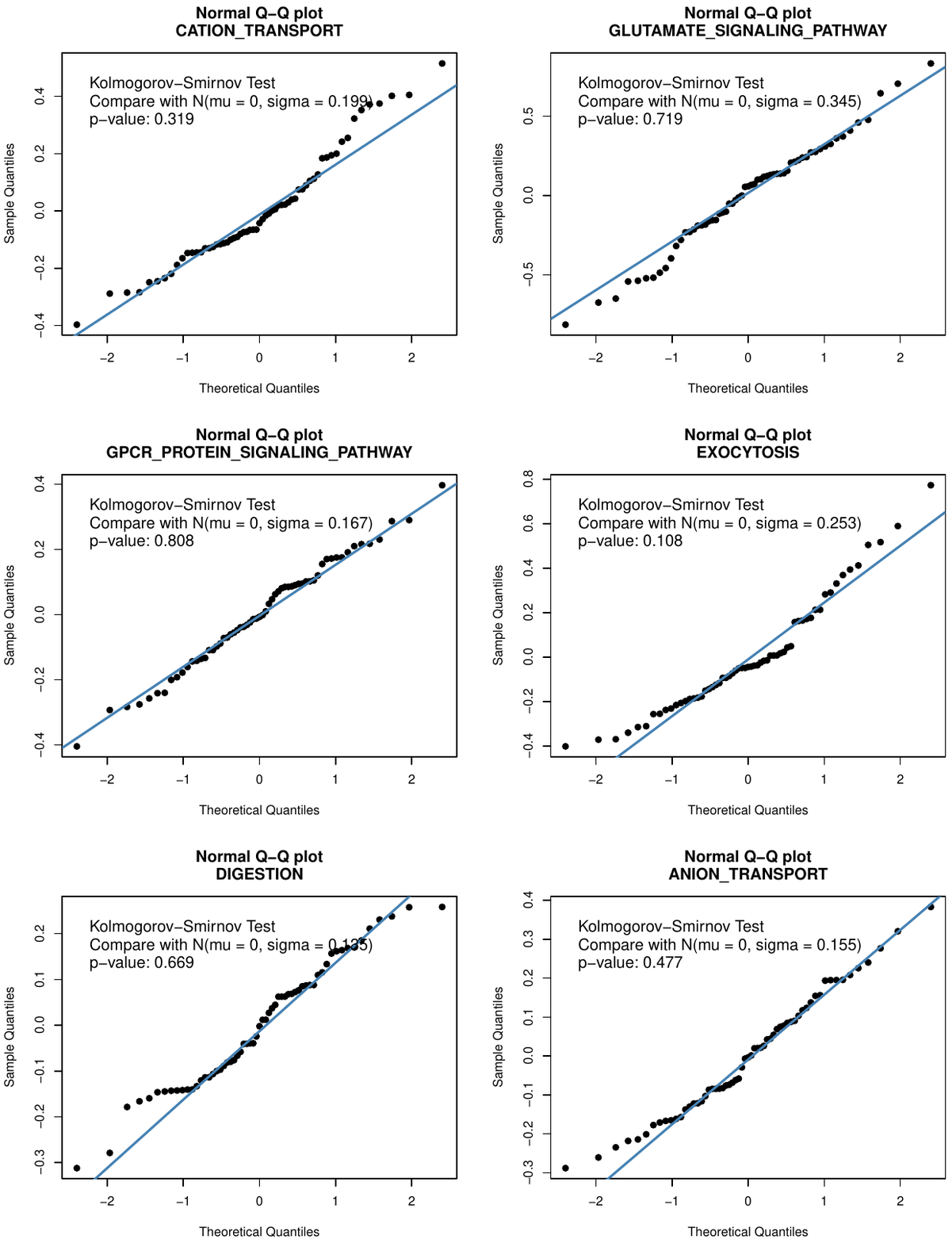}
		}
		\caption{Normal Q-Q plots for the error terms, $\y - X\hat{\bbeta}$, from the RADIOHEAD framework for different C-Pathways.}
		\label{fig: qq3}
	\end{figure}

\begin{figure}[H]
		\centering
		\resizebox{\textwidth}{!}{
			\includegraphics[page=4, trim=0cm 9cm 0cm 0cm, clip]{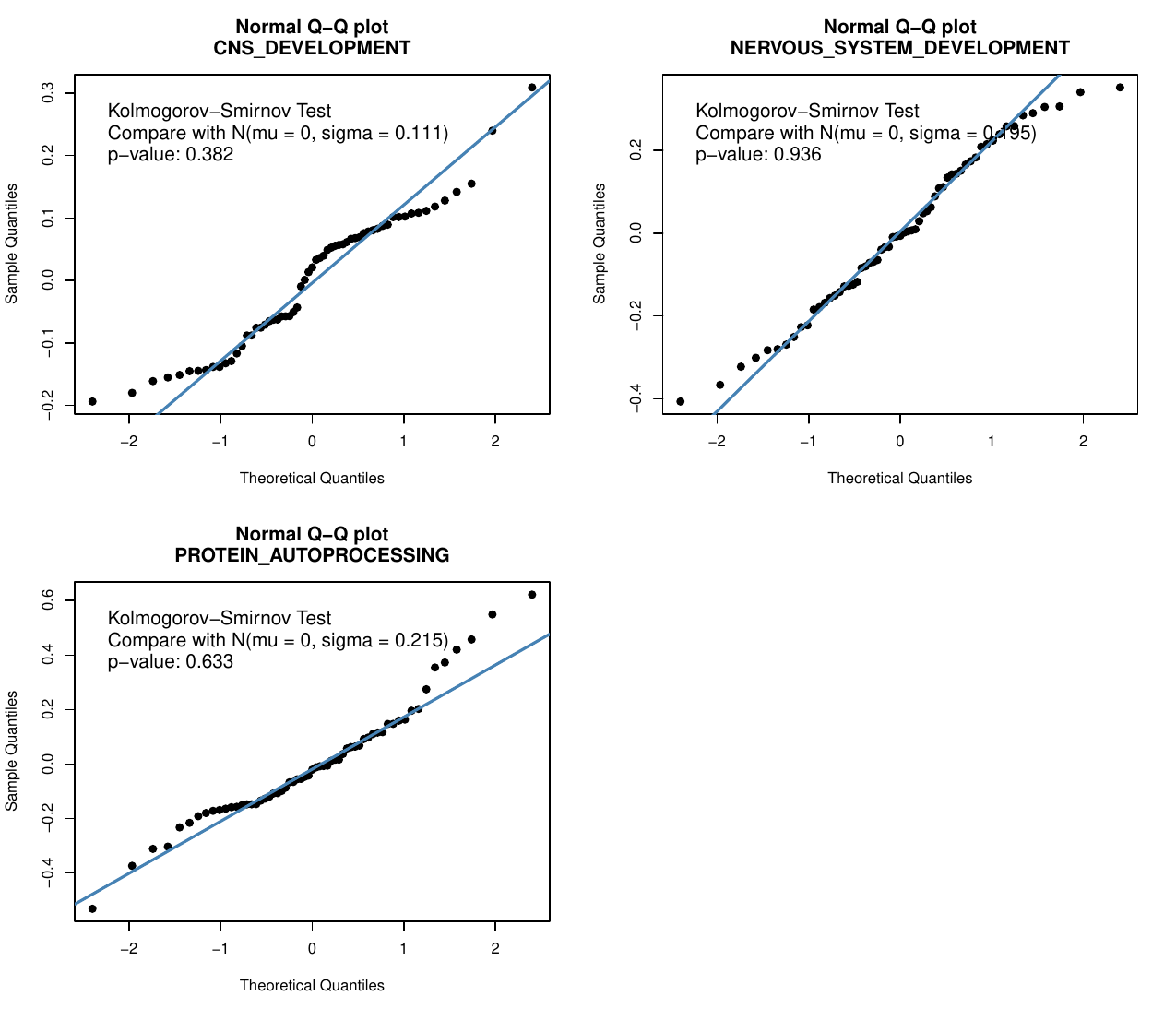}
		}
		\caption{Normal Q-Q plots for the error terms, $\y - X\hat{\bbeta}$, from the RADIOHEAD framework for different C-Pathways.}
		\label{fig: qq4}
	\end{figure}

\begin{figure}[H]
		\centering
		\resizebox{\textwidth}{!}{
			\includegraphics[page=2]{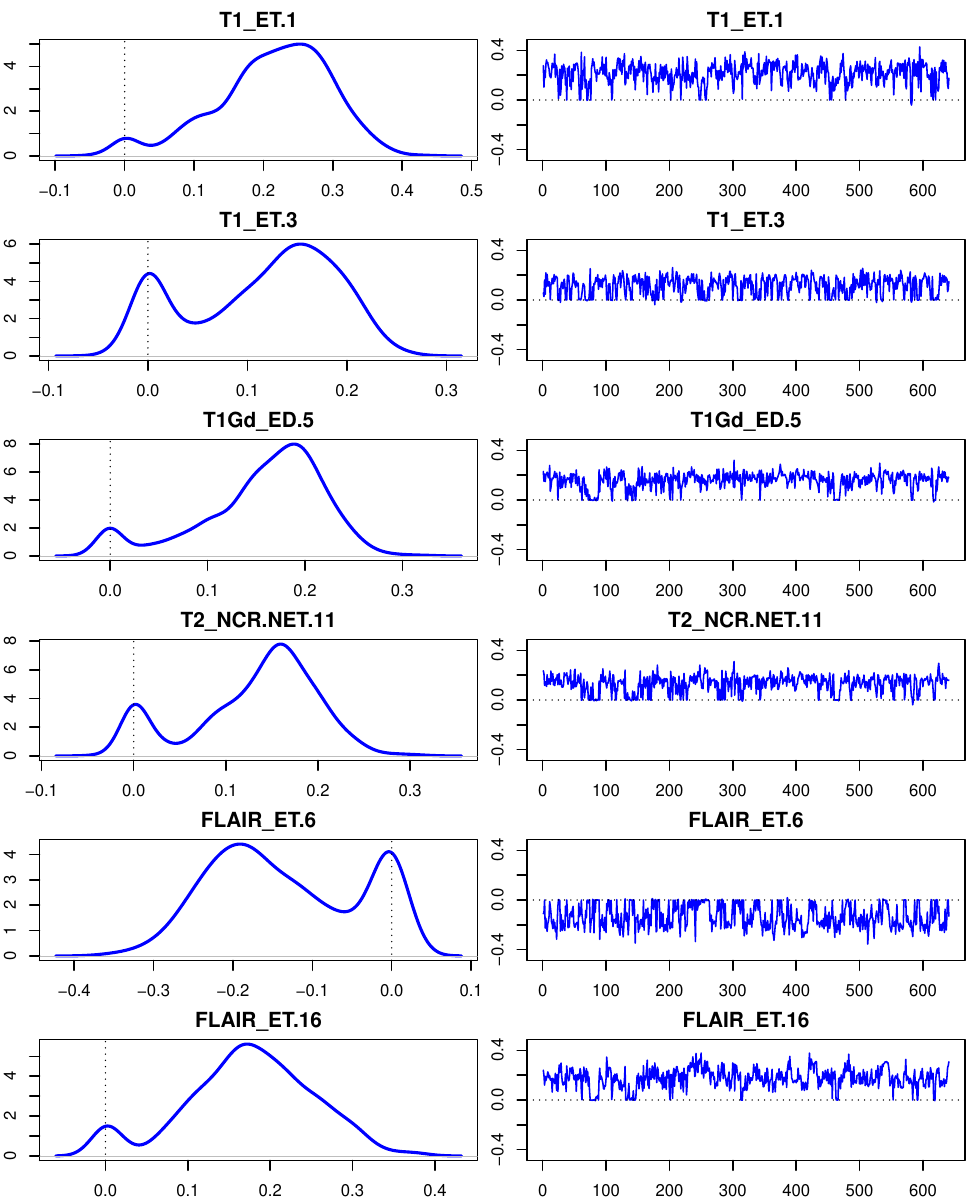}
		}
		\caption{Posterior densities and trace plots corresponding to $\beta_{gk}$ which were selected by the model for the transmission of nerve impulse pathway (pathway score computed with $n=61$ LGG subjects).}
		\label{fig: post_den_s}
	\end{figure}

\begin{figure}[H]
		\centering
		\resizebox{\textwidth}{!}{
			\includegraphics[page=1]{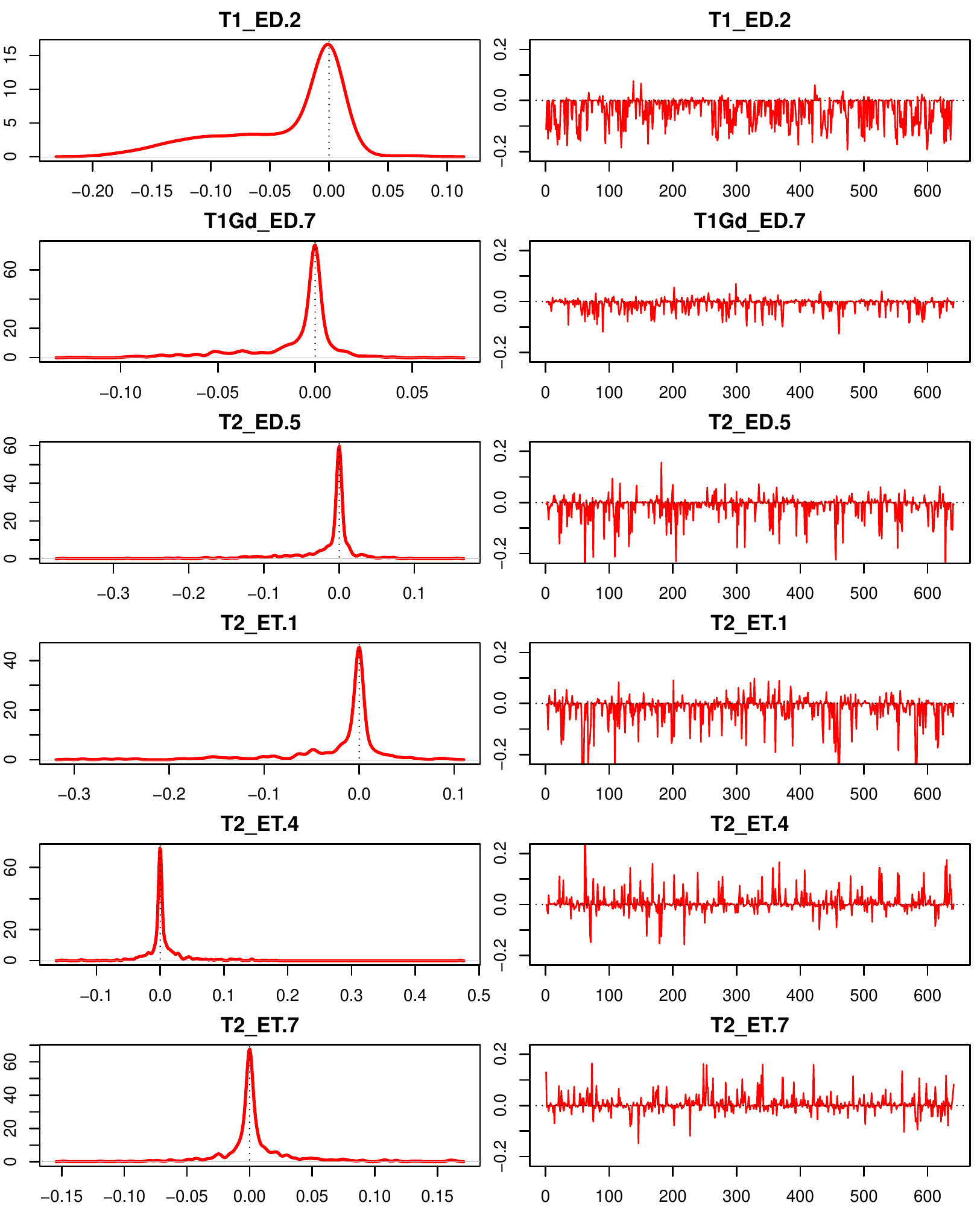}
		}
		\caption{Posterior densities and trace plots corresponding to $\beta_{gk}$ which were not selected by the model for the transmission of nerve impulse pathway (pathway score computed with $n=61$ LGG subjects).}
		\label{fig: post_den_ns}
	\end{figure}

\begin{figure}[H]
		\centering
		\resizebox{\textwidth}{!}{
			\includegraphics{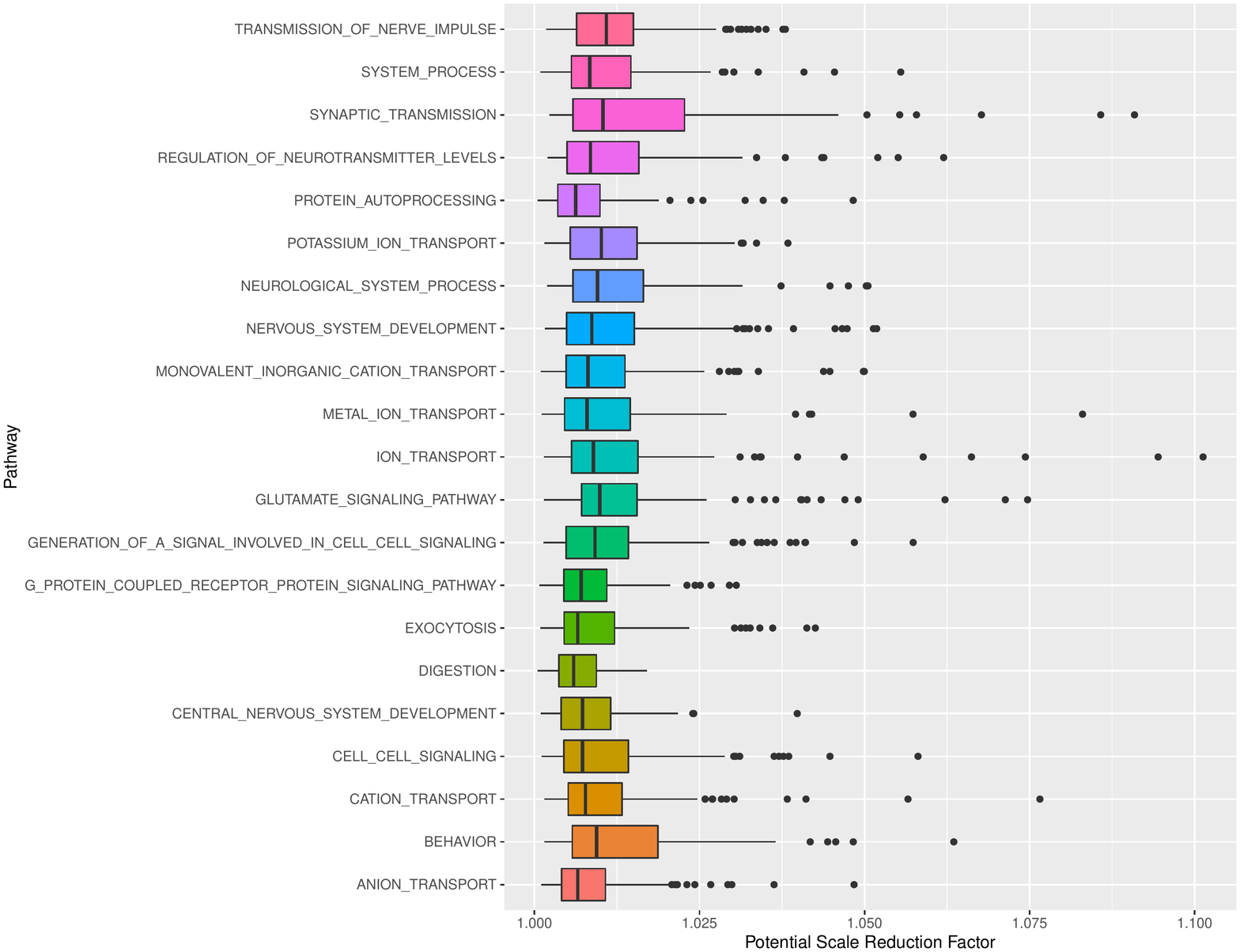}
		}
		\caption{Boxplots of the potential scale reduction factors (PSRF) computed based on the MCMC samples of $\beta_{gk}$ from seven different chains. For each pathway, we run a separate regression using the RADIOHEAD pipeline and the boxplot is constructed using the PSRFs corresponding to all of the coefficients $\beta_{gk}$. Values of PSRF close to 1 (or $<1.2$) indicate agreement across multiple chains.}
		\label{fig: multiple_chain}
	\end{figure}

	Figures \ref{fig: densities_trunc} and \ref{fig: densities} provide the truncated and full densities for all subjects across all four MRI imaging sequences and all three tumor sub-regions. For these densities, an overlaid Karcher mean is shown in Figure \ref{fig: densities_mean} with a truncated $y$-axis for better visualization. Figures S\ref{fig: t1_pc}-S\ref{fig: flair_pc} show the percentage of overall variance explained by the principal components constructed within each imaging sequence across all three tumor sub-regions. Table \ref{tab: pathway_summaries} shows the summary statistics for the pathway scores corresponding to C-pathways. Figures \ref{fig: qq1}-\ref{fig: qq4} show the normal Q-Q plots for the error terms, $\y - X\hat{\bbeta}$, from the RADIOHEAD framework for the C-pathways to validate the normality assumption of our model. Figures \ref{fig: post_den_s} and \ref{fig: post_den_ns} show diagnostic plots of the MCMC chain for randomly chosen $\beta_{gk}$. Specifically, Figure \ref{fig: post_den_s} shows the posterior densities and trace plots for six randomly chosen $\beta_{gk}$ which were selected by the model for the transmission of nerve impulse pathway, and Figure \ref{fig: post_den_ns} shows similar plots for six randomly chosen $\beta_{gk}$ which were not selected by the model for the same pathway. Figure \ref{fig: multiple_chain} shows boxplots of the potential scale reduction factors (PSRF) computed based on the MCMC samples of $\beta_{gk}$ from seven different chains for regression with each pathway separately \citep{gelman1992inference,brooks1998general}.
	
	\begin{landscape}
		\begin{table}[H]
			\centering
			\begin{tabular}{rrrrrrr}
				\hline
				& Min. & 1st Qu. & Median & Mean & 3rd Qu. & Max. \\ 
				\hline
				SYNAPTIC\_TRANSMISSION & -0.443 & -0.288 & -0.098 & -0.013 & 0.277 & 0.549 \\ 
				TRANSMISSION\_OF\_NERVE\_IMPULSE & -0.437 & -0.273 & -0.121 & -0.011 & 0.225 & 0.566 \\ 
				MONOVALENT\_INORGANIC\_CATION\_TRANSPORT & -0.289 & -0.150 & -0.051 & -0.009 & 0.095 & 0.356 \\ 
				NEUROLOGICAL\_SYSTEM\_PROCESS & -0.185 & -0.095 & -0.014 & 0.026 & 0.108 & 0.386 \\ 
				REGULATION\_OF\_NEUROTRANSMITTER\_LEVELS & -0.437 & -0.256 & -0.110 & -0.018 & 0.250 & 0.488 \\ 
				POTASSIUM\_ION\_TRANSPORT & -0.406 & -0.238 & -0.115 & -0.015 & 0.179 & 0.503 \\ 
				METAL\_ION\_TRANSPORT & -0.277 & -0.135 & -0.052 & -0.013 & 0.074 & 0.397 \\ 
				GENERATION\_OF\_A\_SIGNAL\_INVOLVED\_IN\_CELL\_CELL\_SIGNALING & -0.404 & -0.197 & -0.062 & -0.001 & 0.190 & 0.581 \\ 
				ION\_TRANSPORT & -0.242 & -0.114 & -0.047 & -0.011 & 0.074 & 0.341 \\ 
				CELL\_CELL\_SIGNALING & -0.332 & -0.180 & -0.059 & -0.007 & 0.156 & 0.469 \\ 
				SYSTEM\_PROCESS & -0.193 & -0.082 & -0.013 & 0.017 & 0.065 & 0.357 \\ 
				BEHAVIOR & -0.350 & -0.183 & -0.048 & -0.007 & 0.151 & 0.430 \\ 
				CATION\_TRANSPORT & -0.253 & -0.135 & -0.046 & -0.012 & 0.104 & 0.357 \\ 
				GLUTAMATE\_SIGNALING\_PATHWAY & -0.586 & -0.348 & -0.054 & -0.016 & 0.318 & 0.545 \\ 
				G\_PROTEIN\_COUPLED\_RECEPTOR\_PROTEIN\_SIGNALING\_PATHWAY & -0.344 & -0.168 & -0.033 & 0.008 & 0.143 & 0.471 \\ 
				EXOCYTOSIS & -0.333 & -0.174 & 0.004 & -0.005 & 0.115 & 0.385 \\ 
				DIGESTION & -0.232 & -0.088 & 0.004 & 0.033 & 0.171 & 0.354 \\ 
				ANION\_TRANSPORT & -0.209 & -0.096 & -0.033 & -0.004 & 0.096 & 0.340 \\ 
				CENTRAL\_NERVOUS\_SYSTEM\_DEVELOPMENT & -0.268 & -0.100 & 0.036 & -0.003 & 0.094 & 0.261 \\ 
				NERVOUS\_SYSTEM\_DEVELOPMENT & -0.318 & -0.165 & -0.012 & -0.011 & 0.103 & 0.357 \\ 
				PROTEIN\_AUTOPROCESSING & -0.444 & -0.146 & -0.005 & 0.004 & 0.162 & 0.409 \\ 
				\hline
			\end{tabular}
			\caption{Summary statistics for the pathway scores corresponding to C-pathways.}
			\label{tab: pathway_summaries}
		\end{table}
	\end{landscape}
	
	\section{Calibration of Pathway Scores}\label{sec: supp_calibration}
	
\begin{figure}[H]
		\centering
		\resizebox{\textwidth}{!}{%
			\includegraphics[page=1]{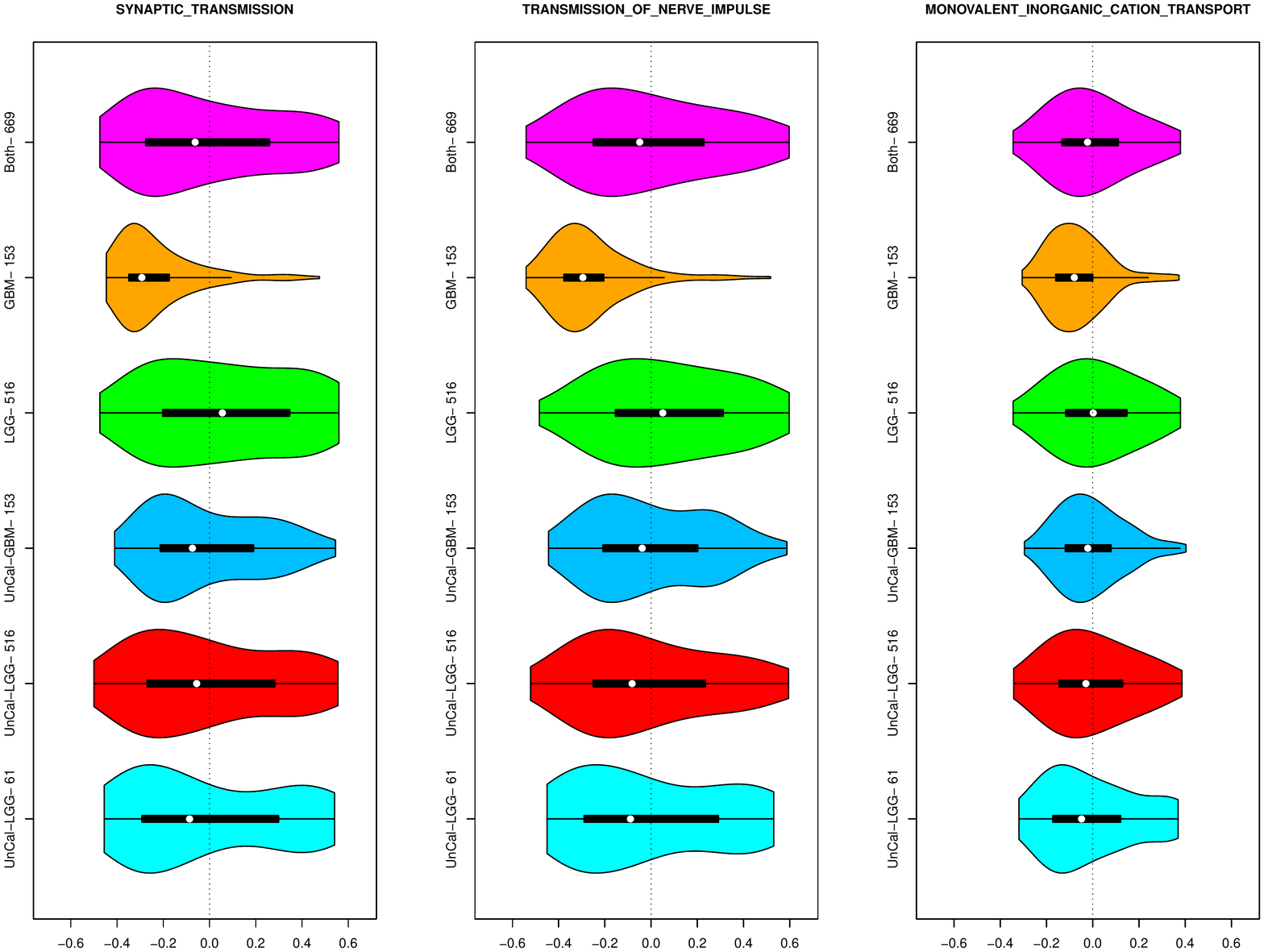}
		}
		\caption{Calibration of pathway scores: Violin plots for the pathway scores (computed by pooling all 669 LGG and GBM subjects) of (a) 516 LGG and 153 GBM subjects (Both-669), (b) 153 GBM subjects (GBM-153) and (c) 516 LGG subjects (LGG-516). Violin plots for pathway scores (computed without pooling) of (d) 153 GBM subjects (UnCal-GBM-153), (e) 153 LGG subjects (UnCal-LGG-516), and (f) 61 LGG subjects for whom the imaging was available (UnCal-LGG-61).}
		\label{fig: calib_1}
	\end{figure}

\begin{figure}[H]
		\centering
		\resizebox{\textwidth}{!}{%
			\includegraphics[page=2]{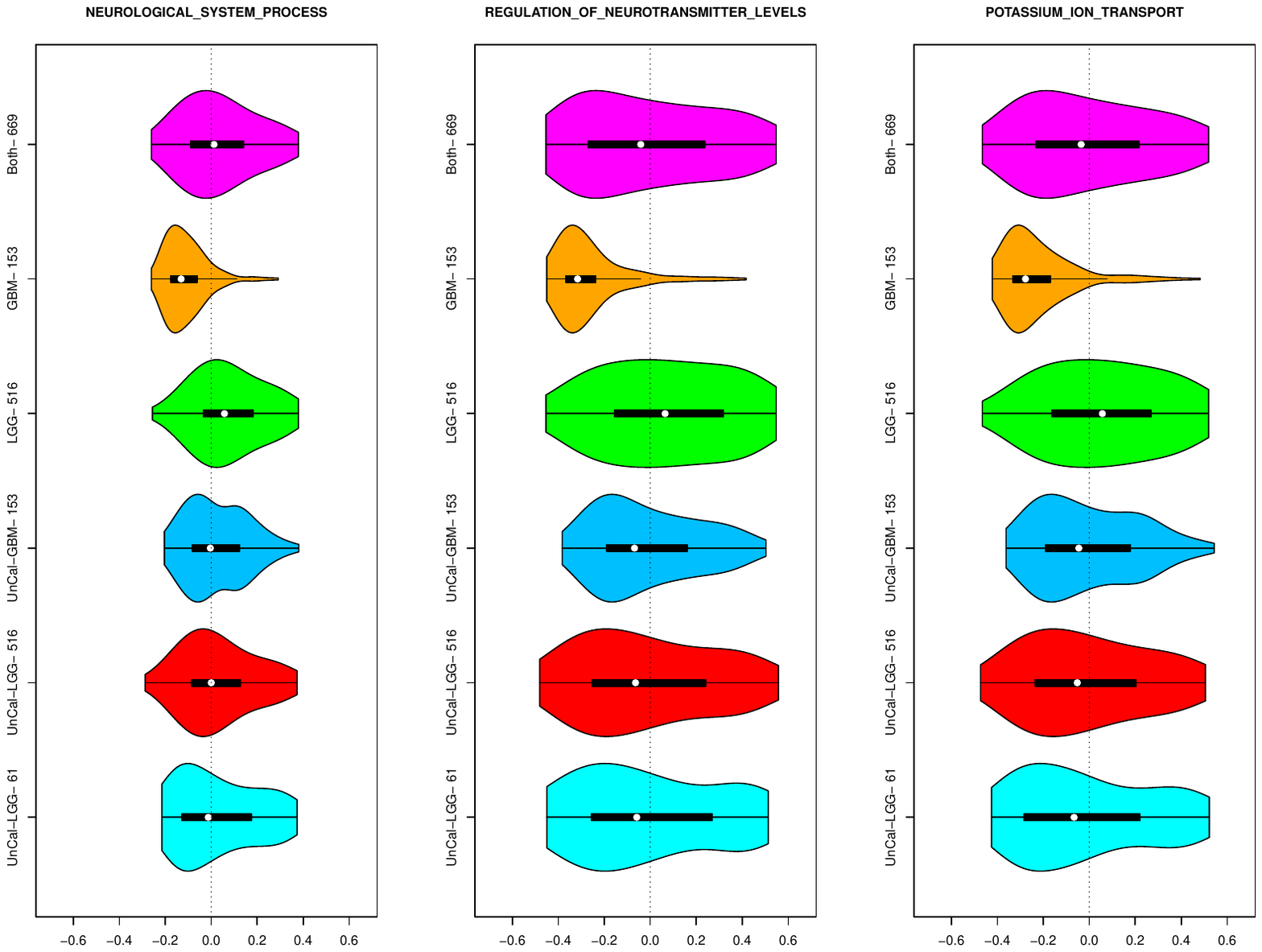}
		}
		\caption{Calibration of pathway scores: Violin plots for the pathway scores (computed by pooling all 669 LGG and GBM subjects) of (a) 516 LGG and 153 GBM subjects (Both-669), (b) 153 GBM subjects (GBM-153) and (c) 516 LGG subjects (LGG-516). Violin plots for pathway scores (computed without pooling) of (d) 153 GBM subjects (UnCal-GBM-153), (e) 153 LGG subjects (UnCal-LGG-516), and (f) 61 LGG subjects for whom the imaging was available (UnCal-LGG-61).}
		\label{fig: calib_2}
	\end{figure}

\begin{figure}[H]
		\centering
		\resizebox{\textwidth}{!}{%
			\includegraphics[page=3]{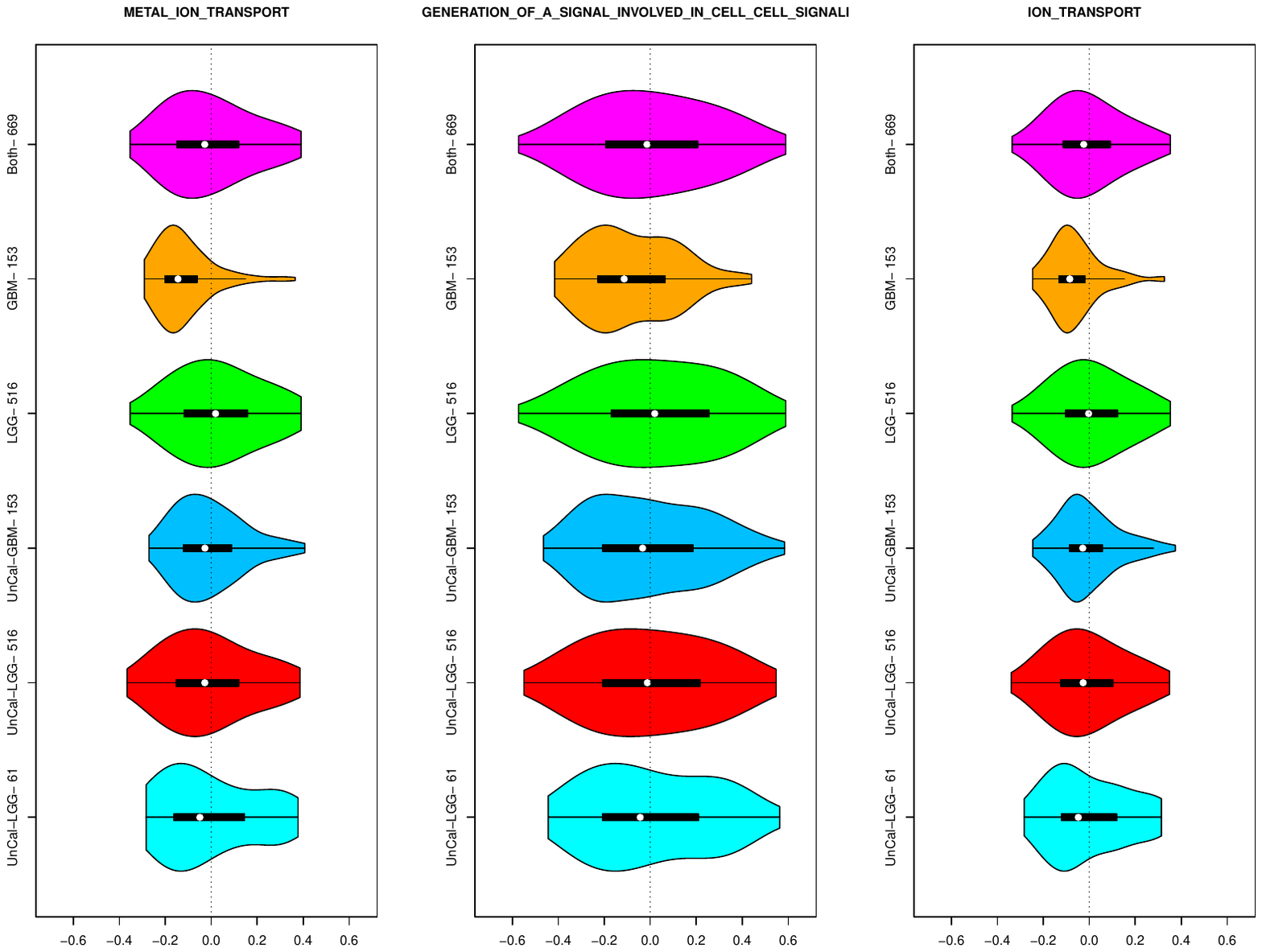}
		}
		\caption{Calibration of pathway scores: Violin plots for the pathway scores (computed by pooling all 669 LGG and GBM subjects) of (a) 516 LGG and 153 GBM subjects (Both-669), (b) 153 GBM subjects (GBM-153) and (c) 516 LGG subjects (LGG-516). Violin plots for pathway scores (computed without pooling) of (d) 153 GBM subjects (UnCal-GBM-153), (e) 153 LGG subjects (UnCal-LGG-516), and (f) 61 LGG subjects for whom the imaging was available (UnCal-LGG-61).}
		\label{fig: calib_3}
	\end{figure}

\begin{figure}[H]
		\centering
		\resizebox{\textwidth}{!}{%
			\includegraphics[page=4]{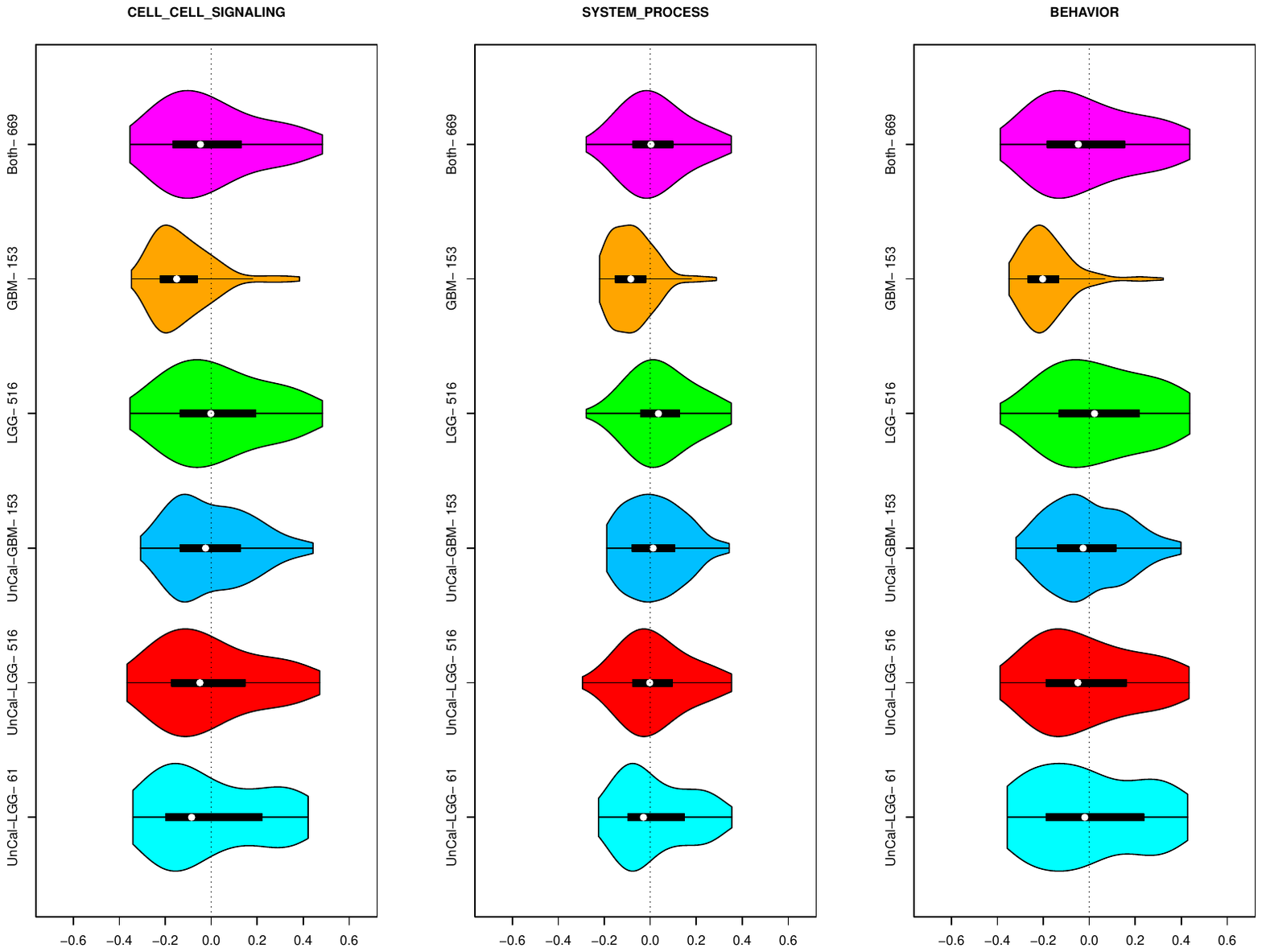}
		}
		\caption{Calibration of pathway scores: Violin plots for the pathway scores (computed by pooling all 669 LGG and GBM subjects) of (a) 516 LGG and 153 GBM subjects (Both-669), (b) 153 GBM subjects (GBM-153) and (c) 516 LGG subjects (LGG-516). Violin plots for pathway scores (computed without pooling) of (d) 153 GBM subjects (UnCal-GBM-153), (e) 153 LGG subjects (UnCal-LGG-516), and (f) 61 LGG subjects for whom the imaging was available (UnCal-LGG-61).}
		\label{fig: calib_4}
	\end{figure}

\begin{figure}[H]
		\centering
		\resizebox{\textwidth}{!}{%
			\includegraphics[page=5]{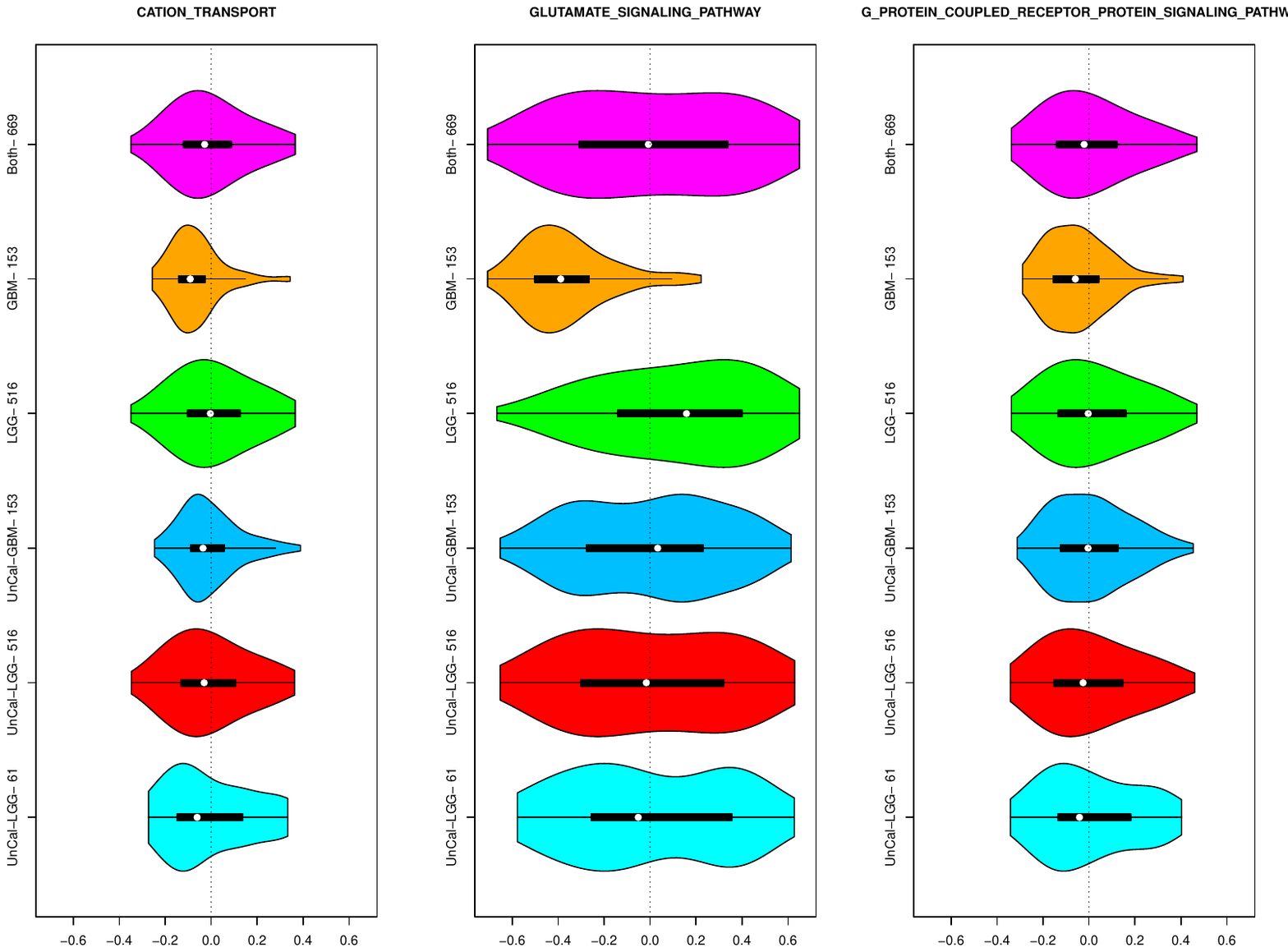}
		}
		\caption{Calibration of pathway scores: Violin plots for the pathway scores (computed by pooling all 669 LGG and GBM subjects) of (a) 516 LGG and 153 GBM subjects (Both-669), (b) 153 GBM subjects (GBM-153) and (c) 516 LGG subjects (LGG-516). Violin plots for pathway scores (computed without pooling) of (d) 153 GBM subjects (UnCal-GBM-153), (e) 153 LGG subjects (UnCal-LGG-516), and (f) 61 LGG subjects for whom the imaging was available (UnCal-LGG-61).}
		\label{fig: calib_5}
	\end{figure}

\begin{figure}[H]
		\centering
		\resizebox{\textwidth}{!}{%
			\includegraphics[page=6]{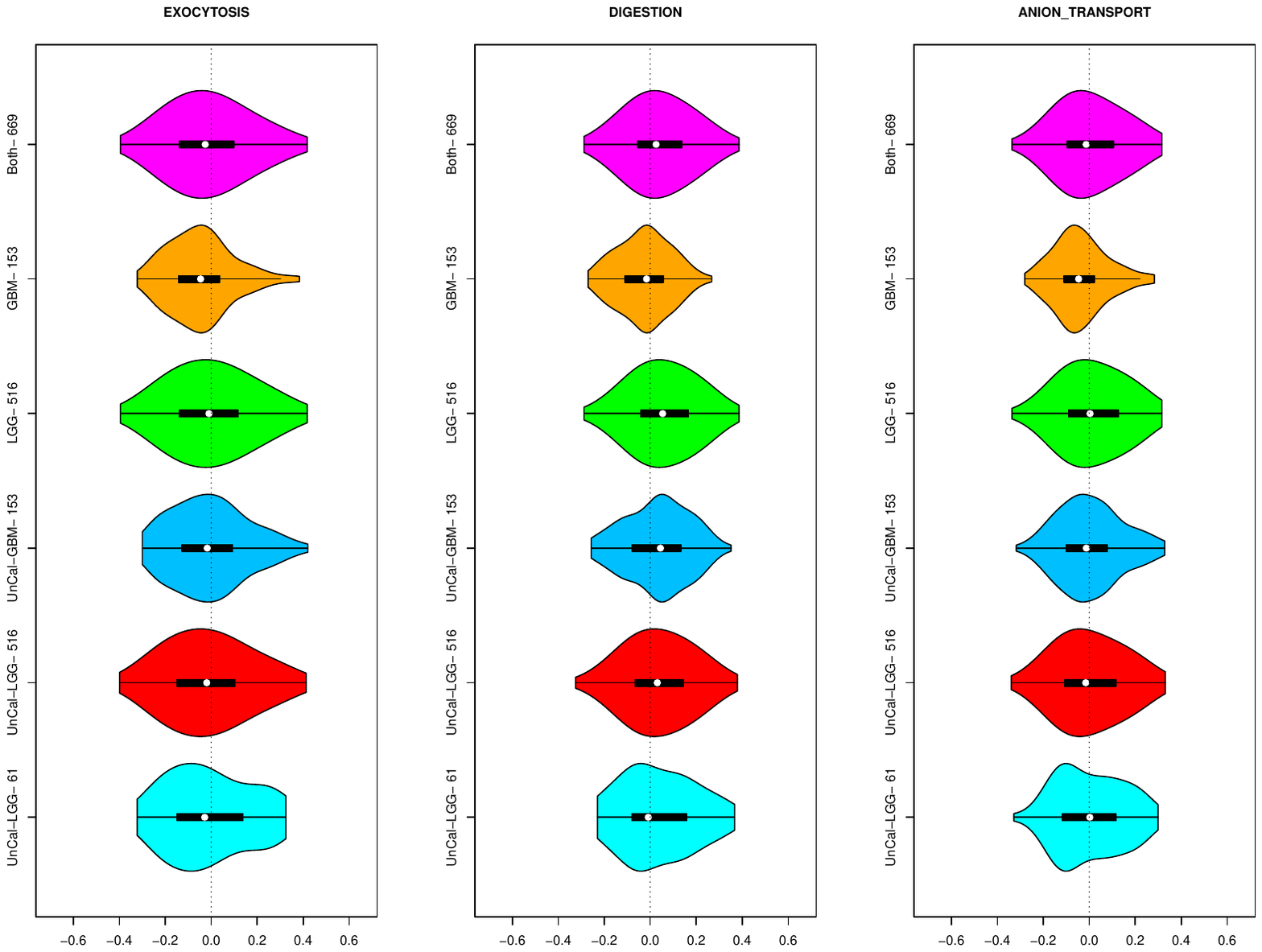}
		}
		\caption{Calibration of pathway scores: Violin plots for the pathway scores (computed by pooling all 669 LGG and GBM subjects) of (a) 516 LGG and 153 GBM subjects (Both-669), (b) 153 GBM subjects (GBM-153) and (c) 516 LGG subjects (LGG-516). Violin plots for pathway scores (computed without pooling) of (d) 153 GBM subjects (UnCal-GBM-153), (e) 153 LGG subjects (UnCal-LGG-516), and (f) 61 LGG subjects for whom the imaging was available (UnCal-LGG-61).}
		\label{fig: calib_6}
	\end{figure}

\begin{figure}[H]
		\centering
		\resizebox{\textwidth}{!}{%
			\includegraphics[page=7]{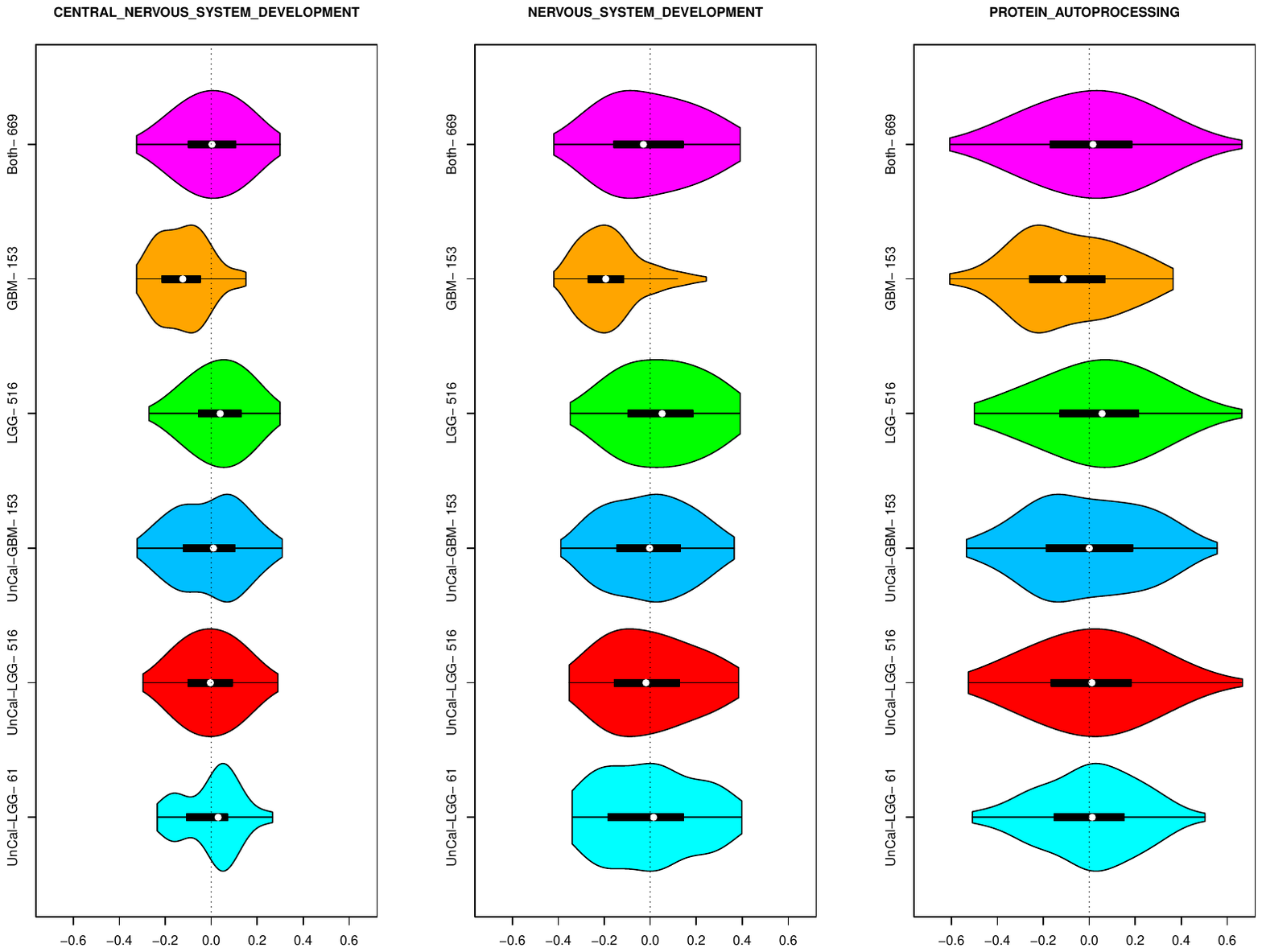}
		}
		\caption{Calibration of pathway scores: Violin plots for the pathway scores (computed by pooling all 669 LGG and GBM subjects) of (a) 516 LGG and 153 GBM subjects (Both-669), (b) 153 GBM subjects (GBM-153) and (c) 516 LGG subjects (LGG-516). Violin plots for pathway scores (computed without pooling) of (d) 153 GBM subjects (UnCal-GBM-153), (e) 153 LGG subjects (UnCal-LGG-516), and (f) 61 LGG subjects for whom the imaging was available (UnCal-LGG-61).}
		\label{fig: calib_7}
	\end{figure}

	The computation of pathway scores can be sensitive to sample composition. To better understand this in our context, we explore their distribution when computed using different sample cohorts. We use the genomic data available for LGG and GBM from TCGA. We have a total of 516 LGG subjects; we use a subset of 61 subjects with matched imaging data in our model. We also have a total of 153 GBM subjects. We compute the pathway scores for these patient cohorts for the C-Pathways based on four different sample compositions: (a) 516 LGG and 153 GBM subjects (Both-669), (b) 153 GBM subjects (UnCal-GBM-153), (c) 516 LGG subjects (UnCal-LGG-516), and (d) 61 LGG subjects with matched imaging data (UnCal-LGG-61). In Figures \ref{fig: calib_1}-\ref{fig: calib_7}, we show violin plots corresponding to the distribution of pathway scores for the sample cohorts in (a)-(d). When the pathway scores are computed for case (a), that is, by pooling subjects from both LGG and GBM, we plot the individual distributions of pathway scores for the LGG patients (LGG-516) and GBM patients (GBM-153).
	
\begin{figure}[H]
		\centering
		\resizebox{\textwidth}{!}{%
			\includegraphics[page=1]{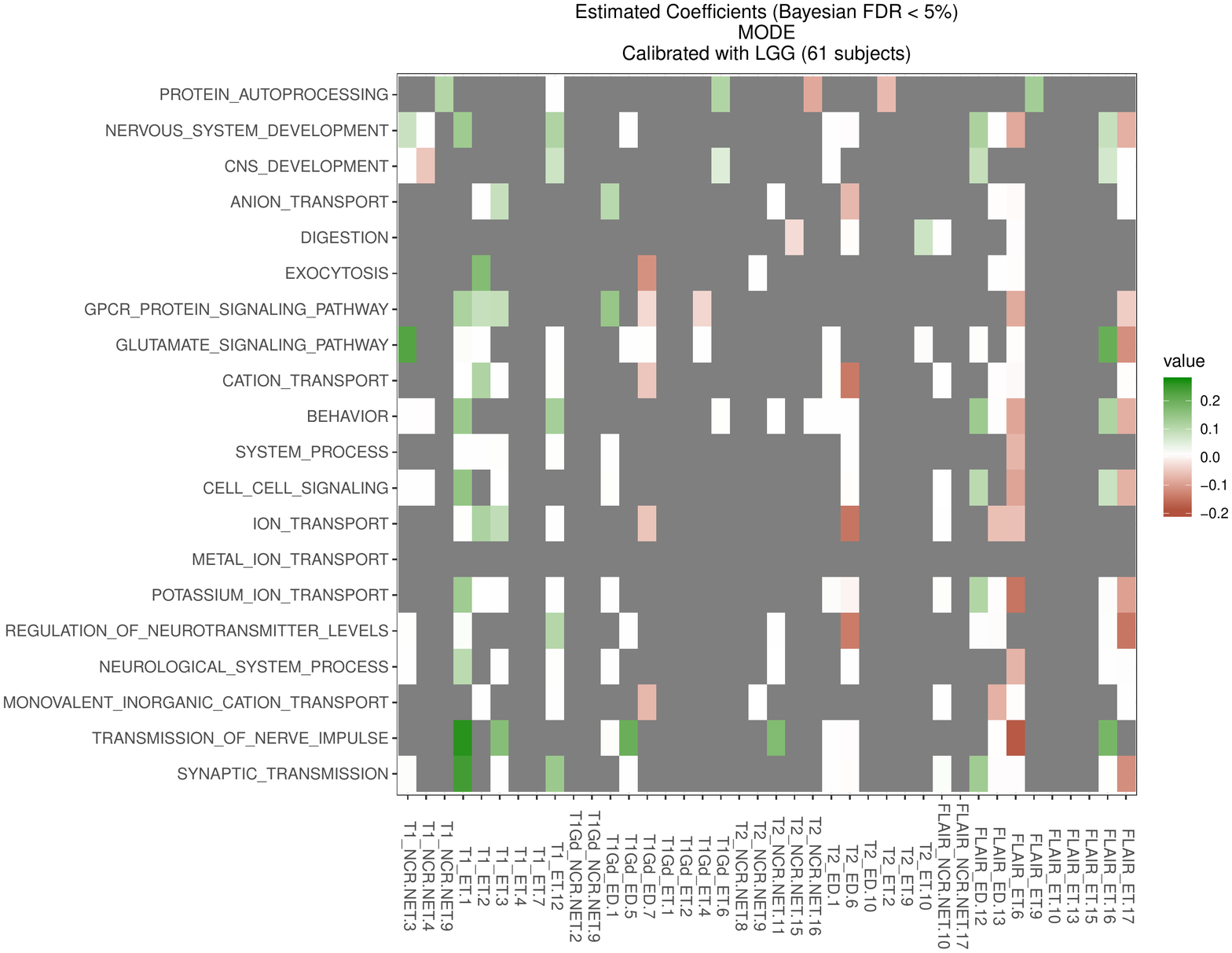}
		}
		\caption{Posterior estimates of $\beta_{gk}$, after FDR-based variable selection at level $0.05$, corresponding to different principal component scores across MRI sequences and tumor sub-regions. Each row corresponds to a pathway from the C-Pathways. The pathway scores for the $61$ subjects in our analysis are computed using $n=61$ LGG samples (no calibration).}
		\label{fig: coef_calib_1}
	\end{figure}

\begin{figure}
		\centering
		\resizebox{\textwidth}{!}{%
			\includegraphics[page=2]{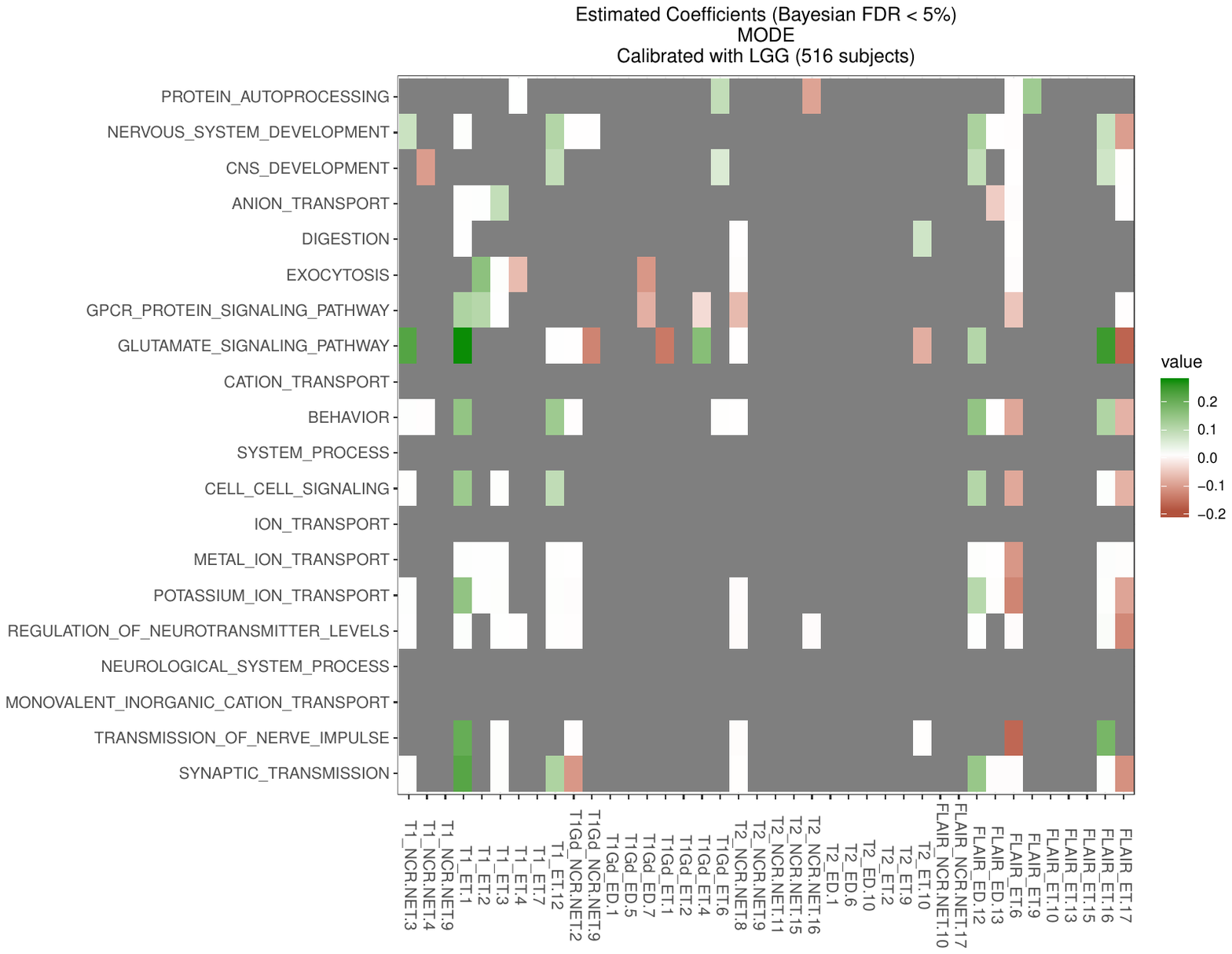}
		}
		\caption{Posterior estimates of $\beta_{gk}$, after FDR-based variable selection at level $0.05$, corresponding to different principal component scores across MRI sequences and tumor sub-regions. Each row corresponds to a pathway from the C-Pathways. The pathway scores for the $61$ subjects in our analysis are computed by calibrating with $n=516$ LGG samples.}
		\label{fig: coef_calib_2}
	\end{figure}

\begin{figure}
		\centering
		\resizebox{\textwidth}{!}{%
			\includegraphics[page=3]{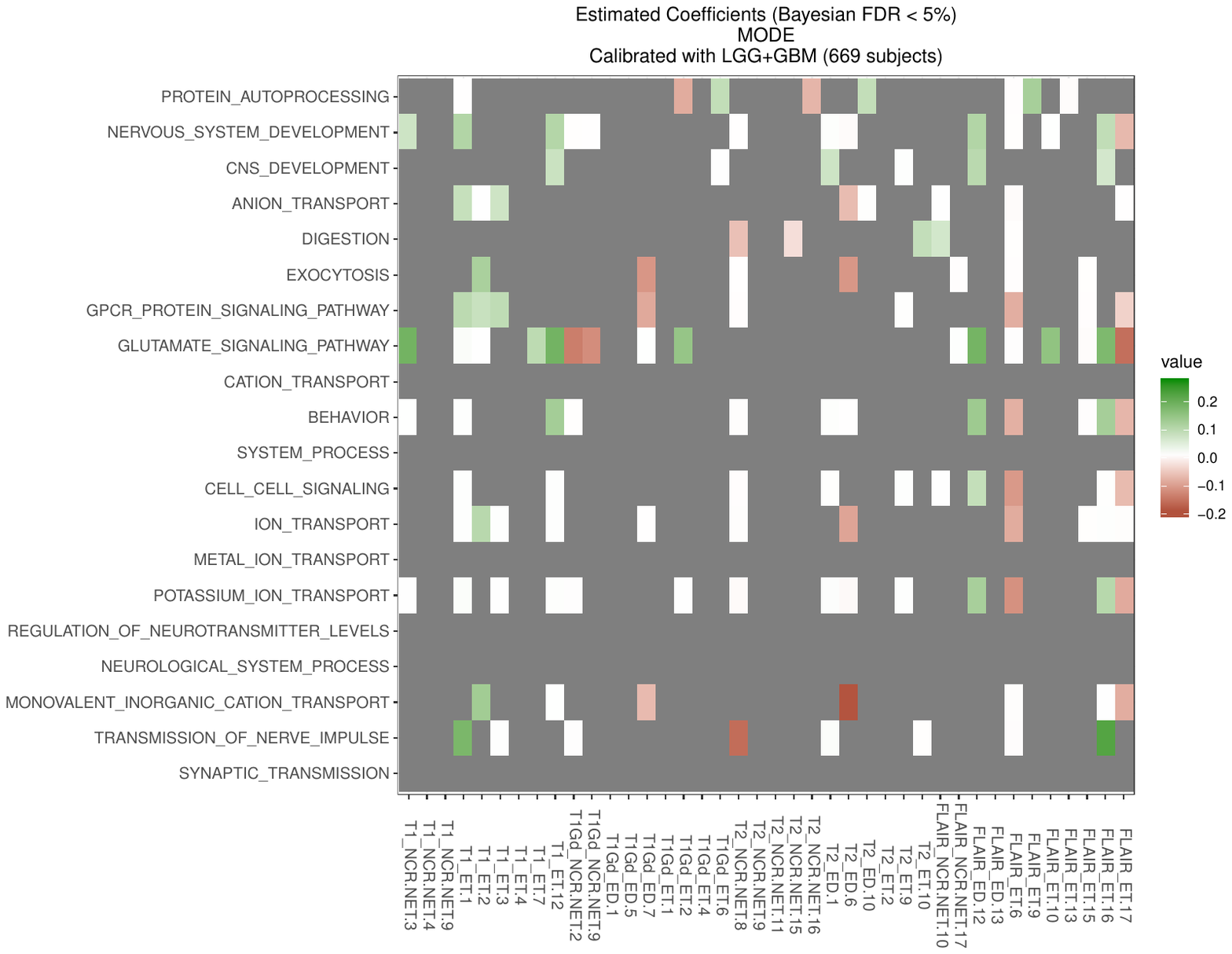}
		}
		\caption{Posterior estimates of $\beta_{gk}$, after FDR-based variable selection at level $0.05$, corresponding to different principal component scores across MRI sequences and tumor sub-regions. Each row corresponds to a pathway from the C-Pathways. The pathway scores for the $61$ subjects in our analysis are computed by calibrating with $n=669$ ($516$ LGG and $153$ GBM) samples.}
		\label{fig: coef_calib_3}
	\end{figure}

	In Figure \ref{fig: coef_calib_1}, we plot posterior estimates of $\beta_{gk}$ corresponding to different principal component scores across MRI sequences and tumor sub-regions. The pathway scores for the $61$ subjects in our analysis are computed using $n=61$ LGG samples (no calibration). Similarly, in Figures \ref{fig: coef_calib_2} and \ref{fig: coef_calib_3}, the pathway scores for the $61$ subjects in our analysis are computed using $n=516$ LGG samples and $n=669$ ($516$ LGG and $153$ GBM subjects) samples respectively.
	
	\newpage
	\section{Utility of Densities as Predictors}\label{sec: utility}
\begin{figure}[H]
		\centering
		\resizebox{\textwidth}{!}{
			\begin{tabular}{|c|}
				\hline
				\begin{tabular}{c|c|c}
					\subfigure[Mean]{\label{fig: evcase1}\includegraphics[trim=0cm 0cm 0cm 0.5cm, clip, width=1in,height=4.25in,page=1]{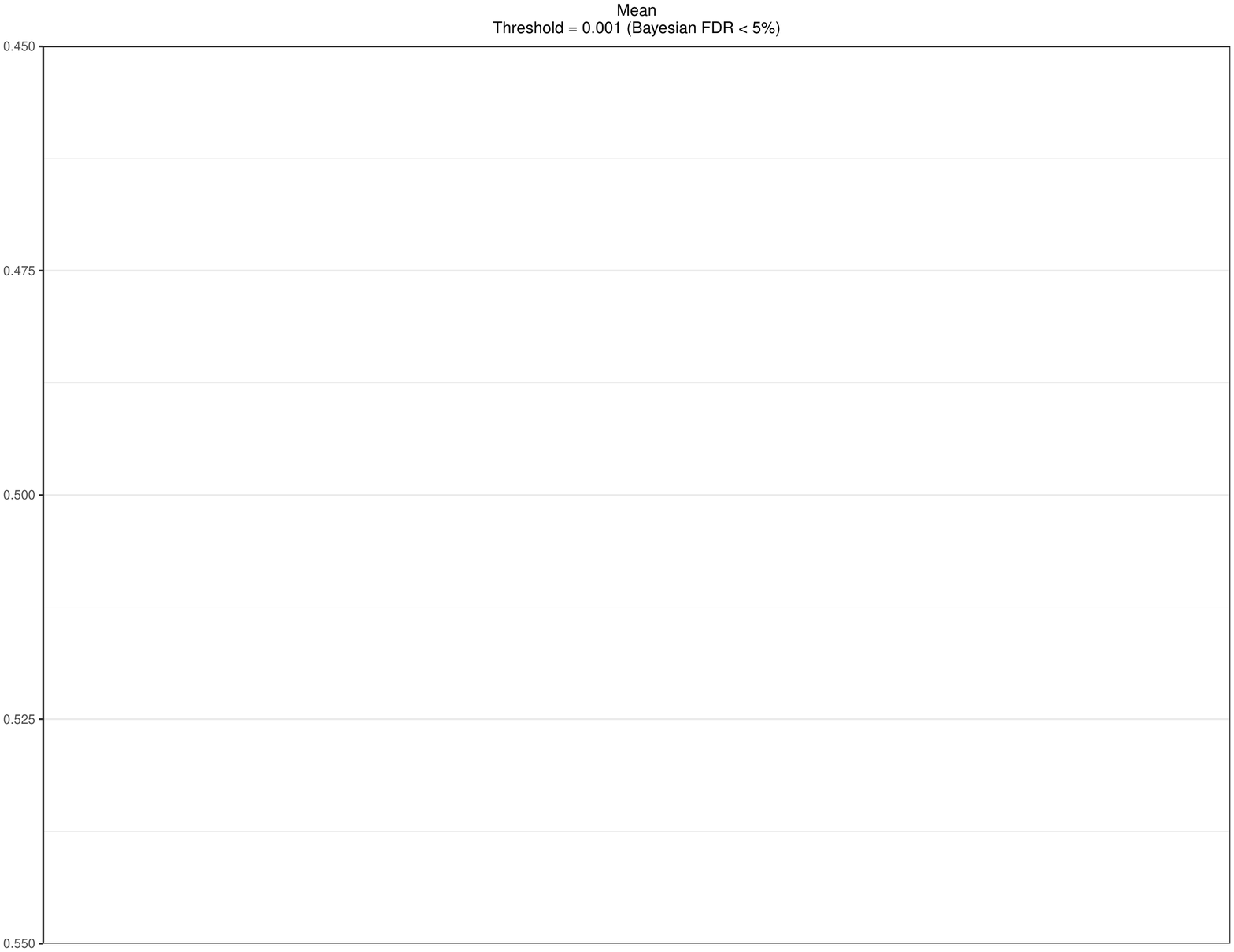}} &
					\subfigure[Mean,Q1,Q3]{\includegraphics[trim=0cm 0cm 0cm 0.5cm, clip, width=1in,height=4.25in,page=2]{plots/coef_quantiles_Mode.pdf}} &
					\subfigure[Five-Number Summary]{\label{fig: evcase3}\includegraphics[trim=0cm 0cm 0cm 0.5cm, clip, width=4.2in,height=4.25in,page=3]{plots/coef_quantiles_Mode.pdf}}
				\end{tabular}\\
				\hline
			\end{tabular}
		}
		\caption{Estimated coefficients when the predictors are (a) mean, (b) first-quartile, mean, and third quartile, (c) five-number summary.}
		\label{fig: evcase2}
	\end{figure}

\begin{figure}[H]
		\centering
		\resizebox{\textwidth}{!}{
			\begin{tabular}{|c|}
				\hline
				\begin{tabular}{c}
					\subfigure[Mean, SD, Skewness, Kurtosis]{\includegraphics[trim=0cm 0cm 0cm 0.5cm, clip, width=6.5in,height=4.1in,page=4]{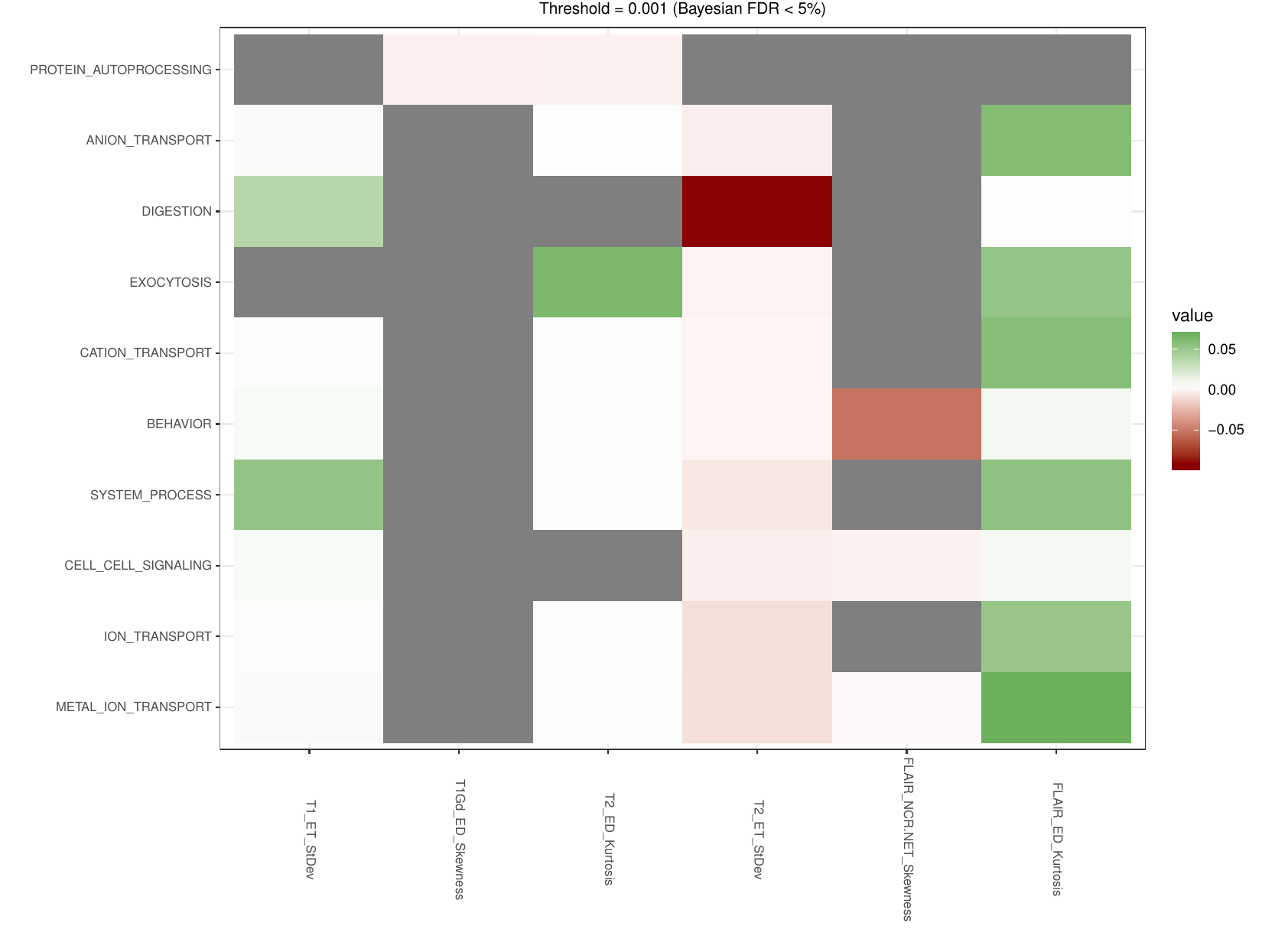}}
				\end{tabular}\\
				\hline
				\begin{tabular}{c}
					\subfigure[Deciles]{\label{fig: evcase5}\includegraphics[trim=0cm 0cm 0cm 0.5cm, clip, width=6.5in,height=4.1in,page=5]{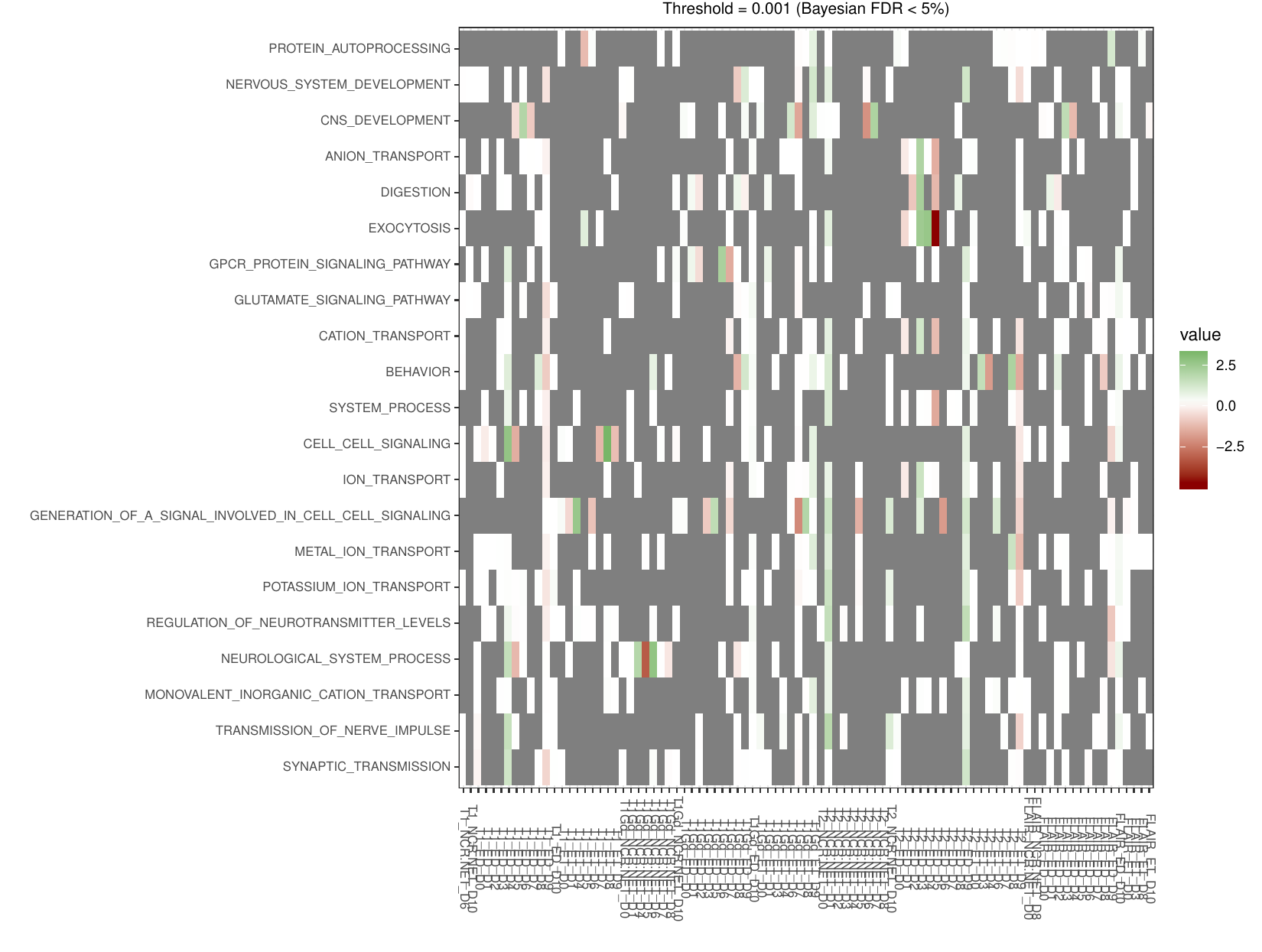}}
				\end{tabular}\\
				\hline
			\end{tabular}
		}
		\caption{Estimated coefficients when the predictors are (a) mean, standard deviation, skewness, and kurtosis, and (b) deciles.}
		\label{fig: evcase4}
	\end{figure}

\begin{figure}[H]
		\centering
		\resizebox{\textwidth}{!}{
			\begin{tabular}{|c|}
				\hline
				\begin{tabular}{c}
					\subfigure[15 Equally Spaced Percentiles]{\includegraphics[trim=0cm 0cm 0cm 0.5cm, clip, width=6.5in,height=4.1in,page=1]{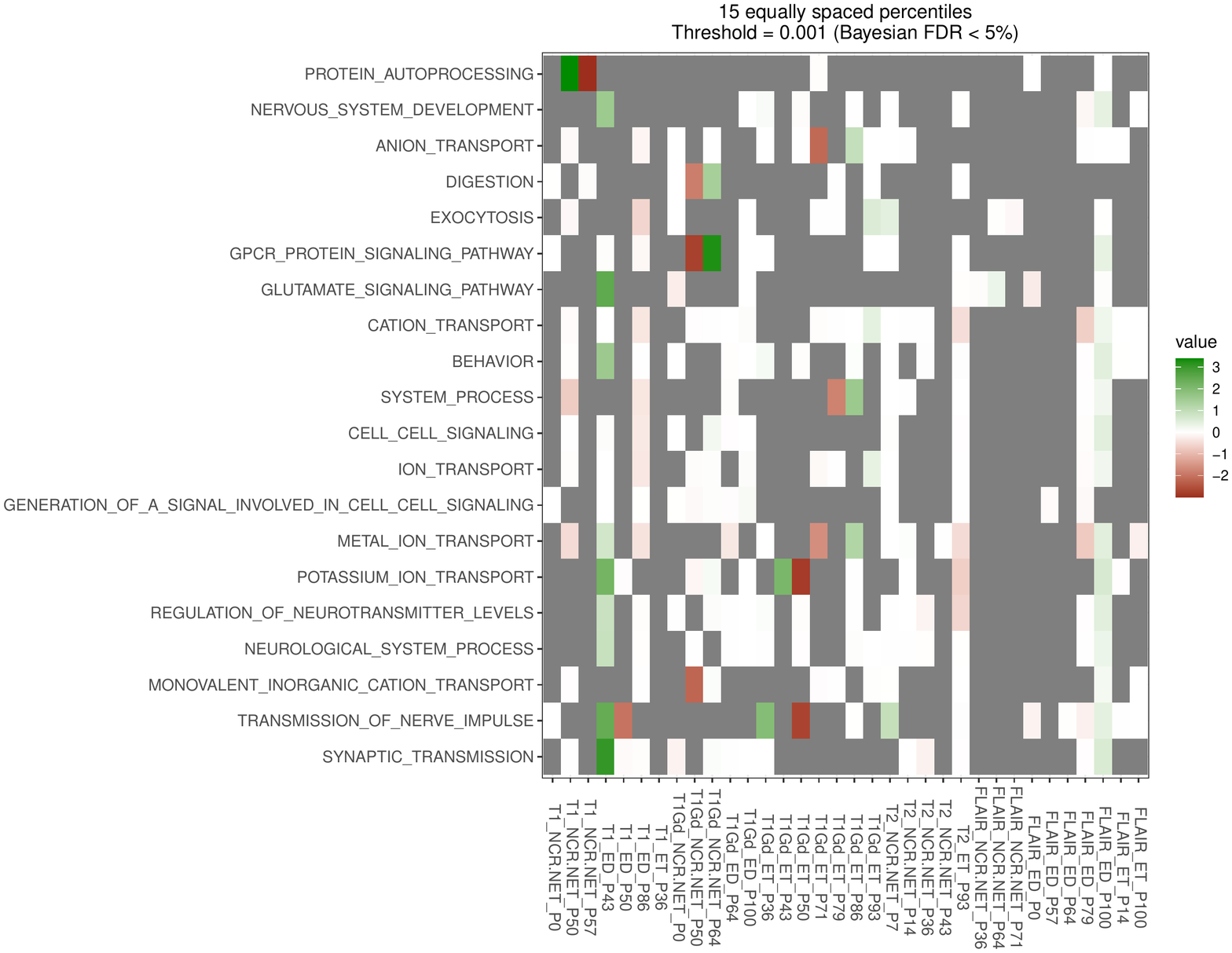}}
				\end{tabular}\\
				\hline
				\begin{tabular}{c}
					\subfigure[20 Equally Spaced Percentiles]{\label{fig: evcase7}\includegraphics[trim=0cm 0cm 0cm 0.5cm, clip, width=6.5in,height=4.1in,page=2]{plots/coef_quantiles_15_Mode.pdf}}
				\end{tabular}\\
				\hline
			\end{tabular}
		}
		\caption{Estimated coefficients when the predictors are (a) 15 equally spaced percentiles, and (b) 20 equally spaced percentiles.}
		\label{fig: evcase6}
	\end{figure}

\begin{figure}[H]
		\centering
		\resizebox{\textwidth}{!}{
			\includegraphics[page=1,trim=0cm 0.5cm 0cm 0cm, clip]{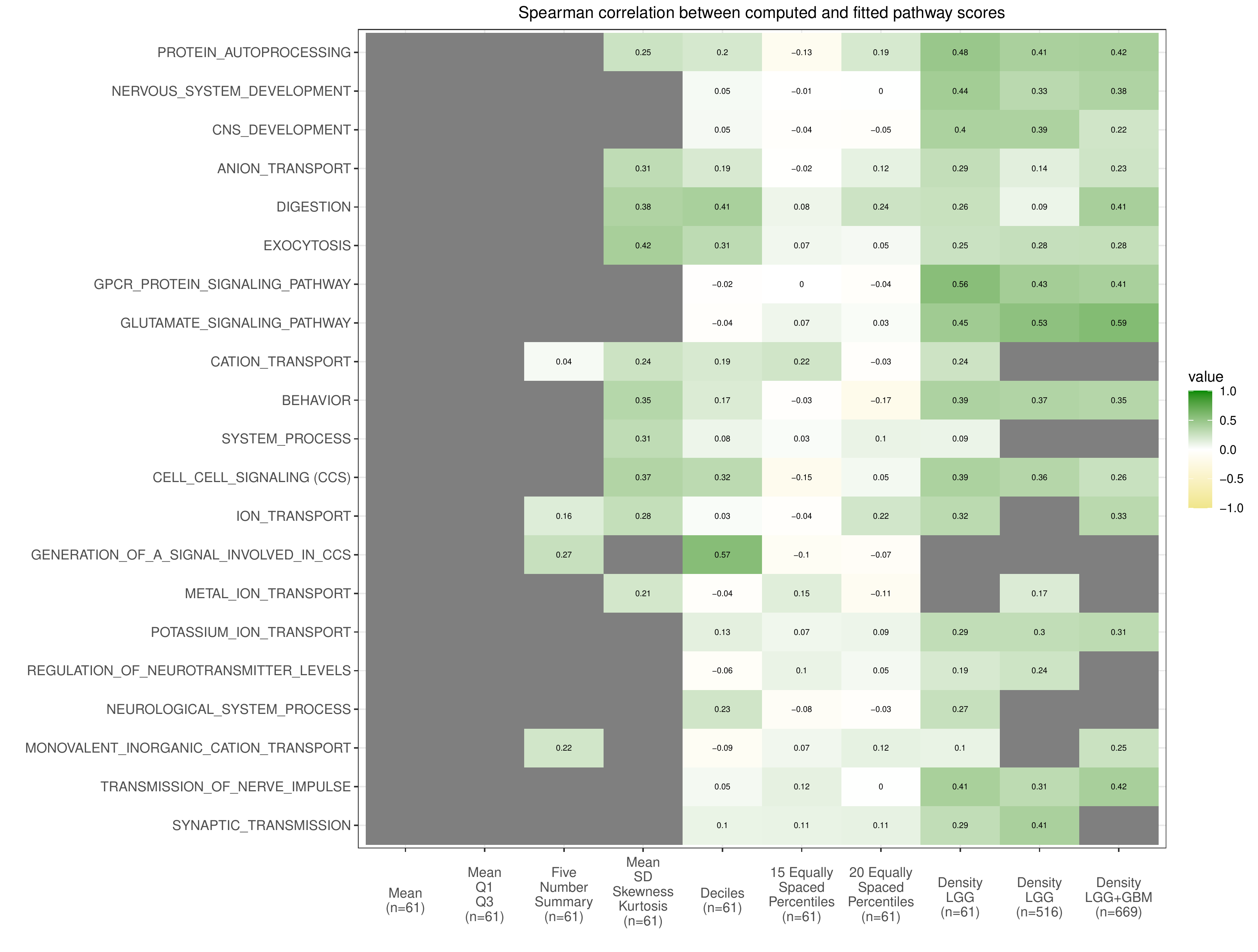}
		}
		\caption{Spearman correlation between observed and predicted pathway scores while various sets of covariates are used.}
		\label{fig: spearman}
	\end{figure}

	We show results of our model using seven different types of potential predictors: (a) mean, (b) mean, first and third quartiles ($Q_1$ and $Q_3$), (c) five-number summary, (d) mean, standard deviation, skewness and kurtosis, (e) deciles, (f) 15 equally spaced percentiles, and (g) 20 equally spaced percentiles. The results based on all of these seven cases are presented in Figures \ref{fig: evcase1}-\ref{fig: evcase7}. In Figure \ref{fig: spearman}, we show the Spearman correlations between the computed (observed) pathway scores and the predicted (using density-based meta-features and the corresponding estimated coefficients after variable selection) pathway scores.
	
	\newpage
	\section{Inference on Group-Level Indicator}\label{sec: group}
	
	We perform a simulation study to compare the performance of RADIOHEAD with (a) a Bayesian group selection (BGS) approach developed for a high-dimensional setting \citep{yang2018consistent}, and (b) the group LASSO (G-LASSO) approach \citep{yuan2006model}. BGS is a Bayesian hierarchical model with a spike-and-slab prior specification to perform group selection in high-dimensional linear regression models. The BGS model formulation is similar to RADIOHEAD, except that the prior on $\bbeta$ is a mixture of a point mass at zero, and a normal distribution which defines the slab. However, unlike RADIOHEAD, the group-level indicator variable in BGS is identifiable, making inference on it feasible. The G-LASSO is devised to select grouped variables (factors) for accurate prediction in regression.
	
	For simplicity, we consider the PC scores constructed for the T1 imaging sequence from our LGG data as predictors. This includes three groups of PC scores corresponding to the NC, ED and ET sub-regions, with $9$, $7$ and $12$ PCs, respectively. Hence we have $X \in \mathbb{R}^{61 \times 28}$ with 61 subjects and the 28 covariates divided into three groups. We standardize the columns of $X$ for further analysis. We define $\bbeta = [\bbeta_1 ~\bbeta_2 ~\bbeta_3] \in \mathbb{R}^{28}$ such that
	\begin{itemize}
		\item $\bbeta_1 = (\beta_{1,1},\ldots,\beta_{1,9}) \in \mathbb{R}^{9}$, where $\beta_{1,k}$ is simulated from a double-exponential distribution with a scale parameter $\theta = 1$,
		\item $\beta_{2,1} = 1$ and $\beta_{2,k} = 0$ for all $k=2,\ldots,7$, and
		\item $\beta_{3,k} = 0$ for all $k=1,\ldots,12$.
	\end{itemize}
	The choice of $\bbeta$ is made so that the first group of covariates is associated with the response, only one component of the second group is associated, and the third group is not associated. We use this $\bbeta$ to simulate the response $\y \in \mathbb{R}^{61}$ from a $N(X\bbeta, \sigma^2\I_n)$ distribution. Here, $\sigma$ is the noise and we define the signal-to-noise ratio (SNR) as $\theta/\sigma$. With $\theta = 1$, we choose different values for $\sigma$ such that the SNR $\in \{10, 1.5, 1, 0.8, 0.6, 0.4, 0.25\}$. The SNR values chosen here include a value comparable to that of the real data analysis where the SNR was close to 1. We use the covariates $X$, their grouping labels and the generated response $\y$, to identify the estimates of group indicators using BGS, and estimate the coefficients $\bbeta_{gk}$ using G-LASSO and RADIOHEAD. Tuning for the hyperparameters in BGS is performed as suggested in \cite{yang2018consistent}. For G-LASSO, the tuning parameter is selected based on minimizing the cross-validated error. We replicate the procedure $50$ times under each setting.
	
	\begin{table}[!h]
		\centering
		\begin{tabular}{crrr}
			\hline
			\textbf{SNR} & \textbf{BGS} & \textbf{G-LASSO} & \textbf{RADIOHEAD} \\ 
			\hline
			10.00 & 0.32 & 1.00 & 1.00 \\ 
			1.50 & 0.32 & 0.96 & 1.00 \\ 
			1.00 & 0.26 & 0.72 & 1.00 \\ 
			0.80 & 0.18 & 0.62 & 1.00 \\ 
			0.60 & 0.14 & 0.38 & 0.88 \\ 
			0.40 & 0.16 & 0.22 & 0.58 \\ 
			0.25 & 0.08 & 0.18 & 0.50 \\ 
			\hline
		\end{tabular}
		\caption{The proportion of times both groups were identified by BGS, G-LASSO and RADIOHEAD algorithms across $50$ replications under different choices of SNR.}
		\label{tab: group}
	\end{table}
	
	The results of this comparison are presented in Table \ref{tab: group}. The first column shows the different choices of SNR. The second column shows the proportion of times BGS selected both groups 1 and 2 across 50 replications. The third column shows the proportion of times at least one coefficient from both groups 1 and 2 was estimated to be non-zero. The fourth column shows the proportion of times RADIOHEAD estimated at least one of $\hat{\beta}_{1,k} \neq 0$ for $k=1,\ldots,9$, and $\hat{\beta}_{2,1} \neq 0$ across 50 replications. Note that the condition to assess the performance of G-LASSO is weaker than that of RADIOHEAD. From these results, we see that RADIOHEAD identifies the true associations almost always under high values of SNR ($>0.8$) and its performance decreases as SNR decreases. However, the performance of BGS lacks severely compared to RADIOHEAD in identifying both groups, even under high-values of SNR. This indicates that, under high within-group sparsity, inference on the group indicator has an inferior performance compared to RADIOHEAD. Similarly, the performance of G-LASSO is weaker for group-level inference when compared to RADIOHEAD, and decreases with increase in SNR.
	
	\section{Sensitivity to $v_0$}\label{sec: sensitivity_v0}
	Details of the prior elicitation and the hyperparameter settings are provided in Section 4 of the manuscript. Note that the choice of priors on $\nu^{-2}$ and $\sigma^{-2}$ are non-informative flat/vague priors since we chose the mean of the Gamma distribution to be 1 with an extremely large variance of $0.001/0.001^2 = 1000$. We present results of sensitivity analysis to the choice of $v_0$. We run the RADIOHEAD pipeline on the LGG data for $v_0 \in \{0.0001, 0.0005, 0.001, \allowbreak 0.005, 0.01, 0.05, 0.1\}$. For each choice of $v_0$, we obtain the estimates $\hat{\beta}_{gk}$. Let $s_{gk}$ denote the standard deviation of the estimates of $\hat{\beta}_{gk}$ across different choices of $v_0$. For each pathway, we report 
	\begin{equation*}
		\text{Mean SD} = \frac{1}{GL} \sum\limits_{g,k} s_{gk} \text{ and  Max SD} = \max\limits_{g,k} s_{gk},
	\end{equation*}
	where $G$ is the number of groups and $L$ is the total number of PCs included across all $G$ groups. In Table \ref{tab: sensitivity}, we present the results from this sensitivity analysis. Each row corresponds to a separate regression using RADIOHEAD for the pathway specified in the first column. The second and third columns show the values for Mean SD and Max SD. We see that most of these values are very close to zero. All of the remaining components (minimum, first and third quartiles) of the five-number summary of $s_{gk}$ were zero. This indicates reasonable consistency (across different choices of $v_0$) in the estimated values of $\hat{\beta}_{gk}$, which are both zero as well as non-zero.
	
	\begin{table}[!t]
		\centering
		\resizebox{\textwidth}{!}{
			\begin{tabular}{rrr}
				\hline
				\textbf{Pathway} & \textbf{Mean SD} & \textbf{Max SD} \\ 
				\hline
				SYNAPTIC\_TRANSMISSION & 0.00373 & 0.08260 \\ 
				TRANSMISSION\_OF\_NERVE\_IMPULSE & 0.00428 & 0.08426 \\ 
				MONOVALENT\_INORGANIC\_CATION\_TRANSPORT & 0.00119 & 0.09903 \\ 
				NEUROLOGICAL\_SYSTEM\_PROCESS & 0.00041 & 0.03226 \\ 
				REGULATION\_OF\_NEUROTRANSMITTER\_LEVELS & 0.00222 & 0.05841 \\ 
				POTASSIUM\_ION\_TRANSPORT & 0.00246 & 0.07810 \\ 
				METAL\_ION\_TRANSPORT & 0.00093 & 0.05012 \\ 
				GENERATION\_OF\_A\_SIGNAL\_INVOLVED\_IN\_CELL\_CELL\_SIGNALING & 0.00171 & 0.06616 \\ 
				ION\_TRANSPORT & 0.00132 & 0.07889 \\ 
				CELL\_CELL\_SIGNALING & 0.00194 & 0.05357 \\ 
				SYSTEM\_PROCESS & 0.00030 & 0.04305 \\ 
				BEHAVIOR & 0.00133 & 0.06125 \\ 
				CATION\_TRANSPORT & 0.00132 & 0.07980 \\ 
				GLUTAMATE\_SIGNALING\_PATHWAY & 0.00232 & 0.08306 \\ 
				G\_PROTEIN\_COUPLED\_RECEPTOR\_PROTEIN\_SIGNALING\_PATHWAY & 0.00205 & 0.07311 \\ 
				EXOCYTOSIS & 0.00115 & 0.07909 \\ 
				DIGESTION & 0.00031 & 0.02433 \\ 
				ANION\_TRANSPORT & 0.00078 & 0.03814 \\ 
				CENTRAL\_NERVOUS\_SYSTEM\_DEVELOPMENT & 0.00060 & 0.01872 \\ 
				NERVOUS\_SYSTEM\_DEVELOPMENT & 0.00210 & 0.05443 \\ 
				PROTEIN\_AUTOPROCESSING & 0.00146 & 0.04687 \\ 
				\hline
			\end{tabular}
		}
		\caption{Sensitivity of the estimates $\hat{\beta}_{gk}$ to the choice of $v_0$.}
		\label{tab: sensitivity}
	\end{table}
	
	\section{Robustness of PC Basis to Sample Composition}\label{sec: robustness}
	We performed sensitivity analysis to assess the effect of sample composition on the computation of PCs by using a leave-one-out approach as follows:
	\begin{itemize}
		\item[1.] Let $g^M_k(R)$ be the $k^{th}$ empirical PC basis function for tumor sub-region $R$ from imaging sequence $M$, where $k = 1,\ldots,L^M_R$. Here, $g^M_k(R)$ is the $k^{th}$ principal component from PCA.
		
		\item[2.] For each $i = 1,\ldots,n$, compute $g^M_{k,-i}(R)$, which is the $k^{th}$ PC basis function for tumor sub-region $R$ and imaging sequence $M$, while leaving out subject $i$ from the computation.
		
		\item[3.] For each $i = 1,\ldots,n$, compute the geodesic distance between $g^M_{k,-i}(R)$ and $g^M_{k}(R)$, that is, $d^{M,R}_{k,i} = \cos^{-1} \langle g^M_{k,-i}(R), g^M_{k}(R) \rangle$, where $\langle \cdot,\cdot\rangle$ denotes the dot product. Note that $g^M_{k,-i}(R)$ and  $g^M_{k}(R)$ are unit norm functions. Since they lie on the unit sphere, the arc length metric is used as the distance. 
	\end{itemize}
	
	\begin{figure}[!t]
		\centering
		\resizebox{\textwidth}{!}{
			\includegraphics{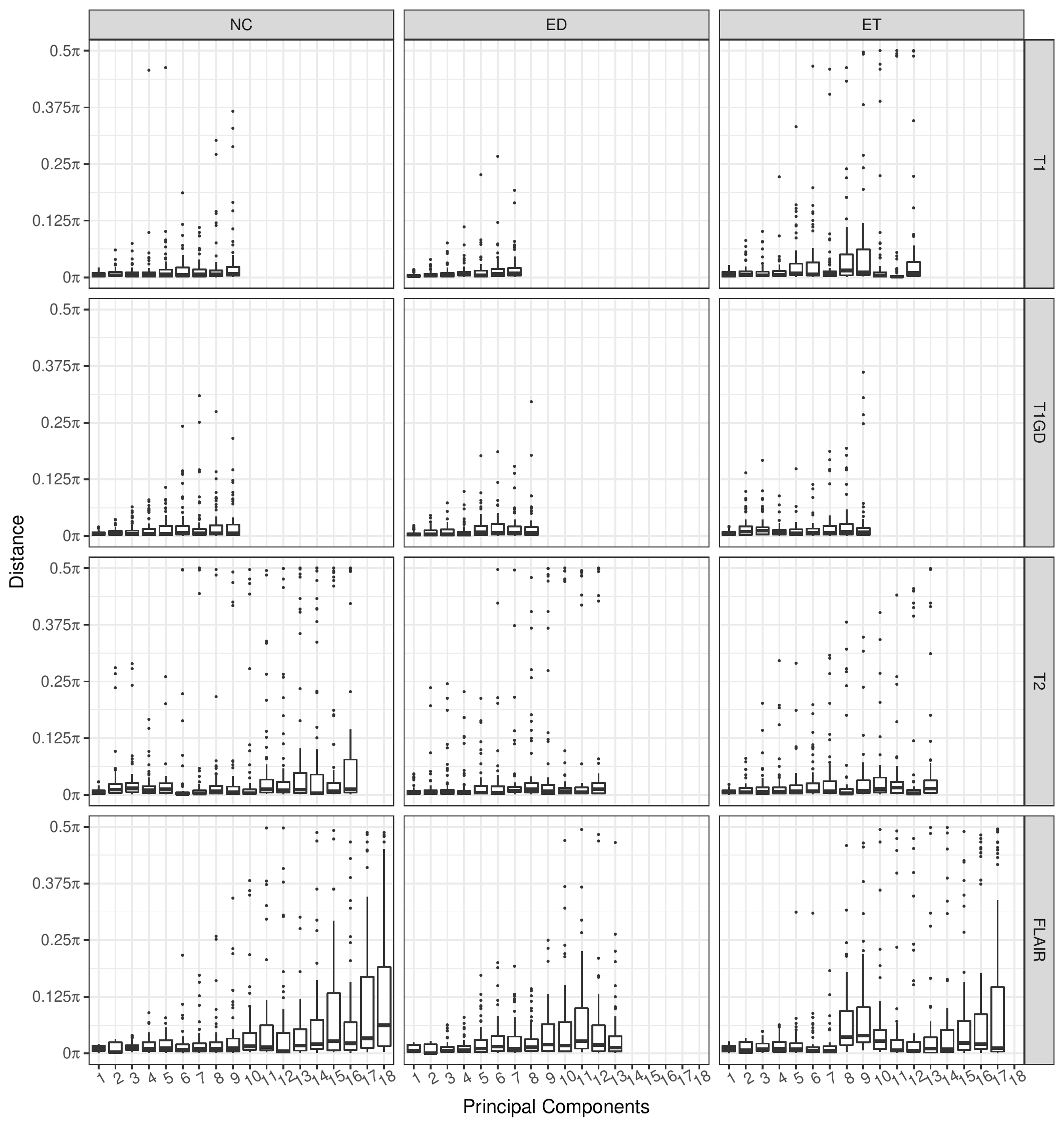}
		}
		\caption{Boxplots of the distance between the PC basis functions when they are computed (a) by pooling all 61 samples together, and (b) by a leave-one-out approach. Note that the maximum distance/angle between PC basis functions is $0.5\pi$.}
		\label{fig: cv-distance}
	\end{figure}
	
	Note that $d^{M,R}_{k,i}$ is the angle between the unit vectors (discretized version of unit norm functions) $g^M_{k,-i}(R)$ and $g^M_{k}(R)$. These basis functions do not have any specific interpretation for positive or negative direction. Hence, we re-calibrate the values of $d^{M,R}_{k,i}$ such that $d^{M,R}_{k,i} \in [0,\pi/2]$. That is, if $\cos^{-1} \langle g^M_{k,-i}(R), g^M_{k}(R)\rangle > \pi/2$, then we set $d^{M,R}_{k,i} = \pi - \cos^{-1} \langle g^M_{k,-i}(R), g^M_{k}(R)\rangle$. For each sequence $M$, each tumor sub-region $R$, and each principal direction $k$, we construct boxplots for the distance between the PC basis vectors ($d^{M,R}_{k,i}$ for all $i=1,\ldots,n$) when they are computed (a) by pooling all 61 samples together, and (b) by a leave-one-out approach. These plots are shown in Figure \ref{fig: cv-distance}. For example, the panel in the top left corresponds to the necrosis region from the T1 sequence. We see that from the pooled PC computation we have included nine PCs in the model. The distances $d^{M,R}_{k,i}$ for all $i=1,\ldots,n$ are extremely close to zero for the first seven of the nine PCs, except for a few outliers. This indicates that the first seven basis functions computed either by pooling all 61 samples together, or by a leave-one-out approach, are similar (in direction). In some cases, we observe large values for distances, which is expected for the PCs with higher value of $k$. Some of the large values, which arise as outliers in the boxplot, indicate that those outliers provide valuable variability in the sample, given the small sample size. These results indicate reasonable consistency in the estimated PC bases under a leave-one-out approach.
	
	More importantly, in our data analysis results, the transmission of nerve impulse pathway and glutamate signaling pathway have the highest magnitude estimates of $\beta_{gk}$. We see that all boxplots for the significantly associated PCs corresponding to highest magnitude coefficients (\texttt{T1\_ET.1} and \texttt{T1\_NCR/NET.3}) are very close to zero. This indicates that these basis functions are capturing similar aspects of the PDFs from both, the pooled and the leave-one-out computation of the PCA basis, and thus demonstrate reasonable robustness to sample composition.
	
	\section{Sensitivity to the Choice of Bandwidth}\label{sec: sensitivity_bandwidth}
	There are several optimal bandwidth selection approaches; however, there is no consensus on which approach works best in general scenarios \citep{zhang2019robust}. In our analysis we use Silverman's approach which is one of the most commonly used approaches that is optimal for normal densities \citep{silverman1986density}. We have performed a sensitivity analysis to assess the differences in the density estimates based on the choice of bandwidth. For comparison we consider two additional bandwidth selection methods (denoted as Scott and BCV) based on two variations proposed by \cite{scott1992multivariate}. Among the various bandwidth selection approaches, these three approaches (Silverman, Scott and BCV) provide reasonably similar estimates of density estimate for our data as shown next. Let $f_i^0, f_i^1$ and $f_i^2$ denote the density estimates for subject $i$ computed with the choice of bandwidth given by Silverman's, Scott and BCV approaches, respectively. Let us denote the geodesic distance between (a) $f_i^0$ and $f_i^1$ as $d_i^1$ and (b) $f_i^0$ and $f_i^2$ as $d_i^2$. Here $d_i^1$ and $d_i^2$ quantify the dissimilarity between the density estimates, that is, Silverman vs Scott and Silverman vs BCV, respectively. Note that these geodesic distances are bounded above by $\pi/2 \approx 1.571$. In Table \ref{tab: bandwidth_distance} below, we present the mean and standard deviations of $d_i^1$ and $d_i^2$ across all of the subjects for each combination of imaging sequence and tumor sub-region. The values of average distances close to zero indicate reasonable consistency in the density estimates computed using the three commonly used bandwidth selection approaches. Hence, inferences will be quite close as well since the PDF is incorporated in the model through the inverse-exponential map at the Karcher mean, which is defined using the geodesic path between the Karcher mean and the PDF under question.
	
	\begin{table}[!t]
		\centering
		\begin{tabular}{lccc}
			\hline
			\textbf{Sequence} & \textbf{Region} & \textbf{Silverman vs Scott} & \textbf{Silverman vs BCV} \\ \hline
			T1 & NC & 0.0069 (0.0100) & 0.0131 (0.0183) \\
			T1 & ED & 0.0034 (0.0036) & 0.0073 (0.0103) \\
			T1 & ET & 0.0110 (0.0218) & 0.0260 (0.0522) \\
			T1Gd & NC & 0.0069 (0.0148) & 0.0133 (0.0209) \\
			T1Gd & ED & 0.0029 (0.0066) & 0.0093 (0.0164) \\
			T1Gd & ET & 0.0149 (0.0186) & 0.0437 (0.0493) \\
			T2 & NC & 0.0584 (0.1993) & 0.0641 (0.1870) \\
			T2 & ED & 0.0317 (0.1179) & 0.0283 (0.0728) \\
			T2 & ET & 0.0315 (0.1054) & 0.0491 (0.1298) \\
			FLAIR & NC & 0.0971 (0.2136) & 0.1018 (0.1859) \\
			FLAIR & ED & 0.0693 (0.1470) & 0.0891 (0.1788) \\
			FLAIR & ET & 0.0990 (0.2119) & 0.1269 (0.2211) \\ \hline
		\end{tabular}
		\caption{Average distance between the density estimates computed using the Silverman's optimal bandwidth versus the Scott's approach and BCV approach.}
		\label{tab: bandwidth_distance}
	\end{table}



\end{document}